\documentclass[lettersize,journal]{IEEEtran}
\usepackage[mathscr]{euscript}
\usepackage{color}
\usepackage{optidef}
\usepackage[algo2e,ruled,vlined]{algorithm2e}
\usepackage{float}
\usepackage{graphicx}
\usepackage{tikz,pgfplots}
\usepackage{upgreek}
\usepackage{multirow}
\usepackage{yfonts}
\usepackage{mathtools}
\usepackage[normalem]{ulem}
\usepackage{latexsym}
\usepackage{url}
\usepackage{amsmath,amssymb,amsbsy,amsfonts}
\usepackage{mathrsfs}
\usepackage{bbm}
\usepackage[export]{adjustbox}

\newcommand\restr[2]{{
  \left.\kern-\nulldelimiterspace 
  {#1}\vphantom{\big|} \right|_{#2}}}

\DeclareMathOperator*{\argmin}{argmin} 

\newcommand{\R}{\mathbb{R}}

\newcommand{\U}{\mathbb{U}}

\newcommand{\D}{\mathcal{D}}

\newcommand{\J}{\mathcal{J}}

\newcommand{\Reg}{\mathcal{R}}

\newcommand{\ipar}{f}

\newcommand{\noise}{\bs e}
\newcommand{\data}{ \bs d }

\newcommand{\bs}[1]{\ensuremath{\boldsymbol{#1}}}

\usepackage{xspace}

\newcommand{\LI}{\mathcal{L}}
\renewcommand{\H}{\mathcal{H}}

\usepackage{cite}
\usepackage{amsmath,amssymb,amsfonts}
\usepackage{graphicx}
\usepackage{textcomp}
\def\BibTeX{{\rm B\kern-.05em{\sc i\kern-.025em b}\kern-.08em
    T\kern-.1667em\lower.7ex\hbox{E}\kern-.125emX}}
\usepackage[math]{cellspace}
\usepackage{graphicx}
\usepackage{algorithm}
\usepackage{algpseudocode}

\usepackage{float}
\usepackage{enumerate}
\usepackage{url}
\usepackage{multirow}

\usepackage{wrapfig,lipsum,booktabs}

\newcommand{\bbr}{\mathbf{r}}

\pgfplotsset{compat = 1.17}
\begin{document} 
\title{A Memory-Efficient Dynamic Image Reconstruction Method using Neural Fields}

\author{Luke Lozenski, \IEEEmembership{Student Member, IEEE}, Mark A. Anastasio, \IEEEmembership{Senior Member, IEEE},\\and Umberto Villa, \IEEEmembership{Member, IEEE}

\thanks{L. Lozenski and U.Villa are with the Department of Electrical and Systems Engineering,
 Washington University in St$.\,$Louis, St$.\,$Louis, MO 63130, USA. e-mail: \texttt{ljlozenski@wustl.edu, uvilla@wustl.edu}.}%
 \thanks{M. Anastasio is with the Department
of Bioengineering, University of Illinois at Urbana-Champaign, Urbana,
IL, 61801 USA. e-mail: \texttt{maa@illinois.edu}.}%
}



\maketitle
\vspace{-.11cm}
\begin{abstract}
Dynamic imaging is essential for analyzing various biological systems and behaviors but faces two main challenges: data incompleteness and computational burden. For many imaging systems, high frame rates and short acquisition times require severe undersampling, which leads to data incompleteness. Multiple images may then be compatible with the data, thus requiring special techniques (regularization) to ensure the uniqueness of the reconstruction. Computational and memory requirements are particularly burdensome for three-dimensional dynamic imaging applications requiring high resolution in both space and time.
Exploiting redundancies in the object's spatiotemporal features is key to addressing both challenges.
This contribution investigates neural fields, or implicit neural representations, to model the sought-after dynamic object. Neural fields are a particular class of neural networks that represent the dynamic object as a continuous function of space and time, thus avoiding the burden of storing a full resolution image at each time frame. Neural field representation thus reduces the image reconstruction problem to estimating the network parameters via a nonlinear optimization problem (training). Once trained, the neural field can be evaluated at arbitrary locations in space and time, allowing for high-resolution rendering of the object. Key advantages of the proposed approach are that neural fields automatically learn and exploit redundancies in the sought-after object to both regularize the reconstruction and significantly reduce memory storage requirements. The feasibility of the proposed framework is illustrated with an application to dynamic image reconstruction from severely undersampled circular Radon transform data.

\end{abstract}



\section{Introduction}

Many biomedical imaging applications, ranging from functional imaging of large-scale organs to molecular imaging for tracking the diffusion of a contrast agent, seek to capture time-varying features of the target. There is a trade-off between temporal resolution and the number of measurements taken at each time frame for several imaging modalities \cite{LuisRazansky14, Liang2007, Plenge12, Mickan00, Liu13}. In such cases, data incompleteness means that reconstructions are not unique and highly sensitive to measurement noise. Furthermore, dynamic image reconstruction incurs a significant computational burden. For dynamic imaging of three-dimensional objects, na\"ive reconstruction approaches, which discretize the sought-after image as a collection of space-time voxel values, can quickly exceed current hardware memory capabilities when high resolution in space and time is required.

Therefore, an effective dynamic image reconstruction strategy must exploit redundancies in the object's features \cite{zontak2011internal} to address these twin challenges of incomplete data and computational burden. Several dynamic image reconstruction methods invoke a space-time semiseparable approximation of the sought-after image\cite{Liang2007, Lingalaetal11, Zhaoetal12, WangVillaThompsonEtAl21}. Under the semiseparable approximation,  a small set of functions represent the dynamic image. Algebraically, this is equivalent to assuming a low-rank matrix structure whose columns represent the time-snapshot of the dynamic object. However, this is often not the case, and many basis functions are required to capture dynamics accurately.

This contribution explores a class of deep learning methods, neural fields, to represent a dynamic object as a continuous function of space and time. The dynamic image reconstruction problem then reduces to estimating a finite number of network parameters by solving a nonconvex optimization problem\cite{choromanska2015loss}. Unlike other deep learning reconstruction methods, the proposed method is self-supervised and does not require a dataset of ground truth images for training. Furthermore, neural fields allow for memory-efficient representation of the object \cite{MartelLindellLinetal2021}  and provide implicit regularization due to the class of functions that can be approximated \cite{SitzmannMartelBergmanetal2020}. Neural field approaches have also been investigated in the specific area of limited-view computed tomography problems to infer missing data\cite{SunLiuXieEtAl2021, SidkyKaoPan2006},  to describe the mechanical deformation of a dynamic object\cite{ReedKimAnirudhetal2021}, and learning static 3D, complex-valued refractive indexes \cite{LiuSunetal2021}. 

Three numerical experiments illustrate the approach's feasibility. The first experiment investigates the approximation properties (representation power in computer vision's parlance) of neural fields for anatomically realistic dynamical objects. The second and third studies are inspired by photoacoustic tomography imaging of small animals \cite{WangXiaetal14}. The forward operator in these studies is a severely undersampled circular Radon transform \cite{ReddinNewsam2001, AmbartsoumianKuchment06 }, which plays a role in several imaging modalities such as photoacoustic tomography image reconstruction and synthetic aperture radar \cite{Poudel_2019, HALTMEIERetal07}. One of these experiments analyzes the performance of neural field image reconstruction with varying levels of data completeness; the other explores the sensitivity to noise. These experiments include comparisons in terms of the accuracy and memory requirements of the proposed approach to those of two classical reconstruction methods, in which the sought-after image is discretized using space-time pixel values. In these studies, a two-dimensional slice was extracted from a dynamic three-dimensional extended cardiac-torso murine numerical phantom~\cite{SegarsTsuiFreyEtAl04,SegarsTsui2009}.

Neural fields have been implemented with various neural architectures incorporating a series of nonlinear activation functions and trainable weights. However, traditional multi-layer perceptron (MLP) architectures lead to poor quality, low-resolution representations and are unable to capture high-frequency components\cite{Rahamanetal19}. Positional encoding is often applied to address this issue and uses a fixed (i.e. non-trainable) nonlinear transformation to increase the dimension of the input space of the neural field. For spatiotemporal images, a common practice is to use Fourier feature encoding, where the fixed nonlinear transformation is a set of sinusoidal functions with different frequencies \cite{TancikPratuletal20}. Examples of architectures applying positional encoding are in  Sun et al. \cite{SunLiuXieEtAl2021}, which applied an architecture with a series of skip connections on Fourier encoded inputs, and in spatially adaptive progressive encoding (SAPE) presented in Hertz et al. \cite{hertz2021sape}, which gradually unmasks signal components with increasing frequencies as a function of time and space. To represent images in the gigapixel range, Martell et al. \cite{MartelLindellLinetal2021} proposes a neural field architecture with positional encoding that implements an adaptive multiscale octree to divide the image into regions of fine and coarse detail. Sitzmann et al. \cite{SitzmannMartelBergmanetal2020} develops neural fields with sinusoidal activations as an alternative to Fourier encoding, noting that Fourier encoding is equivalent to a single layer hidden layer with sinusoidal activations\cite{XieTakikawaSaitoEtAl21}.

This paper develops the partition of unity network (POUnet) \cite{LeeTraskPateletal21} for solving inverse problems. The specific contributions of this paper are: the first implementation of the POUnet for image reconstruction, the proposal of a sparsity enforcing term in the POUnet training process, the implementation of the POUnet with sinusoidal activation functions \cite{SitzmannMartelBergmanetal2020}, and the introduction of grid-free continuous-to-discrete imaging operators for the circular Radon transform using neural fields. Finally, The approach and methods presented in this work can be easily extended to the case of other linear imaging operators beyond the circular Radon transform.

The remainder of this paper is structured as follows. In Section \ref{sec:background}, the dynamic image reconstruction problem is formulated as an optimization problem and classical solution approaches based on a discretization of the dynamic object as a linear combination of spatiotemporal basis functions are surveyed. In Section \ref{sec:method}, a solution method for the dynamic image problem is presented that relies on training a neural field. In Section \ref{sec:numerical_studies}, the design of three numerical studies is presented: the first is tests the representation power of a neural field compared to semiseparable approximations; the second study compares the results of image reconstruction problems with a varying number of views per frames using a neural field approach and two classical approaches; the third study compares the results of image reconstruction problems with varying noise levels using a neural field approach and two classical approaches. In Section \ref{sec:results}, the results of these three numerical studies are presented. Section \ref{sec:conclusion} presents the conclusions drawn from these results and future extensions.

\section{Background}
\label{sec:background}
\subsection{Dynamic Image Reconstruction}
Image reconstruction aims at estimating an object $\ipar$ given a set of possibly noisy and incomplete measurements $\data$. The object $\ipar$ and the measurements $\data$ are associated via a continuous-to-discrete imaging operator $\H:\U \rightarrow \D$
\begin{equation}\label{eqn:imaging_eq}
    \data = \H(\ipar) + \noise,
\end{equation}
where $\noise$ represents additive noise. The imaging operator $\H$ is a, possibly nonlinear, mapping from an infinite-dimensional Hilbert space $\U$, representing the object domain, to a finite-dimensional Euclidean space $\D$, representing the data domain \cite{scherzer2009variational}. In many cases, $\H$ is not injective, i.e. multiple distinct objects can lead to identical measurements. Lack of injectivity means there is no direct inverse for $\H$ and nonunique solutions to the image reconstruction problem. Furthermore, the presence of noise $\noise$ implies that measurements will not exactly correspond to the underlying image. In many cases, $\H$ may be sensitive to noise, and small changes in the data can lead to significant changes in the reconstructed image. Nonuniqueness and sensitivity to noise mean that reconstructing $\ipar$ is an ill-posed problem.

In a dynamic imaging problem the sought-after image can be described as a scalar function $\ipar(\mathbf{x})$. The input variable $\mathbf{x} = (\bbr, t)$ represents the spatiotemporal coordinates of a point lying in a domain  $\Omega_T = \Omega \times [0,T]$ composed of a field of view $\Omega \subset \mathbb{R}^d$ (d=2,3) and an image acquisition time $T$. At each imaging frame $k$ the object $\ipar$ and measurement data $\data_k$ are associated via relationship

\begin{equation*}
    \data_k = \H_k(\ipar(\cdot, t_k)) + \noise_k, \quad k=1, \ldots, K,
\end{equation*}
where $K$ denotes the total number of imaging frames, $\ipar(\cdot, t_k)$ denotes the object at the time $t_k \in [0, T]$ when the $k$-th frame was acquired, $\H_k$ denotes the imaging operator associated with frame $k$, and $\noise_k$ represents measurement noise. For several applications of practical relevance \cite{Liang2007}, $\H_k$ is different for each frame and severely undersampled.

Setting
\begin{equation*}
        \data  = \begin{pmatrix} \data_1 \\ \vdots \\ \data_K \end{pmatrix}, \, 
        \H(\ipar)  =  \begin{pmatrix} \H_1(\ipar(\cdot, t_1)) \\ \vdots \\ \H_K(\ipar(\cdot, t_K)) \end{pmatrix}, \,
        \noise  = \begin{pmatrix} \noise_1 \\ \vdots \\ \noise_K \end{pmatrix}
\end{equation*}
recovers the imaging relationship in Eq. \eqref{eqn:imaging_eq}.

Directly inverting the imaging relationship in Eq. \eqref{eqn:imaging_eq} is generally severely ill-posed due to data incompleteness. In these cases, a unique reconstruction can be found through the solution of a penalized least squares problem of the form
 \begin{equation}\label{eqn:Objective}
    \min_{\ipar} \J(\ipar) := \LI(\ipar)  + \Reg(\ipar),
 \end{equation}
where 
\begin{equation*}
    \LI(\ipar) = \frac{1}{2\sigma^2} \| \H(\ipar) - \data\|^2 = \frac{1}{2\sigma^2} \sum_{k=1}^K \| \H_k(\ipar(\cdot, t_k)) - \data_k\|^2
\end{equation*} denotes the data fidelity term and $\Reg(\ipar)$ is a convex penalization functional, such as Tikhonov \cite{TikhonovArsenin77} or total variation regularization\cite{RudinOsherFatemi92}, used to ensure uniqueness of the solution and is not necessarily smooth. Above, the weight $\sigma^2$ associated to data fidelity fidelity represent the variance of the additive noise $\boldsymbol{e}$.

\subsection*{Classical Dynamic Reconstruction Methods}

Dynamic imaging problems are classically formulated in a discrete form as a large-scale optimization problem. The continuous object $\ipar$ is approximated with a linear combination of a finite number of basis functions $\{\beta_n(\mathbf{r})\}_{n=1}^N$ of the form

\begin{equation*}
    \ipar_{N,K} (\mathbf{r}, t_k) = \sum_{n=1}^N \alpha_{n}^k \beta_n(\mathbf{r}) \quad k=1,\ldots,K.
\end{equation*}
The coefficients $\alpha_n^k$ can be collected into an $N\times K$ matrix representation
\begin{equation}\label{eq:DiscreteParam}
    \boldsymbol{F} = \begin{pmatrix} \alpha_1^1 & \hdots & \alpha_1^K \\ 
    \vdots & \ddots & \vdots \\ 
    \alpha_N^1 & \hdots & \alpha_N^K \end{pmatrix},
\end{equation}
where the $k$-th column $\boldsymbol{F}$ represents the dynamic object at a time frame $t_k$, and the $n$-th row is the time activity associated with the basis function $\beta_n(\mathbf{r})$.

 Applying this discrete representation then transforms the imaging operator into a discrete-to-discrete operator $\mathbf{H}:\R^{N\times K} \rightarrow \mathcal{D}$. The imaging equation is then expressed in a discrete form 

\begin{equation*}
    \data = \mathbf{H}(\boldsymbol{F}) + \noise = \H(\ipar_{N,K}) + \noise.
\end{equation*}
This discrete representation of the imaging equation leads to discrete representations of the data fidelity, regularization and cost functionals. The image reconstruction problem can be formulated as an $N\times K$-dimensional optimization problem, 
 
\begin{equation}\label{eqn:classical_approach}
  \min_{\boldsymbol{F} \in \R^{N \times K}} \mathbf{J}(\boldsymbol{F})=\mathbf{L}(\boldsymbol{F}) + \mathbf{R}(\boldsymbol{F}), 
\end{equation}
where discrete data fidelity and regularization terms are respectively given by 
$$\mathbf{L}(\boldsymbol{F}) = \frac{1}{2\sigma^2}||\mathbf{H}(\boldsymbol{F}) - \data||^2$$ 
and 
$$\mathbf{R}(\boldsymbol{F}) = \Reg(\ipar_{N,K}).$$

Having a large $N$ and $K$ is necessary to ensure the discrete image $\boldsymbol{F}$ is a high fidelity representation of the continuous object $\ipar$. When $N$ and $K$ are large, this becomes an optimization problem over a high-dimensional space. In such cases, storing $\boldsymbol{F}$ may exceed memory allocations, or the optimization process may become intractable. 
Two classical regularization approaches for dynamic imaging, spatiotemporal total variation and nuclear norm regularization, are surveyed below and will be used in Section \ref{sec:results} to assess image quality and memory requirements of the proposed method. 

\paragraph{Total Variation Regularization}

Total variation is useful for image reconstruction because of its ability to preserve sharp edges in the reconstructed object estimates \cite{RudinOsherFatemi92}. This approach then used the regularization 

\begin{equation}\label{eqn:total_variaiton}
    \mathbf{R}(\boldsymbol{F}) = \gamma \operatorname{TV}_{D}( \boldsymbol{F}),
\end{equation} where $\gamma > 0$ is a regularization weight. Total variation regularization is also notably nonsmooth and not everywhere differentiable.

\paragraph{Nuclear Norm Regularization}

The nuclear norm is a form of regularization used to reduce the number of necessary parameters for representing a spatiotemporal image or video signal \cite{Gu_2014_CVPR}. It exploits space-time redundancies in the objects and promotes space-time semiseparability by penalizing the rank of the discrete representation $\boldsymbol F$ of the dynamic object. Furthermore, by promoting this low-rank structure, nuclear norm regularization effectively promotes dynamic objects with fewer parameters, thus reducing the computational burden and storage requirements.

Specifically, the nuclear norm regularization \cite{Gu_2014_CVPR} is given by

\begin{equation}\label{eqn:nuclear_norm}
    \mathbf{R}(\boldsymbol{F}) = \gamma \operatorname{Nuc}(\boldsymbol F)=\gamma\sum_{i=1}^{\min(N,K)} \sigma_i,
\end{equation}
where $\gamma > 0$ is the regularization weight and $\{\sigma_i\}_{i=1}^{\min(N,K)}$ denotes the set of singular values of $\bs{F}$.

 Since total variation and the nuclear norm are not everywhere differentiable, the optimization problem in  Eq. \eqref{eqn:classical_approach} cannot be solved with gradient-based methods. An alternative to the gradient operator for nondifferentiable functions is the proximal operator. The proximal operator shares many common characteristics with the projection operators and gives rise to proximal splitting methods \cite{Combettes2011}. This problem is well suited for proximal splitting methods because $\mathbf{L}$ is differentiable and $\mathbf{R}$ is convex.

\section{Method}
\label{sec:method}

This section presents the main contribution of this work: the development of an alternative to classical discretization-based approaches to dynamic image reconstruction that is feasible for high-dimension dynamic imaging problems. The proposed method represents the sought-after dynamic object via a neural field. Neural fields, also often referred to as implicit neural representations, are a novel non-parametric approach to describe continuous objects, such as images, videos, and audio signals. Neural fields have found success in many applications, including image generalization, hybrid representation, image representation, and volumetric rendering \cite{XieTakikawaSaitoEtAl21}.

\subsection*{Neural Fields}

A neural field is a neural network that maps the coordinates of an arbitrary point $\mathbf{x}_i \in  \Omega_T$ to a numerical value $\ipar_i = \ipar(\mathbf{x}_i)$ representing the object at that point. In what follows, $\Phi_{\boldsymbol \xi}$ denotes the neural field with trainable parameters $\boldsymbol \xi$. Given training pairs $\{(\mathbf{x}_i, \ipar_i)\}_{i=1}^I$, the parameters $\boldsymbol \xi$ can be estimated by minimizing the distance between the output of the network $\Phi_{\boldsymbol \xi}(\mathbf{x}_i)$ and the data $\{\ipar_i\}_i$.
That is, the the parameters $\boldsymbol \xi$ are found by solving the so-called embedding problem
\begin{equation}\label{eqn:embedding}
    \min_{\boldsymbol \xi}  
    \frac{1}{2} \sum_{i=1}^I |\Phi_{\boldsymbol \xi}(\mathbf{x}_i) - \ipar_i|^2.
\end{equation}

Once trained, the neural field can be evaluated at arbitrary points in $\Omega_T$,  allowing the object to be rendered or volumetrically represented at arbitrary resolution. Furthermore, this allows an approximation of the object's spatial and temporal gradients because the neural field $\Phi_{\boldsymbol \xi}(\mathbf{x})$  can be differentiated with respect to its input $\mathbf{x}$ \cite{SitzmannMartelBergmanetal2020}.

\subsection*{Dynamic Image Reconstruction Using Neural Fields}
In practice, pointwise evaluations $\ipar_i$ of the object are seldom available, instead only indirect and possibly noisy measurement $\mathbf{d}$ can be collected via an imaging operator as in Eq. \eqref{eqn:imaging_eq}. Replacing the spatiotemporal object $\ipar$ in Eq. \eqref{eqn:Objective} with the neural representation $\Phi_{\boldsymbol \xi}$ leads to an optimization problem involving the parameters $\boldsymbol \xi$ of the form
\begin{equation}\label{eqn:INR_obj}
    \min_{\boldsymbol \xi}  \J(\Phi_{\boldsymbol \xi}) = 
    \frac{1}{2\sigma^2} \| \H(\Phi_{\boldsymbol \xi}) - \data\|^2 + \Reg(\Phi_{\boldsymbol \xi}).
\end{equation}
For a fixed network architecture and a given sought-after object $\ipar(\mathbf{x})$, the results of the embedding problem in Eq. \eqref{eqn:embedding} can be used as a lower bound for the lowest possible error for the neural field image reconstruction in Eq. \eqref{eqn:INR_obj}.

Key advantages of solving Eq. \eqref{eqn:INR_obj} compared to the classical approach in Eq. \eqref{eqn:classical_approach} are 1) reduced memory requirements for optimization and 2) convenient and computationally efficient evaluation of the cost functional and its derivatives using automatic differentiation. In fact, neural fields can provide high-resolution rendering and capture fine features of the object using a small number of trainable parameters\cite{MartelLindellLinetal2021}, thus overcoming the memory requirement challenges posed by dynamic imaging. Also, solving Eq. \eqref{eqn:INR_obj} no longer requires an explicit discretization of the imaging operator $\H$ or regularization functional $\Reg$. With a neural field, spatial and temporal derivatives of the object can be computed exactly using automatic differentiation, and Monte Carlo approximations can be used to evaluate integrals. For example, the total variation functional used as a regularization term for the numerical results in Section \ref{sec:CRT_recons} can be readily evaluated as follows
\begin{equation}\label{eqn:stochastic_tv}
     \operatorname{TV}_\mathcal{Y}(\Phi_{\boldsymbol \xi}) = \frac{1}{|\mathcal{Y}|} \sum_{\mathbf{y} \in \mathcal{Y}} || \nabla \Phi_{\boldsymbol \xi}(\mathbf{y})||,
\end{equation}
where $\mathcal{Y} = \{ \mathbf{y}_i \}_{i=1}^S$ is a set of $|\mathcal{Y}|=S$ randomly sampled uniformly distributed points $\mathbf{y}_i \in \Omega_T$. The set $\mathcal{Y}$ is sampled at each call of the functional.

\subsection*{Partition of Unity Network}

A possible challenge in solving Eq. \eqref{eqn:INR_obj} is that, even in cases in which the imaging operator $\H$ is linear, the objective function in Eq. \eqref{eqn:INR_obj} is not guaranteed to be convex, thus requiring special techniques to ensure convergence to a global minimum. 

In this work, a particular neural field architecture,  the partition of unity network (POUnet)\cite{LeeTraskPateletal21} is used to address this issue and ensure robust training. The POUnet relies on an adaptive basis \cite{CyrGulianPateletal2019} that forms a partition of unity (POU). A POU is a function $\boldsymbol \Psi: \Omega_T \rightarrow \R^P, $ $ \boldsymbol \Psi(\mathbf{x}) = (\boldsymbol \Psi^1(\mathbf{x}), \hdots , \boldsymbol \Psi^P(\mathbf{x}) )^T$ such that for every $\mathbf{x} \in \Omega_T$

\begin{equation}
\label{eq:define_POU}
    \boldsymbol \Psi(\mathbf{x})^T\mathbbm{1} = \sum_{j=1}^P\boldsymbol \Psi^j(\mathbf{x}) = 1, \boldsymbol \Psi^j(\mathbf{x}) \geq 0.
\end{equation}

Partitions of unity are useful for constructing and approximating functions that are piecewise continuous and piecewise differentiable. Such functions can be written in the form

\begin{equation*}
    f(\mathbf{x}) = \boldsymbol \Psi(\mathbf{x})^T\boldsymbol C\boldsymbol B(\mathbf{x}),
\end{equation*}
where $\boldsymbol C \in \R^{P \times M}$ is a matrix of interpolation coefficients and $\boldsymbol B:\Omega_T \rightarrow \R^M$ is a set of globally defined basis functions, such as polynomial or sinusoidal functions. Furthermore, approximations of this form can be constructed to arbitrary accuracy dependent on the span of functions used and support of the partitions, see Theorem 1 from Lee et al. \cite{LeeTraskPateletal21}.

The POUnet uses a neural network $\boldsymbol \Psi = \boldsymbol \Psi_{\boldsymbol \eta}$ with training parameters $\boldsymbol \eta$ to define an image adaptive partition of unity, leading to the neural field representation  
\begin{equation*}
    \Phi_{\boldsymbol \xi}(\mathbf{x}) = \boldsymbol \Psi_{\boldsymbol \eta}(\mathbf{x})^T \boldsymbol C \boldsymbol B(\mathbf{x}),
\end{equation*}
where $\boldsymbol \xi = (\boldsymbol \eta ,\boldsymbol C)$ denotes the trainable parameters. 

A block coordinate descent optimizer was proposed in \cite{PatelTraskGulianCyr2020} to ensure robust training of the POUnet when solving the embedding problem in Eq. \eqref{eqn:embedding}. It uses Adam optimizer \cite{KingmaBa2014} to alternatively minimize Eq. \eqref{eqn:embedding} with respect to $\boldsymbol \eta$ and $\boldsymbol C$.  The alternating optimization strategy is advantageous because, thanks to the linearity of the POUnet with respect to $\boldsymbol C$, the embedding problem in Eq. \eqref{eqn:embedding} is strictly convex with respect to $\boldsymbol C$. Thus, the optimal coefficient $\boldsymbol C$ can be found for a fixed partition of unity. 

This work generalizes the training of the POUnet to problems other than the embedding problems and proposes an improved formulation of the block coordinate descent optimizer in \cite{PatelTraskGulianCyr2020} that is attuned to solving dynamic image reconstruction problems. In the proposed method the objective function $\J(\Phi_{(\boldsymbol \eta, \boldsymbol C)})$ in Eq. \eqref{eqn:INR_obj} is augmented with two additional regularization terms acting on the POUnet trainable parameters $\boldsymbol \xi =(\boldsymbol \eta, C)$ and reads
\begin{equation*}
   \J_{\rho,\gamma}(\Phi_{(\boldsymbol \eta,  \boldsymbol C)}) =  \J(\Phi_{(\boldsymbol \eta,  \boldsymbol C)}) + \frac{\rho}{2}\| \boldsymbol  C\|_F^2 + \tau \|  \boldsymbol \Psi_{\boldsymbol \eta} \|_{q,\epsilon}^q.
\end{equation*} Above $\| \boldsymbol  C\|_F$ denotes the Frobenious norm of the matrix $C$ and $\|  \boldsymbol \Psi_{\boldsymbol \eta} \|_{q,\epsilon} = (\int_{\Omega_T} \sum_{p=1}^P (\boldsymbol \Psi^p_{\boldsymbol \eta}(\mathbf{x}) + \epsilon)^{q} d\mathbf x)^{1/q}$ denotes the smoothed $q$-norm ($q \in (0,1)$) of the POU $\boldsymbol \Psi_{\boldsymbol \eta}$, where $\epsilon \ll 1$ ensures differentiability of $\|  \boldsymbol \Psi_{\boldsymbol \eta} \|_{q,\epsilon}^q$.
 The regularization weights $\rho$ and $\tau$ are steadily decreased after each outer iteration and eventually set to zero to recover the original optimization problem in Eq. \eqref{eqn:INR_obj}.

The regularization term $\|\boldsymbol {C}\|_F^2$ was proposed in \cite{LeeTraskPateletal21} to prevent overfitting in early iterations and distribute coefficients evenly among partitions. This work introduces the regularization term with $\|  \boldsymbol \Psi_{\boldsymbol \eta} \|_{q,\epsilon}$ to promote sparsity in the partitions \cite{xuetal12} and minimize the overlap among the partitions' supports. Note that to promote sparsity $q$ must be chosen strictly less than $1$. In fact, $||\boldsymbol \Psi_{\boldsymbol \eta}(\mathbf{x})||_{1, 0} = \Omega_T$ for any choice of $\boldsymbol \eta$ due to the definition of POU in Eq. \eqref{eq:define_POU}.

The proposed method for solving a dynamic image reconstruction problem using the POUnet is provided in Algorithm \ref{alg:inr_recon}. In the proposed implementation, the inner problems in Eq.s \eqref{eq:AdamUpdateC}-\eqref{eq:AdamUpdateEta} are solved using the Adam optimizer and a randomized estimator of the data fidelity term $\mathcal{L}(f)$. In particular, at each Adam update, the randomized approximation
$$ \mathcal{L}_{ \mathcal{K}_b}(f) = \frac{1}{2\sigma^2}\frac{K}{|\mathcal{K}_b|}\sum_{k \in \mathcal{K}_b} \| \mathcal{H}_k(f(\cdot, t_k)) - \data_k\|^2$$
of the data fidelity term is used to estimate the gradient direction. Above $\{\mathcal{K}_b\}_{b=1}^{n_b}$ denotes a partition of the set $\mathcal{K}=\{k=1,\cdots,K\}$ of all imaging frames. The number of elements $|\mathcal{K}_b|$ in each subset $\mathcal{K}_b$ is approximately constant and equal to $K/n_b$. The subsets $\{\mathcal{K}_b\}_{b=1}^{n_b}$ are kept fixed and cyclically selected selected during the solution of the inner optimization problems for $\boldsymbol C$ and $\boldsymbol{\eta}$.

\begin{algorithm2e}[t]
\SetAlgoLined
Given measurements $\data$.\\
Randomly initialize coefficients $\boldsymbol{C}^0$.\\
Randomly initialize partition of unity with weights $\boldsymbol \eta^0$

 \For{$i =1,\hdots, {\rm outer\_max\_iter}$}{
    Update coefficients for fixed partition by solving 
    
    \begin{equation}
    \label{eq:AdamUpdateC}   
    {\boldsymbol C}^{i} = \argmin_{{\boldsymbol C}} \J(\Phi_{({\boldsymbol C},\boldsymbol \eta^{i-1})}) + \frac{\rho}{2}\|{\boldsymbol C}\|_F^2.
    \end{equation}
    
    Update partition of unity for fixed coefficients by solving
    
    \begin{equation}
    \label{eq:AdamUpdateEta} 
    \boldsymbol \eta^{i} = \argmin_{\boldsymbol \eta} \J(\Phi_{({\boldsymbol C}^i, \boldsymbol \eta)}) + \tau \|  \boldsymbol \Psi_{\boldsymbol \eta} \|_{q,\epsilon}^q
    \end{equation}

    Decrease $\rho$ and $\tau$, or set to zero after a sufficient number of iterations.
}
\caption{Algorithm for neural field estimation of dynamic image}
\label{alg:inr_recon}
\end{algorithm2e}

\section{Numerical Studies}\label{sec:numerical_studies}

The numerical studies used to assess the proposed method utilized a stylized imaging operator in two spatial dimensions inspired by dynamic photoacoustic computed tomography (PACT) imaging of small animals \cite{MatthewsAnastasio2017}. PACT is a hybrid modality that combines endogenous contrast of optical imaging with the high-resolution of ultrasound detection technologies to provide maps of total hemoglobin content and oxygen saturation within tissue \cite{WangYao16}. The process begins with a short laser pulse in the infrared range illuminating the object of interest. The underlying material then absorbs this optical energy and generates heat, resulting in a local increase in pressure. This pressure distribution then propagates through the object, and pressure waveform data are recorded over a measuring surface. Assuming that the object is acoustically homogeneous and non-attenuating, the relationship between waveform measurements and the initial pressure distribution can be mathematically modeled using the circular Radon transform \cite{WangAnastasio15}. Image reconstruction from PACT data can be viewed as a two-step procedure. The first step involves reconstructing the initial pressure distribution given measurements on the boundary of the domain. The second step involves reconstructing the optical properties of the tissue based on the initial pressure distribution. The numerical studies presented in this work focus on the first step, that is to reconstruct the initial pressure distribution based on measurements collected on a circular aperture surrounding the object.  

\subsection{Construction of the Dynamic Object}

\begin{figure*}
    \centering
\includegraphics[width = 0.95 \textwidth]{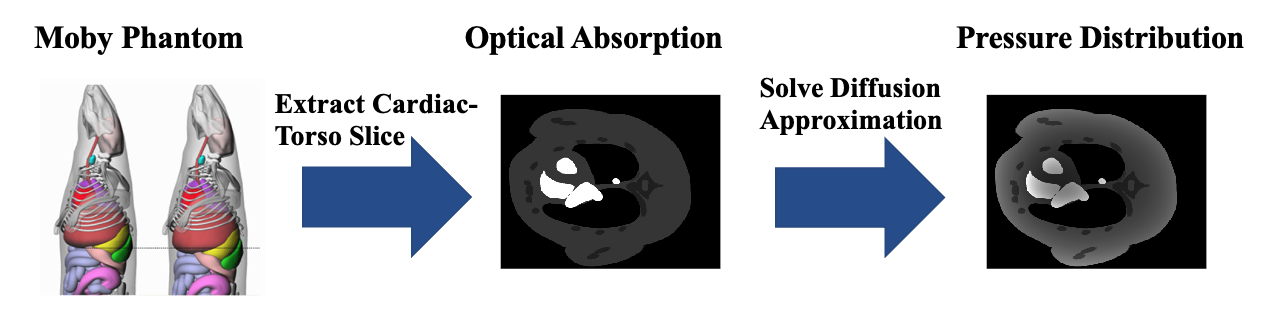}
    \caption{Construction of the dynamic object. A cardio-torso slice is extracted from the 3D mouse whole body (MOBY) phantom. A map of optical absorption is defined based on the anatomical structure of the selected slice. The initial pressure distribution map is then computed from the solution of the diffusion approximation to the radiative-transport equations. An animation of the dynamic object (optical absorption and pressure distribution maps) is available in the supplemental multimedia materials (Video 1).}
    \label{fig:moby}
\end{figure*}

The dynamic object used in these studies is a spatiotemporal map of the induced pressure distribution in a two-dimensional (2D) cardio-torso slice extracted from an anatomically realistic whole-body murine numerical phantom (MOBY) \cite{SegarsTsuiFreyEtAl04,SegarsTsui2009}. MOBY comprises detailed three-dimensional (3D) anatomical structures featuring tens of organs and models physiologically realistic cardiac and respiratory motions. The 3D anatomy of the MOBY phantom is displayed in Fig. \ref{fig:moby}. A 2D cardiac-torso slice was extracted from MOBY. Each organ was then assigned the optical absorption coefficient reported in Table \ref{tab:organ}\cite{DogdasStoutChatz07}. The optical scattering coefficient was assumed constant for all tissues and organs with a value of $0.5 \ mm^{-1}$ \cite{KloseLarsen06}. A uniform illumination pattern was assumed for this computation. The open-source finite element library FEniCS \cite{LoggMardalWells12} was used to solve the diffusion approximation equation.

\begin{table}
    \centering
    \caption{Optical Absorption Coefficient ($mm^{-1}$) by Organs}
    \begin{tabular}{c|c}
        Organ & Absorption Coefficient  \\
        \hline 
        Skin & 0.0191 \\ 
        Liver & 0.0720 \\ 
        Heart & 0.0240 \\ 
        Lung & 0.0000\\
        Skeleton &  0.0136 \\ 
        Intestines & 0.0000 \\ 
        Bladder & 0.0240
    \end{tabular}
    
    \label{tab:organ}
\end{table}

\subsection{Definition of the Virtual Imaging System}
\label{sec:d2d}

\begin{figure}
    \centering
    \includegraphics[width=.6 \columnwidth]{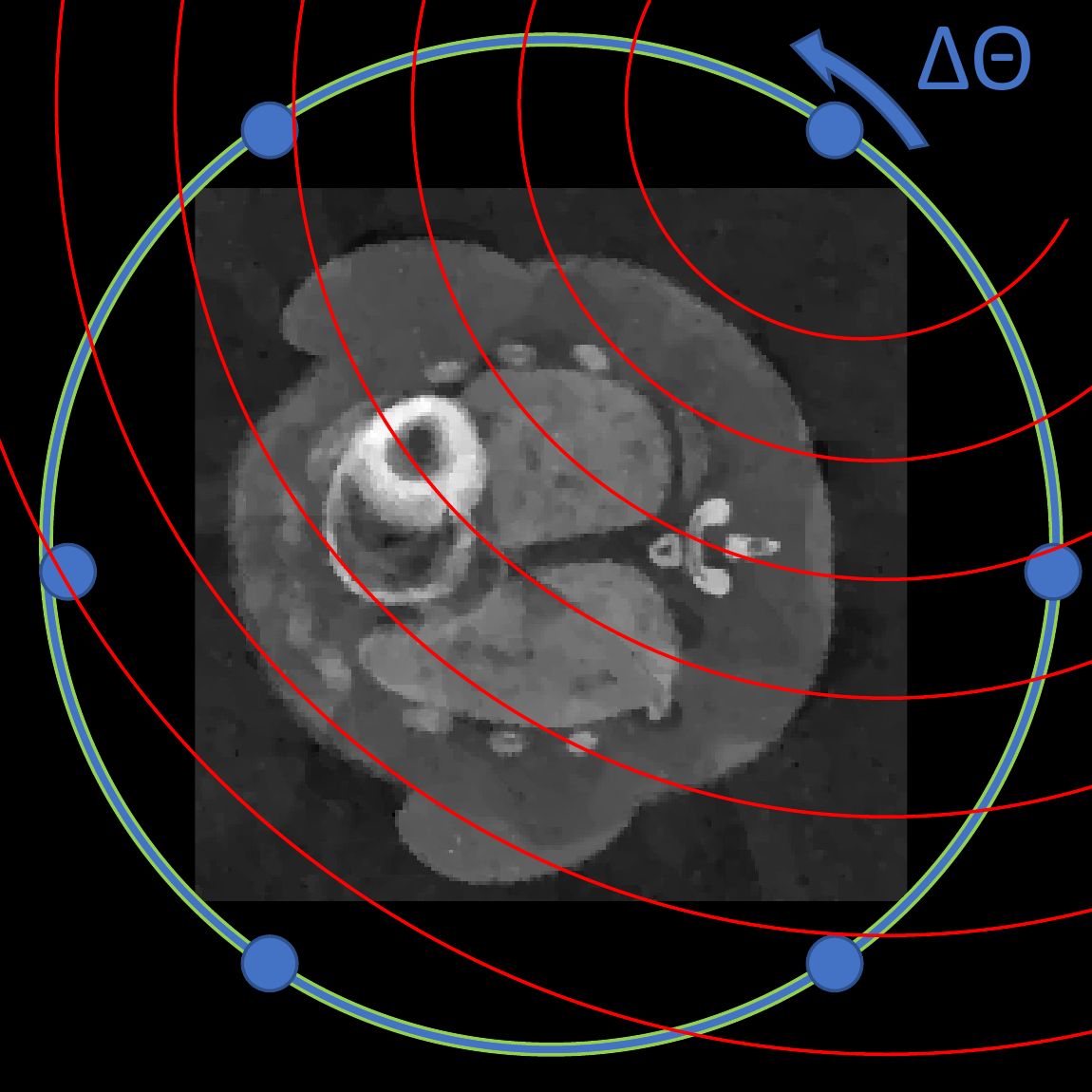}
    \caption{Illustration of the virtual system for dynamic imaging using a CRT model.}
    \label{fig:schema}
\end{figure}

Figure \ref{fig:schema} illustrates the virtual imaging system used in the numerical studies. The measurement geometry was a circle of radius $R$ (shown in green), along which $S$ equispaced idealized point-like ultrasound transducers are distributed. The object was placed at the center of the imaging system and fully contained within a square region of size $L$ (field of view). At each imaging frame, CRT data were collected by each sensor corresponding to the integral along $I$ concentric arcs (shown in red). Sensors rotated an angle $\Delta \Theta$ after each frame. The dynamic changes of the object were assumed to be slow enough such that the object was roughly constant during the acquisition of each imaging frame. The imaging systems parameters are displayed in Table \ref{Tab:is_parameters}.

Under the above assumptions, the circular Radon transform was used to compute the data $\boldsymbol{d}_k \in \mathbb{R}^M$ acquired at time frame $t_k$ as 
\begin{align}
    [\boldsymbol{d}_k]_{(s-1)I + i} & = [\mathcal{H}_k \ipar(\mathbf{r}, t_k)]_{(s-1)I + i} \nonumber\\
    & = \int_{|\mathbf{r} - \mathbf{r}_{k}^{s}| = l_i} \ipar(\mathbf{r}, t_k) d\mathbf{r}, \label{eq:cd_crt}
\end{align}
for $k=1,\ldots,K$, $s=1,\dots,S$, $i=1,\dots,I$. Above $[\boldsymbol{d}_k]_{m}$ denotes the $m$-th component of $\boldsymbol{d}_k$, $\mathbf{r}_{k}^{s}$ denotes the location of the $s$-th sensor at the $k$-th time frame, and $l_i$ denotes the radius of the $i$-th arc.

To numerically evaluate the continuous-to-discrete imaging operator acting on a neural field representation $\Phi_{\boldsymbol \xi}(\mathbf{r}, t)$ of the sought-after object, the integral along the $i$-th arc is approximated using numerical integration. In particular, the use of the mid-point quadrature rule provides the approximation
\begin{equation*}
[\mathcal{H}_k \Phi_{\boldsymbol \xi}(\mathbf{r}, t_k)]_{(s-1)I + i} \approx \sum_{q=1}^{Q} h^{s,i,q}_{k} \Phi_{\boldsymbol \xi}(\mathbf{r}^{s,i,q}_{k},t_k),
\end{equation*}
where the quadrature points $\mathbf{r}^{s,i,q}_{k}$ and weights $h^{s,i,q}_{k} $ are given by
$$\mathbf{r}^{s,i,q}_{k} = \mathbf{r}^s_{k} + l_i\begin{bmatrix}\cos\phi_q,\\ \sin\phi_q \end{bmatrix}, \text{ with }  \phi_q = 2\pi \frac{q}{Q} $$ 
and 
$$
h^{s,i,q}_{k} = \begin{cases} 2 \pi\frac{l_i}{Q} & \mathbf{r}^{s,i,q}_{k} \in \Omega \\ 0 & \mathbf{r}^{s,i,q}_{k} \not \in \Omega \end{cases}.$$

\begin{table}
\vspace{-4mm}
\centering
\caption{Imaging system parameters} 
\label{Tab:is_parameters}
\begin{tabular}{ll}
\hline
\multicolumn{2}{c}{Imaging system}\\
FOV diameter $L$ & 2.9 cm\\
Number of pixels $N = N_s^2$ & $200^2$ \\ 
Aperture $R$ & 2.05 cm \\
Acquisition time $T$ &  5 s \\ 
Number of frames $K$ & 180 \\ 
Rings per view $I$ & 283 \\
Rotation angle $\Delta\Theta$ & $2^\circ$ \\
\hline
Relative Noise & $0.025$\\
Number of views per frame $S$ & $2,4,8,16,32$ \\
\hline 
Relative Noise & $0.0125, 0.025, 0.05, 0.1, 0.2$\\
Number of views per frame $S$ & $4$\\

\hline
\end{tabular}
\vspace{-4mm}
\end{table}

Finally, the discrete-to-discrete counterpart of the CRT imaging operator in Eq. \eqref{eq:cd_crt} is introduced. The discrete-to-discrete CRT operator is used in the implementation of the classical reconstruction methods introduced in Section \ref{sec:background}, against which the performanced of the proposed method is compared in Section \ref{sec:results}. In particular, the dynamic object is approximated as piecewise constant on a spatiotemporal Cartersian grid. That is, the basis functions $\{\beta_n(\textbf{r})\}_{n=1}^N$ in Eq. \eqref{eq:DiscreteParam} are chosen as the set of indicator functions for each pixel, or equivalently the object is approximated using pixel-wise constant functions.  The discrete-to-discrete CRT imaging operator at frame $k$ is represented by the matrix $\mathbf{H}_k$, those entry $[\mathbf{H}_k]_{m,n}$ is proportional to the length of the intersection the $m$-th arc ($m=(s-1)I + i)$ with the $n$-th pixel. In the numerical results presented below, the discrete-to-discrete CRT imaging operator was implemented using the AirToolsII MatLab toolbox \cite{HansenJorgensen2018} and stored in the form of a compressed sparse column (CSC) matrix \cite{gilbert1992sparse}.  

\subsection{Study Design}

The first experiment addresses the approximation properties (representation power) achieved by the neural field representation using the POUnet architecture. The neural field parameters are found by solving the embedding problem defined in Eq. \eqref{eqn:embedding}. The approximation properties of the neural field are compared to that achieved by a truncated singular value decomposition (SVD) of the dynamic object $\boldsymbol{F}$ in terms of relative root mean square error (RRMSE), structural self-similarity index (SSIM, \cite{WangBoviketall04}), and number of parameters. 

For solving the dynamic image reconstruction problem, three approaches were considered. The first was a pixel-wise implementation with total variation regularization defined in Eq. \eqref{eqn:total_variaiton} (PW-TV), calculated with finite differences. The second reconstruction approach was a pixel-wise implementation using the nuclear norm defined in Eq. \eqref{eqn:nuclear_norm} as regularization (PW-NN). The third approach implemented the proposed neural field approach using the stochastic total variation regularization in Eq. \eqref{eqn:stochastic_tv} (NF-TV). 

Forward data were generated by applying the discrete-to-discrete CRT imaging operator $\boldsymbol{H}_k$ to a discretization of the object using a Cartesian grid with $400\times 400$ spatial pixels and 180 time frames. Measurements were then perturbed with additive white Gaussian noise with zero mean and standard deviation $\sigma$.

The regularization parameter $\gamma$ for each reconstruction was selected using Morozov's discrepancy principle \cite{Morozov66}. This criterion selects the largest regularization parameter $\gamma$ such that the difference between the expected measurements and the true measurements has an average magnitude less than the noise standard deviation. That is, Morozov's discrepancy principle select the largest $\gamma$ such that
$$ ||\H(\ipar^\gamma) - \data||^2 \leq \sigma ^2 N_m,$$
where $N_m$ is the dimension of $\data$ and $\ipar^\gamma$ solves
$$\ipar^\gamma = \argmin_\ipar \frac{1}{2\sigma^2} ||\H(\ipar) - \data||^2 + \gamma \Reg(\ipar).$$ 

Two image reconstruction experiments were conducted. The first image reconstruction experiment varied the number of views per frame with a fixed noise level. This experiment investigates the performance of the neural field reconstruction approach with varying levels of data incompleteness compared to classical methods based on a discretization of the object on a spatiotemporal Cartesian grid. The second image reconstruction experiment used a varying noise level and a fixed number of views per frame. This experiment seeks to analyze the sensitivity to noise of the proposed neural fields reconstruction method and compares it to that of classical methods.

To compare the RRMSE and SSIM achieved by the neural field reconstruction approach with that achieved by classical reconstruction methods, the neural field representation was rendered at the same resolution as the reconstructed images obtained by classical methods ($200\time 200$ spatial pixels and 180 times frames).

\section{Results}\label{sec:results}

All numerical studies were performed on a workstation with two Intel Xeon Gold 5218 Processors (16 core, 32 threads, 2.3 Ghz, 22 MB cache each), 384 GB of DDR4 2933Mhz memory, and one NVidia Titan RTX 24GB graphic processing unit (GPU).
The neural field method was implemented using PyTorch, an open-source machine learning framework \cite{NEURIPS2019_9015}, and executed on the GPU. The parameters of the image and the neural field are presented in Table \ref{Tab:rep_parameters}. The POUnet architecture consisted of 4 hidden layers with sinusoidal activation functions \cite{SitzmannMartelBergmanetal2020} fed into an output layer of size $P = 40$ with a softmax activation. The basis functions $B$ are the tensor product of polynomials of total degree 3 in space and time. An initial guess for the POUnet parameters $\boldsymbol{\boldsymbol \eta}$ and $\boldsymbol{C}$ was obtained by performing embedding of the time-averaged object for the studies in Section \ref{sec:embed} and by solving a static image reconstruction problem on time-averaged measurements for the studies in Section \ref{sec:CRT_recons}. Following this initialization, the neural field was trained following the strategy outlined in Algorithm \ref{alg:inr_recon} with batch optimization.  
The regularization implemented was the stochastic approximation of total variation outlined in Eq. \eqref{eqn:stochastic_tv}. The internal updates for $\boldsymbol C$ were accomplished with an Adam optimizer. The learning rate of this Adam optimizer was reduced multiplicatively across updates of $\boldsymbol C$ and reduced in total by $10^{-3}$ throughout all iterations. This optimizer performed at most 20 epochs, where an epoch is a single pass through all the imaging frame subsets $\{\mathcal{K}_b\}_{b=1}^{n_b}$. The batch size $|{\mathcal{K}_b}|$ was chosen according to the available GPU memory. Similarly, the internal updates for $\boldsymbol \eta$ were accomplished using a second Adam optimizer, with 20 epochs and a slowly shrinking learning rate. The sparsity enforcing regularization $\|\boldsymbol \Psi_{\boldsymbol \eta}\|_{q, \epsilon}$ used the parameters $q = \frac{1}{2}$ and $\epsilon = 10^{-2}$. The initial training regularization weights were $\rho = \frac{10^{-4}}{\sigma^2}$ and $\tau = 1$. Across the first 10 epochs, $\rho$ and $\tau$ were gradually reduced until they reached a thousandth of their original values. For the second 10 epochs, $\rho$ and $\tau$ were set to zero to recover the cost function in Eq. \eqref{eqn:INR_obj}. The stopping criterion for this optimization process was when $\J$ did not decrease between two global updates of $\boldsymbol \xi$.

The classical, discrete version of the image reconstruction problem in Eq. \eqref{eqn:classical_approach} were solved using Matlab on the workstation CPU. Specifically, the Fast Iterative Shrinkage Algorithm (FISTA) \cite{BeckTeboulle09} as implemented in UNLocBoX, a MATLAB toolbox for proximal-splitting methods \cite{PerraudinKalofoliasShuman2014}, was used for the reconstructions.

\subsection{Embedding Problem}\label{sec:embed}
\begin{figure}
    \centering
    \includegraphics[width = 0.8 \columnwidth]{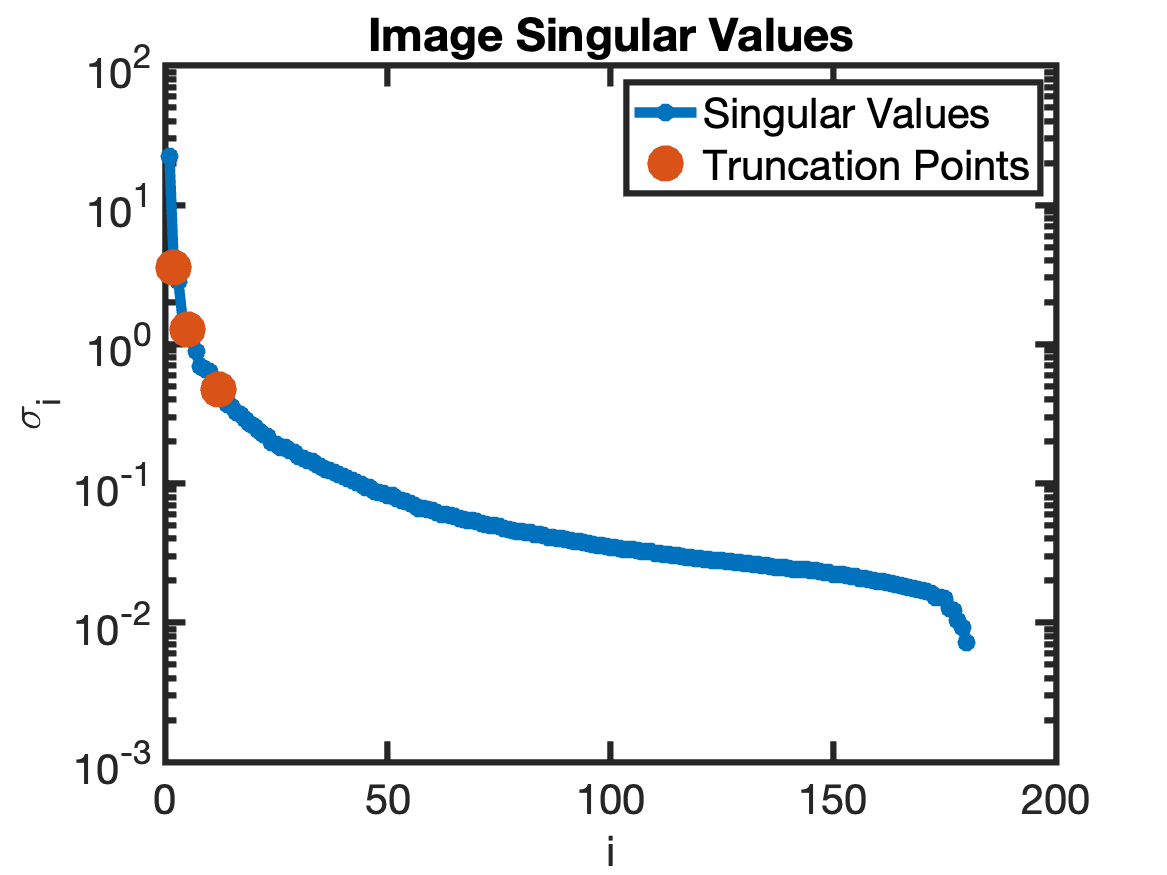}
    \caption{Singular value of dynamic object on semilog scaled in y-axis. Singular values 2,5, and 12 are highlighted in orange. Rapid decay of singular values implies that this image can be accurately approximated with a semiseparable approximation.}
    \label{fig:obj_svd}
\end{figure}

In this section, neural field (NF) representation and the semiseparable approximation with rank $r$ ((SS($r$)) approaches are compared under three conditions: approximately same number of parameters ($r=2$), approximately same value of SSIM ($r=5$), approximately same RRMSE ($r=12$). The singular value spectra of the discretized dynamic object $\boldsymbol{F}$ is displayed in Fig. \ref{fig:obj_svd}. The rapid decay of the singular vectors indicates that the dynamic object can be accurately represented using a semiseparable approximation.   
\begin{table}
\centering
\vspace{-4mm}
\caption{Image and neural field parameters} 
\label{Tab:rep_parameters}
\begin{tabular}{ll}
\hline
\multicolumn{2}{c}{Image parameters}\\
Number of pixels $N = N_s^2$ & $200^2$ \\ 
Number of frames $K$ & 180 \\ 
Number of parameters  & $7.2 \, 10^6$ \\

\hline
 \multicolumn{2}{c}{Neural field reconstruction} \\
Number of partitions $P$ & 40 \\
Number of basis functions $M$ & 40 \\ 
Number of parameters & 86,020\\
Network Width & 140\\
Network Depth & 4\\
\hline
\end{tabular}
\vspace{-4mm}
\end{table}

\begin{figure*}
    \centering
    \resizebox{.95\textwidth}{!}{\includegraphics[page=1, angle = 270, trim={0 0cm 2cm 4cm },clip]{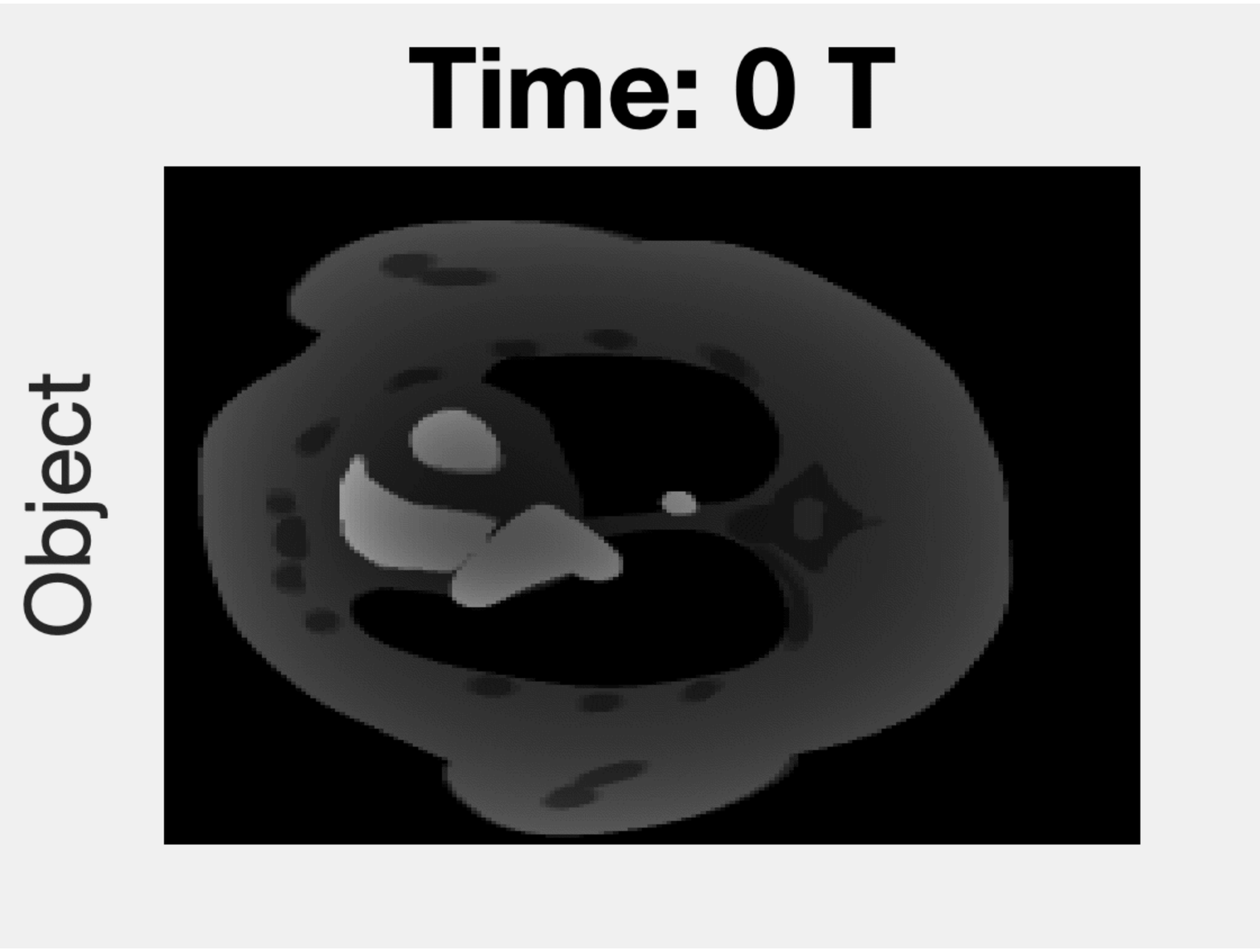}\includegraphics[page =  2, angle = 270, trim={0 3.5cm 2cm 4cm },clip]{Figures/Phantom.pdf}\includegraphics[page =  3, angle = 270, trim={0 3.5cm 2cm 4cm },clip]{Figures/Phantom.pdf} \includegraphics[page =  4, angle = 270, trim={0 3.5cm 2cm 4cm },clip]{Figures/Phantom.pdf}\includegraphics[page =  5, angle = 270, trim={0 3.5cm 2cm 4cm },clip]{Figures/Phantom.pdf}\includegraphics[page =  6, angle = 270, trim={0 3.5cm 2cm 4cm },clip]{Figures/Phantom.pdf}}
    
    \resizebox{.95\textwidth}{!}{\includegraphics[page=1, angle = 270, trim={4.5cm 0cm 2cm 4cm },clip]{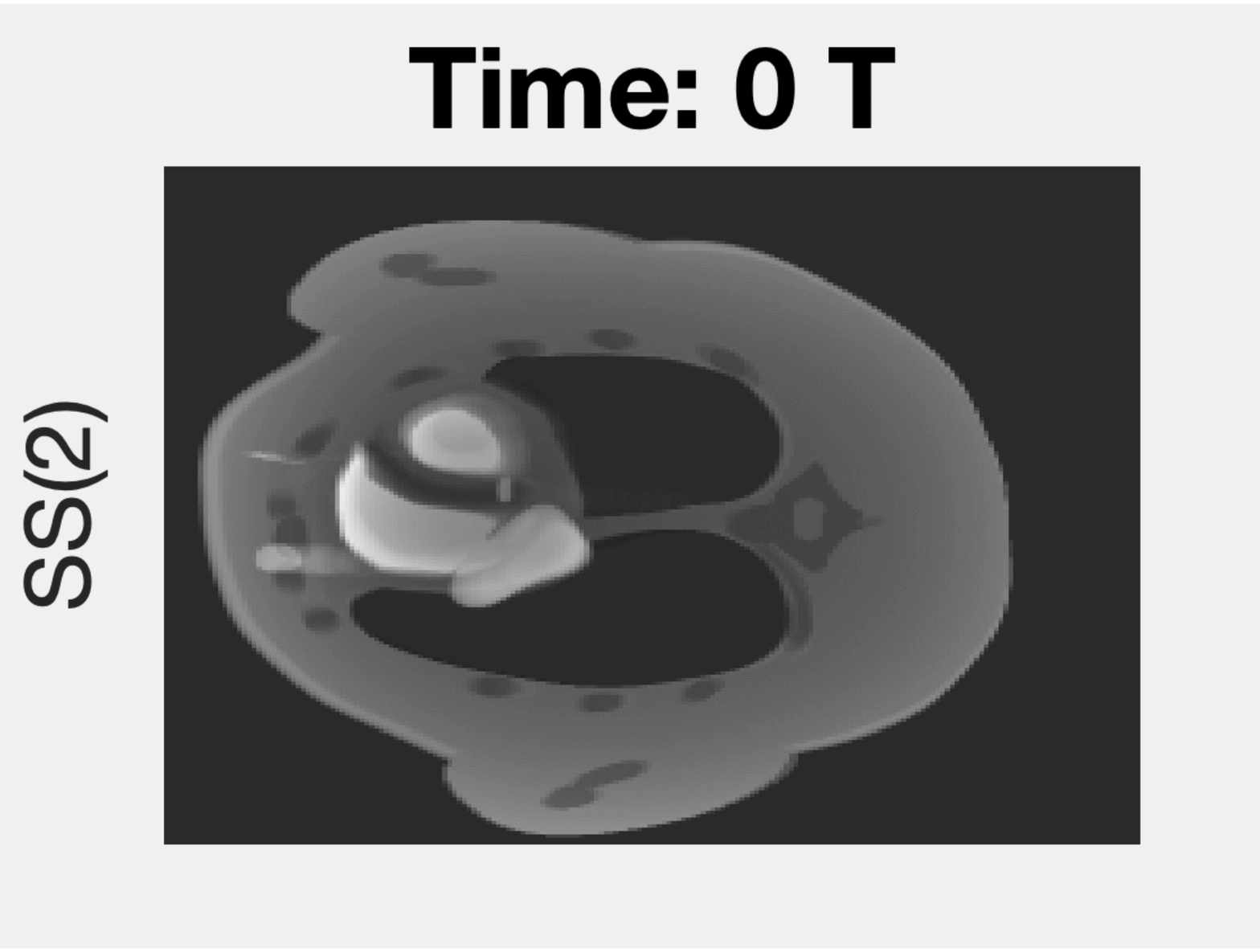}\includegraphics[page =  2, angle = 270, trim={4.5cm 3.5cm 2cm 4cm },clip]{Figures/low_rank.pdf}\includegraphics[page =  3, angle = 270, trim={4.5cm 3.5cm 2cm 4cm },clip]{Figures/low_rank.pdf} \includegraphics[page =  4, angle = 270, trim={4.5cm 3.5cm 2cm 4cm },clip]{Figures/low_rank.pdf}\includegraphics[page =  5, angle = 270, trim={4.5cm 3.5cm 2cm 4cm },clip]{Figures/low_rank.pdf}\includegraphics[page =  6, angle = 270, trim={4.5cm 3.5cm 2cm 4cm },clip]{Figures/low_rank.pdf}}
    
    \resizebox{.95\textwidth}{!}{\includegraphics[page=1, angle = 270, trim={4.5cm 0cm 2cm 4cm },clip]{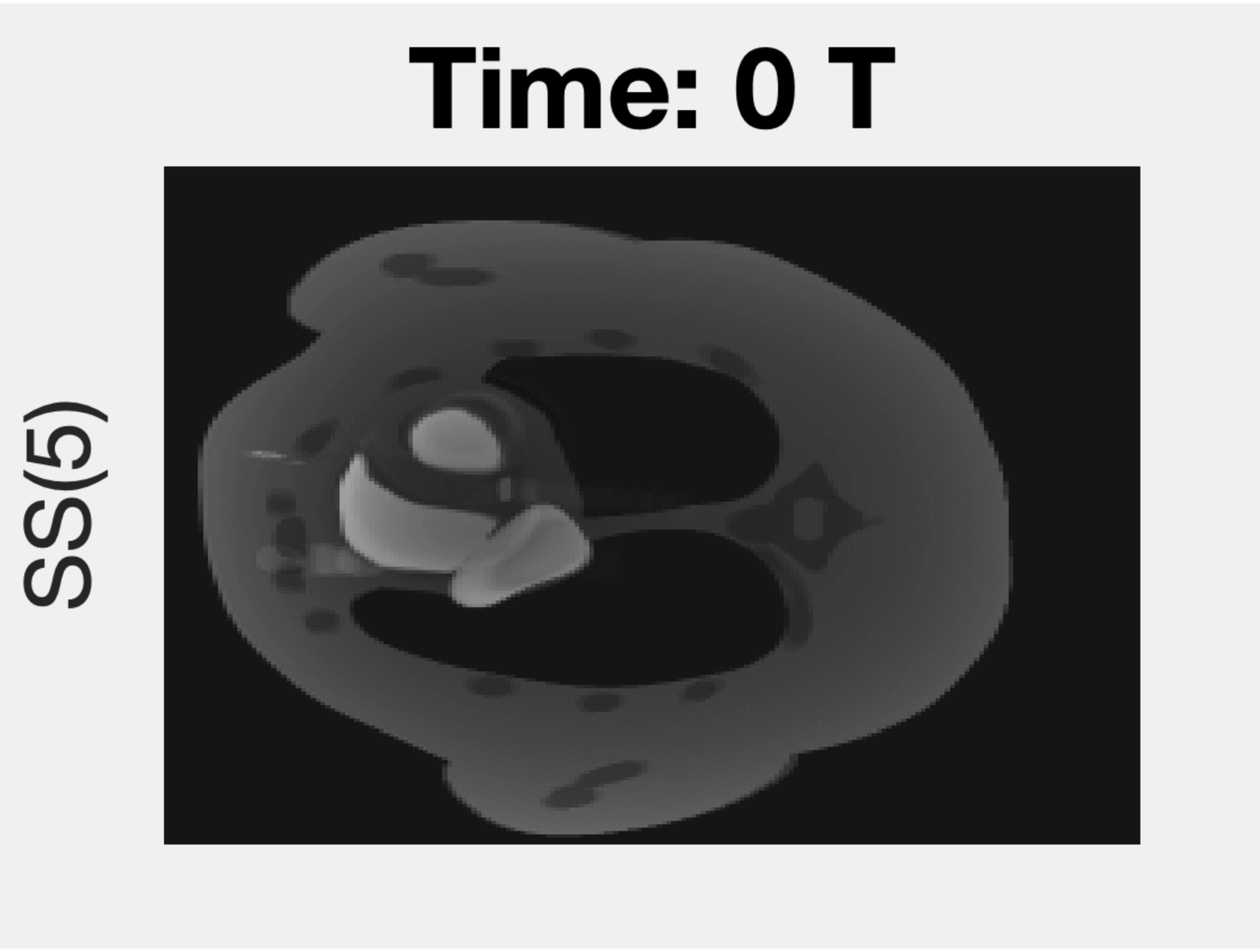}\includegraphics[page =  2, angle = 270, trim={4.5cm 3.5cm 2cm 4cm },clip]{Figures/ss5.pdf}\includegraphics[page =  3, angle = 270, trim={4.5cm 3.5cm 2cm 4cm },clip]{Figures/ss5.pdf} \includegraphics[page =  4, angle = 270, trim={4.5cm 3.5cm 2cm 4cm },clip]{Figures/ss5.pdf}\includegraphics[page =  5, angle = 270, trim={4.5cm 3.5cm 2cm 4cm },clip]{Figures/ss5.pdf}\includegraphics[page =  6, angle = 270, trim={4.5cm 3.5cm 2cm 4cm },clip]{Figures/ss5.pdf}}
    
    \resizebox{.95\textwidth}{!}{\includegraphics[page=1, angle = 270, trim={4.5cm 0cm 2cm 4cm },clip]{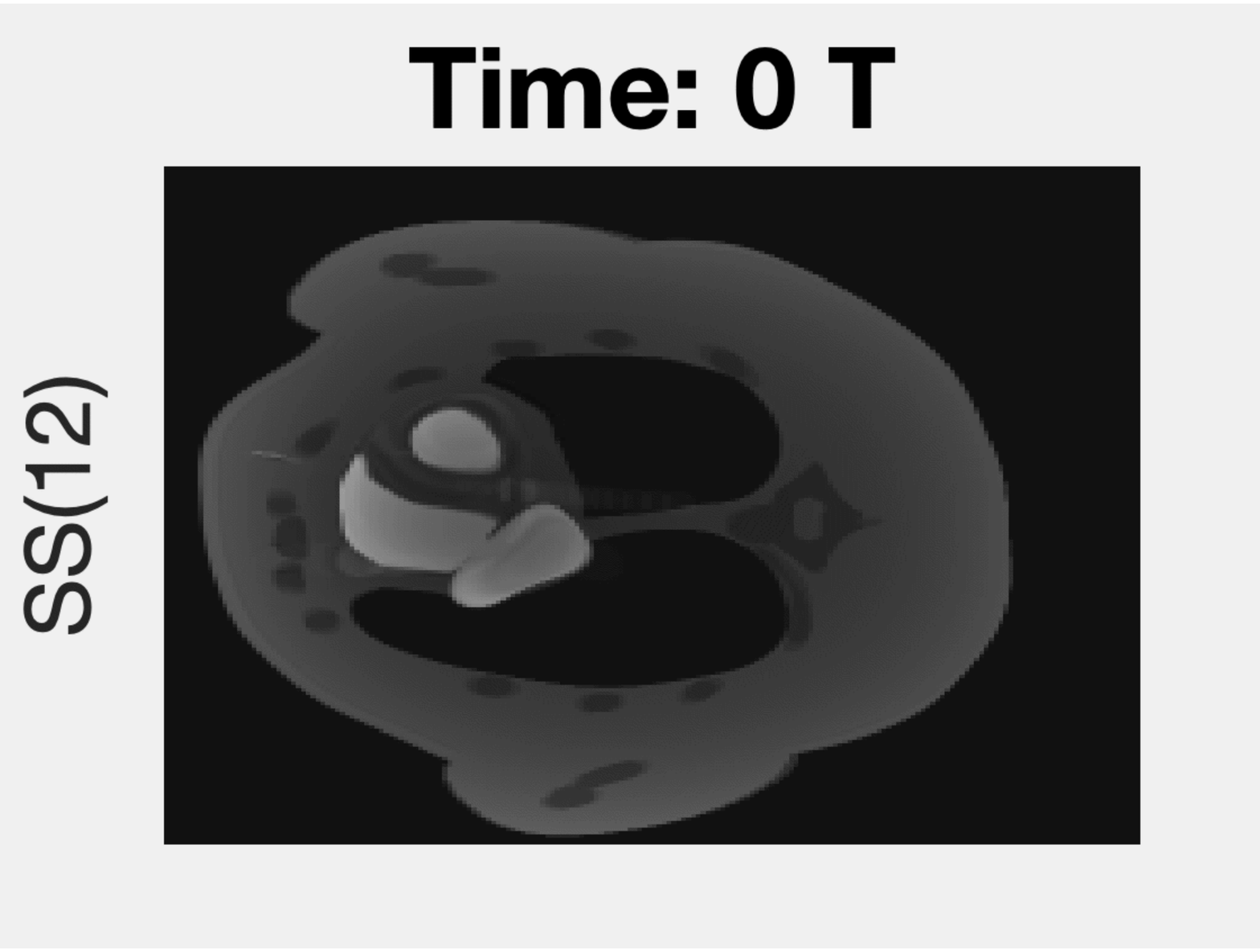}\includegraphics[page =  2, angle = 270, trim={4.5cm 3.5cm 2cm 4cm },clip]{Figures/ss12.pdf}\includegraphics[page =  3, angle = 270, trim={4.5cm 3.5cm 2cm 4cm },clip]{Figures/ss12.pdf} \includegraphics[page =  4, angle = 270, trim={4.5cm 3.5cm 2cm 4cm },clip]{Figures/ss12.pdf}\includegraphics[page =  5, angle = 270, trim={4.5cm 3.5cm 2cm 4cm },clip]{Figures/ss12.pdf}\includegraphics[page =  6, angle = 270, trim={4.5cm 3.5cm 2cm 4cm },clip]{Figures/ss12.pdf}}
    
    \resizebox{.95\textwidth}{!}{\includegraphics[page=1, angle = 270, trim={4.5cm 0cm 2cm 4cm },clip]{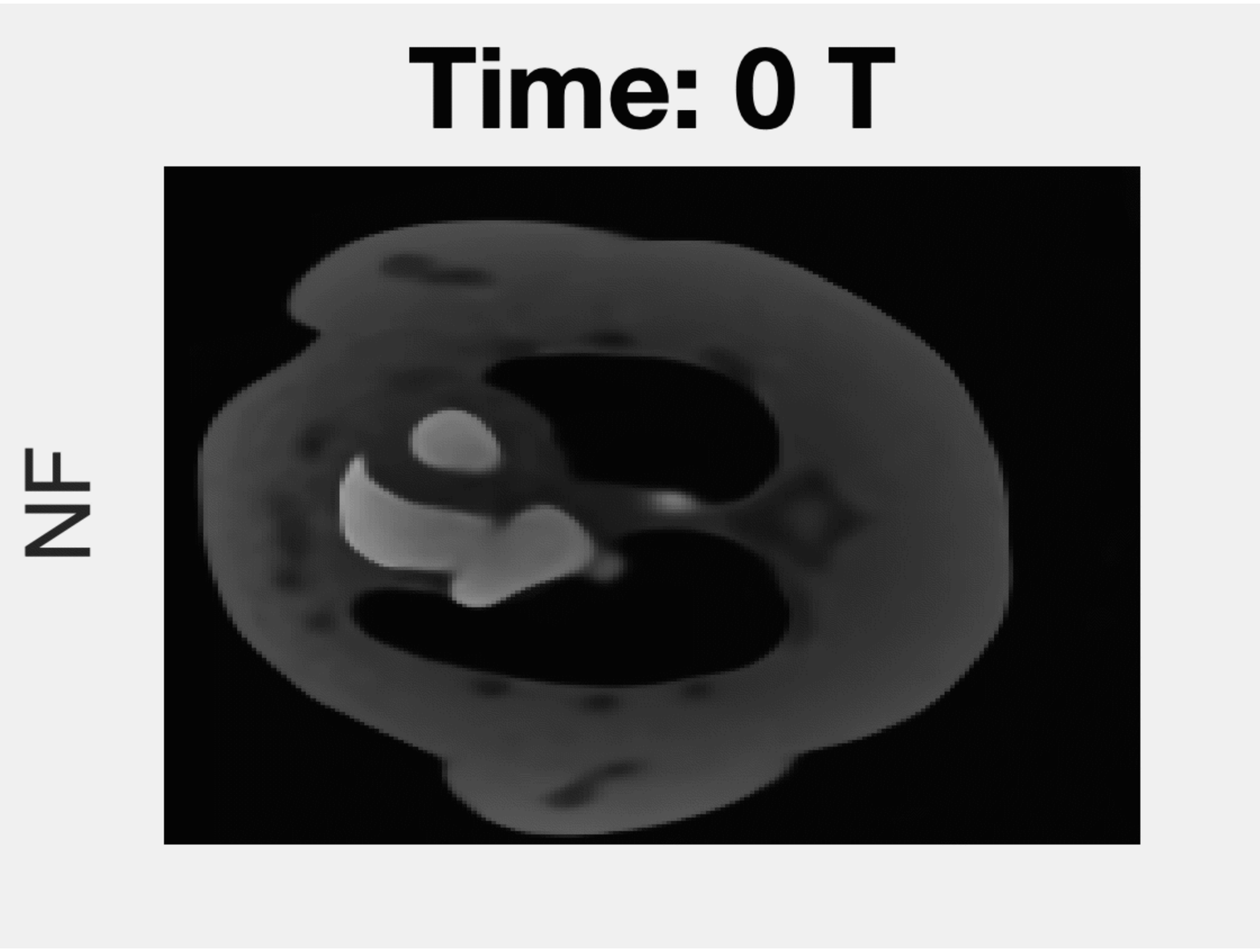}\includegraphics[page =  2, angle = 270, trim={4.5cm 3.5cm 2cm 4cm },clip]{Figures/Representation.pdf}\includegraphics[page =  3, angle = 270, trim={4.5cm 3.5cm 2cm 4cm },clip]{Figures/Representation.pdf} \includegraphics[page =  4, angle = 270, trim={4.5cm 3.5cm 2cm 4cm },clip]{Figures/Representation.pdf}\includegraphics[page =  5, angle = 270, trim={4.5cm 3.5cm 2cm 4cm },clip]{Figures/Representation.pdf}\includegraphics[page =  6, angle = 270, trim={4.5cm 3.5cm 2cm 4cm },clip]{Figures/Representation.pdf}}

    \caption{Embedding problem: Dynamic object (top), semiseparable approximations SS($r$) for $r=2,5,$ and $12$ (middle rows), and neural field NF (bottom). NF outperforms (lower RRMSE, higher SSIM) the semiseparable approximation SS(2), which has approximately the same number of parameters as NF. NF uses less than half the number of parameters of SS(5) while achieving comparable SSIM and less than a fifth of the number of parameters of SS(12) while achieving comparable RRMSE. An animation of the dynamic object and its representations using the SS and NF approaches is available in the supplemental multimedia materials (Video 2).}
    
    \label{fig:Representation_Test}
    \centerline{\includegraphics[width = \textwidth]{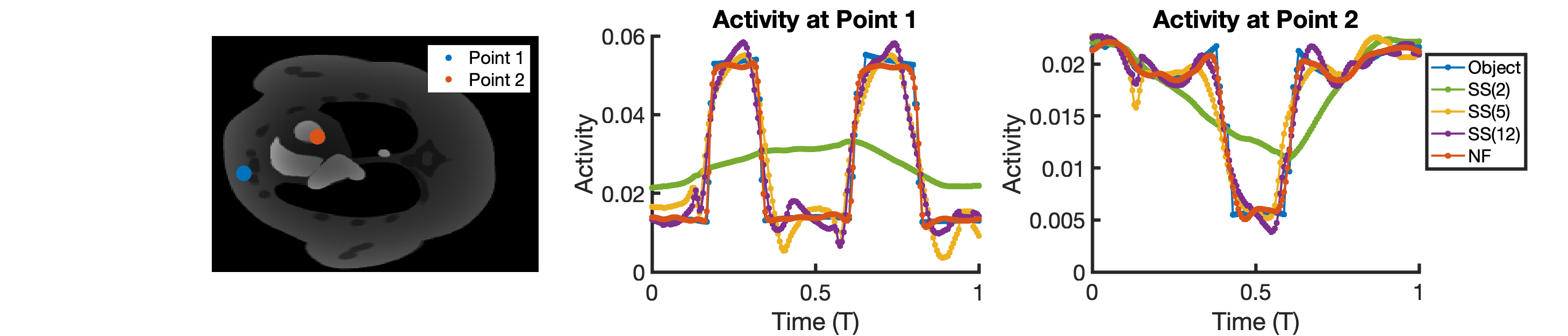}}
    \caption{Embedding problem: Time-activity curves at two points in space for dynamic object, neural field embedding (NF), and semiseparable approximations (SS(2), SS(5), SS(12)).  NF matches the dynamic behavior of the Object at the two points closely with less oscillatory behavior than SS(5) and SS(12). SS(2) largely smooths over the sharp dynamic features displayed in the time profile. }
    \label{fig:embedding_activity}
    
\end{figure*}

\begin{figure*}
    \centering
    \resizebox{.95\textwidth}{!}{\includegraphics[page=1, angle = 270, trim={0 3.5cm 2cm 4cm },clip]{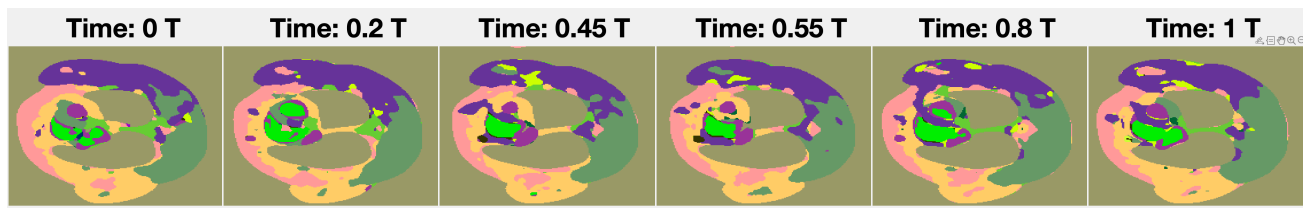}\includegraphics[page =  2, angle = 270, trim={0 3.5cm 2cm 4cm },clip]{Figures/partitions.pdf}\includegraphics[page =  3, angle = 270, trim={0 3.5cm 2cm 4cm },clip]{Figures/partitions.pdf} \includegraphics[page =  4, angle = 270, trim={0 3.5cm 2cm 4cm },clip]{Figures/partitions.pdf}\includegraphics[page =  5, angle = 270, trim={0 3.5cm 2cm 4cm },clip]{Figures/partitions.pdf}\includegraphics[page =  6, angle = 270, trim={0 3.5cm 2cm 4cm },clip]{Figures/partitions.pdf}}
    \caption{Embedding problem: The learned partition of unity adapts itself to the features of the dynamic object. Partitions, labeled using a different color in the plots, conform themselves to the anatomical structures of the object and adapt to the object's motion in time. An animation of the time evolution of the POUnet partitions is available in the supplemental multimedia materials (Video 3).}
    \label{fig:partitions}
\end{figure*}

The results of the embedding experiment are shown in Fig. \ref{fig:Representation_Test} with the dynamic object in the top row, the semiseparable approximations SS(2), SS(5), and SS(12) in the middle rows, and the NF in the bottom row. Figure \ref{fig:embedding_activity} displays the time-activity curve for the true object, the semiseparable approximations SS($r$),  and the NF at two points in space. The partition of unit learned by the POUnet is displayed in Fig. \ref{fig:partitions} with each partition labeled with a different color. From the figure, it can be observed how these learned partitions conform to the anatomical structures of the object and adapt in time following the object's motion.

Table \ref{tab:embeding_acc} reports the number of parameters used by each approach and the corresponding RRMSE and SSIM. For approximately the same number of parameters, NF leads to an over 3-fold reduction in RRSME compared to SS(2) and an increase in SSIM from $0.9416$ to $0.9713$. For approximately the same SSIM, NF allows for a 2.5 reduction in the number of parameters compared to SS(5). Similarly, for approximately the same RRMSE, NF achieves a 5-fold reduction in the number of parameters.

\begin{table}[thb]
    \caption{Embedding problem: Number of parameters, RRMSE, and SSIM achieved by the neural field representation NF and the rank $r$ semiseparable approximation SS($r$)}.
    \label{tab:embeding_acc}
    \centering
    \begin{tabular}{l|ccc}
    \hline
       SS(2)  & 80K & $0.1790$ & $0.9416$ \\
       SS(5)  & 200K & 0.1008 & 0.9749\\
       SS(12) & 480K & 0.0565  & 0.9898\\
       NF     & 86K & $0.0568$ & $0.9713$\\
    \hline
    \end{tabular}
\end{table}

\subsection{ Image Reconstruction with CRT}\label{sec:CRT_recons}

Two experiments were conducted involving image reconstruction from incomplete dynamic CRT data. The first investigates the effect of the degree of undersampling (determined by the number of views per frame) on the quality (RRMSE and SSIM) of the reconstructed images; the second investigates the effect of different noise levels. 
 
\subsubsection{Varying number of views per frame}\label{sec:vpt}

\begin{figure*}
    \centering
    \resizebox{.95\textwidth}{!}{\includegraphics[angle = 270, page = 2, trim={0 0 3cm 4cm },clip]{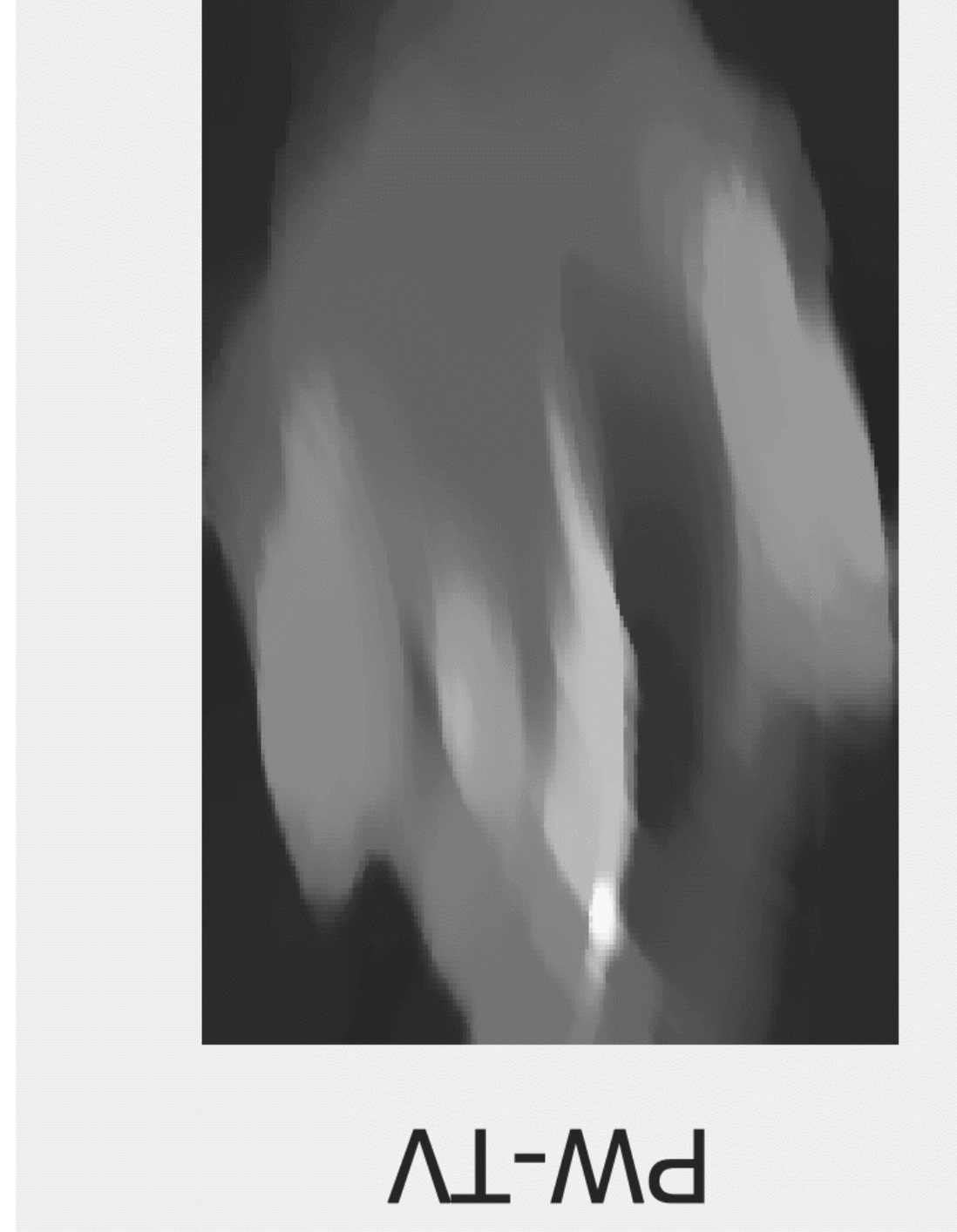}\includegraphics[angle = 270, page = 2, trim={0 3.5cm 3cm 4cm },clip]{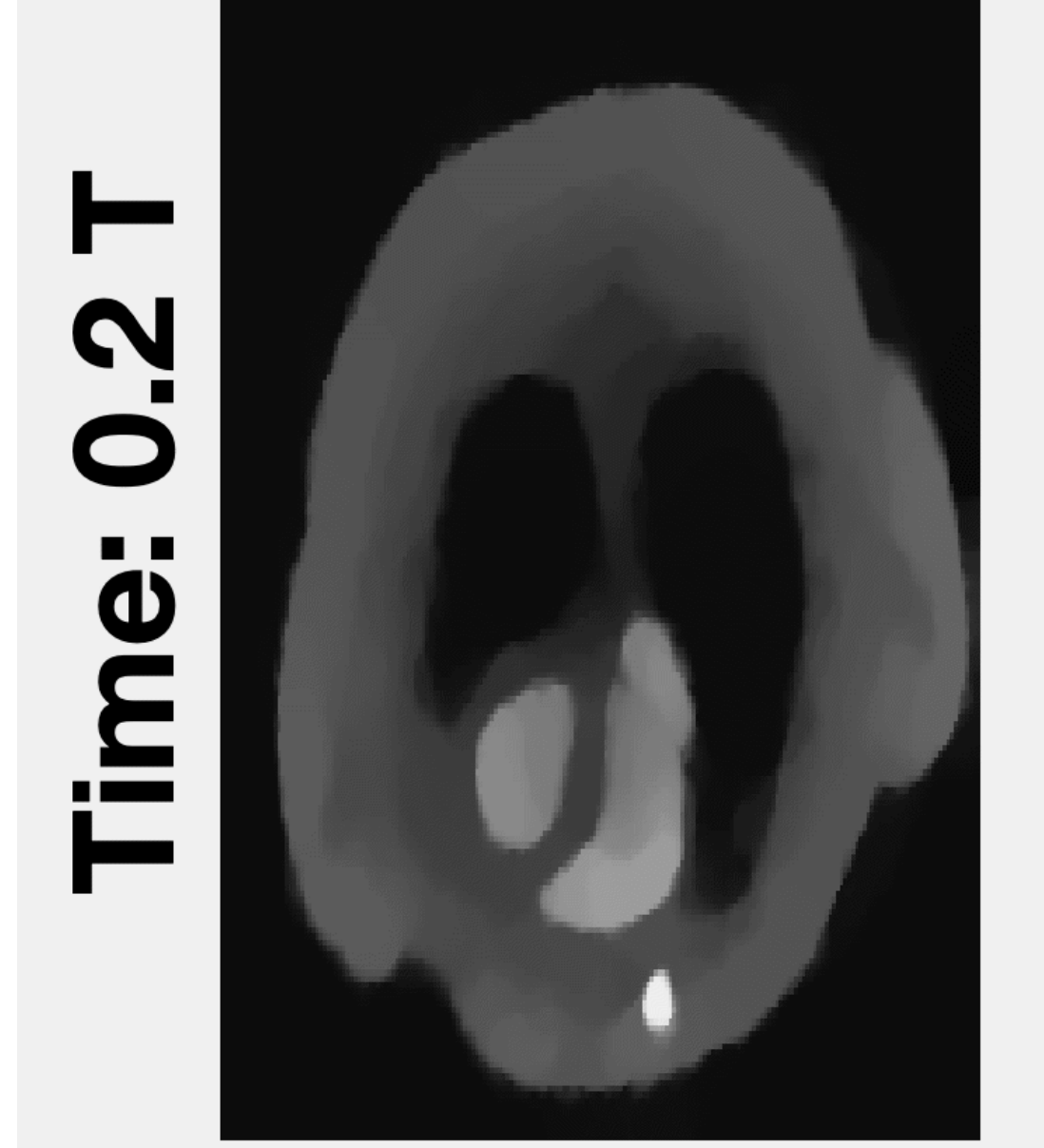} \includegraphics[angle = 270, page =2, trim={0 3.5cm 3cm 4cm },clip]{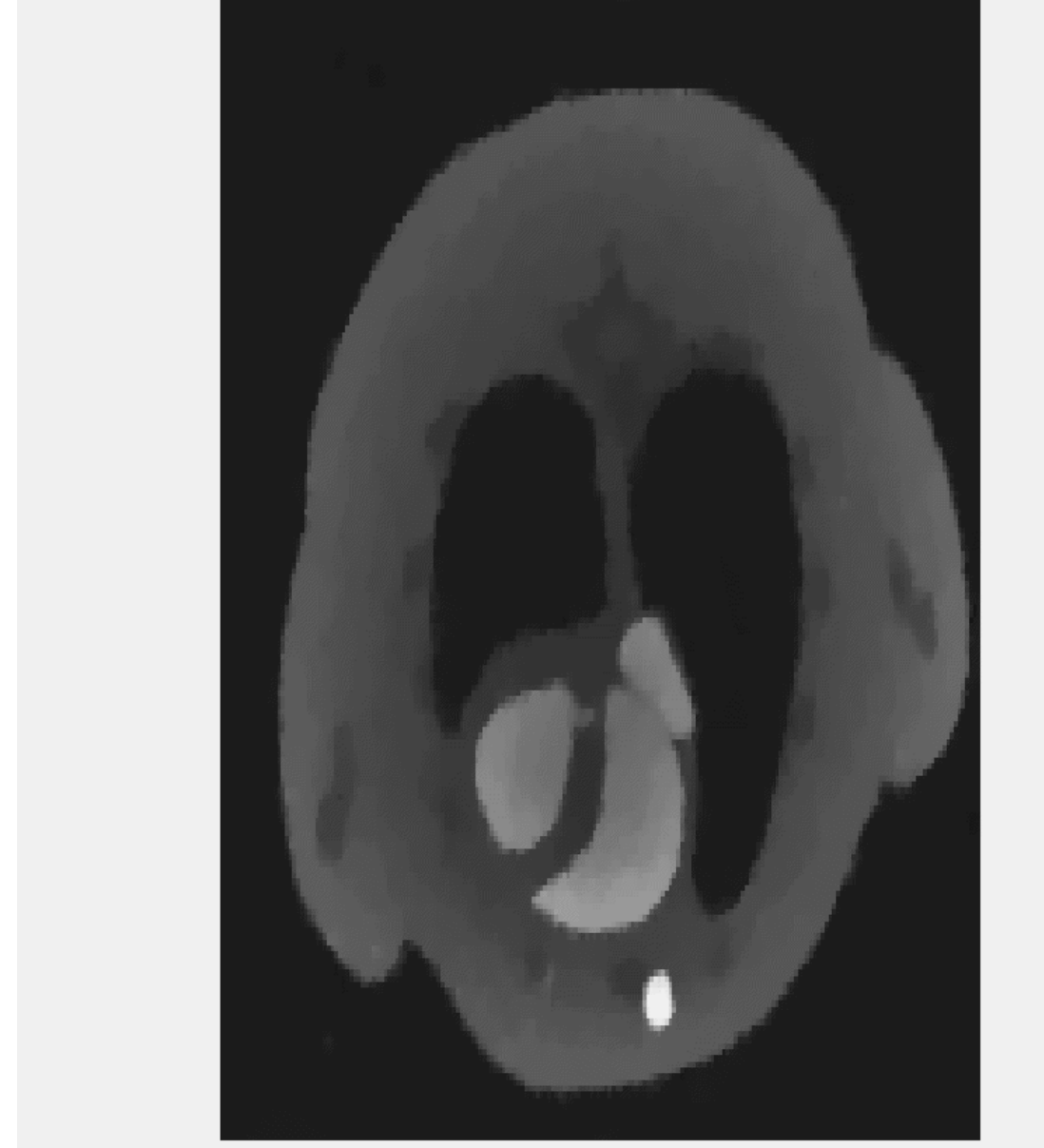} \hspace{4cm} \includegraphics[angle = 270, page = 5, trim={0 3.5cm 3cm 4cm },clip]{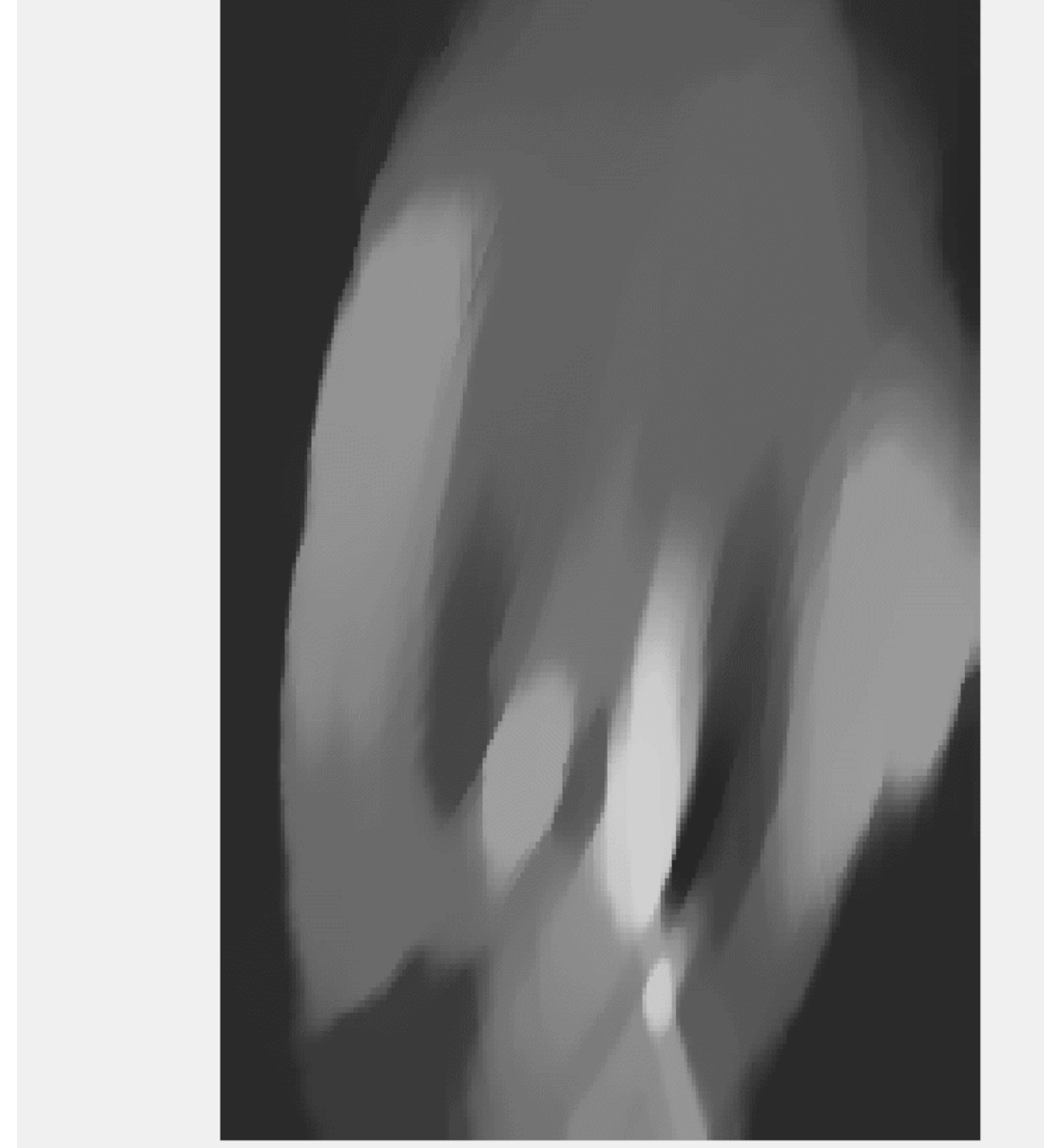}\includegraphics[angle = 270, page = 5, trim={0 3.5cm 3cm 4cm },clip]{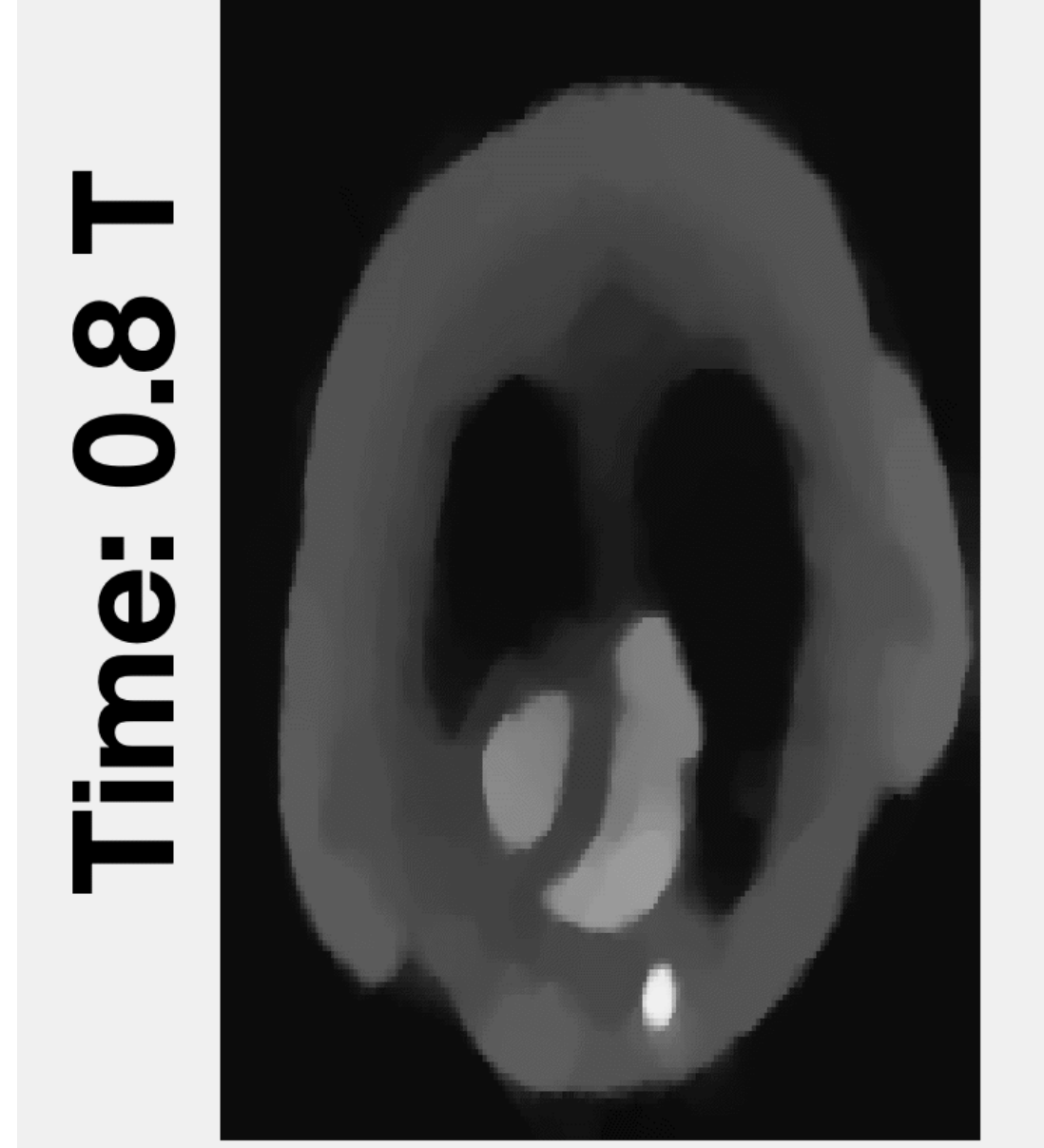}\includegraphics[angle = 270, page = 5, trim={0 3.5cm 3cm 0 },clip]{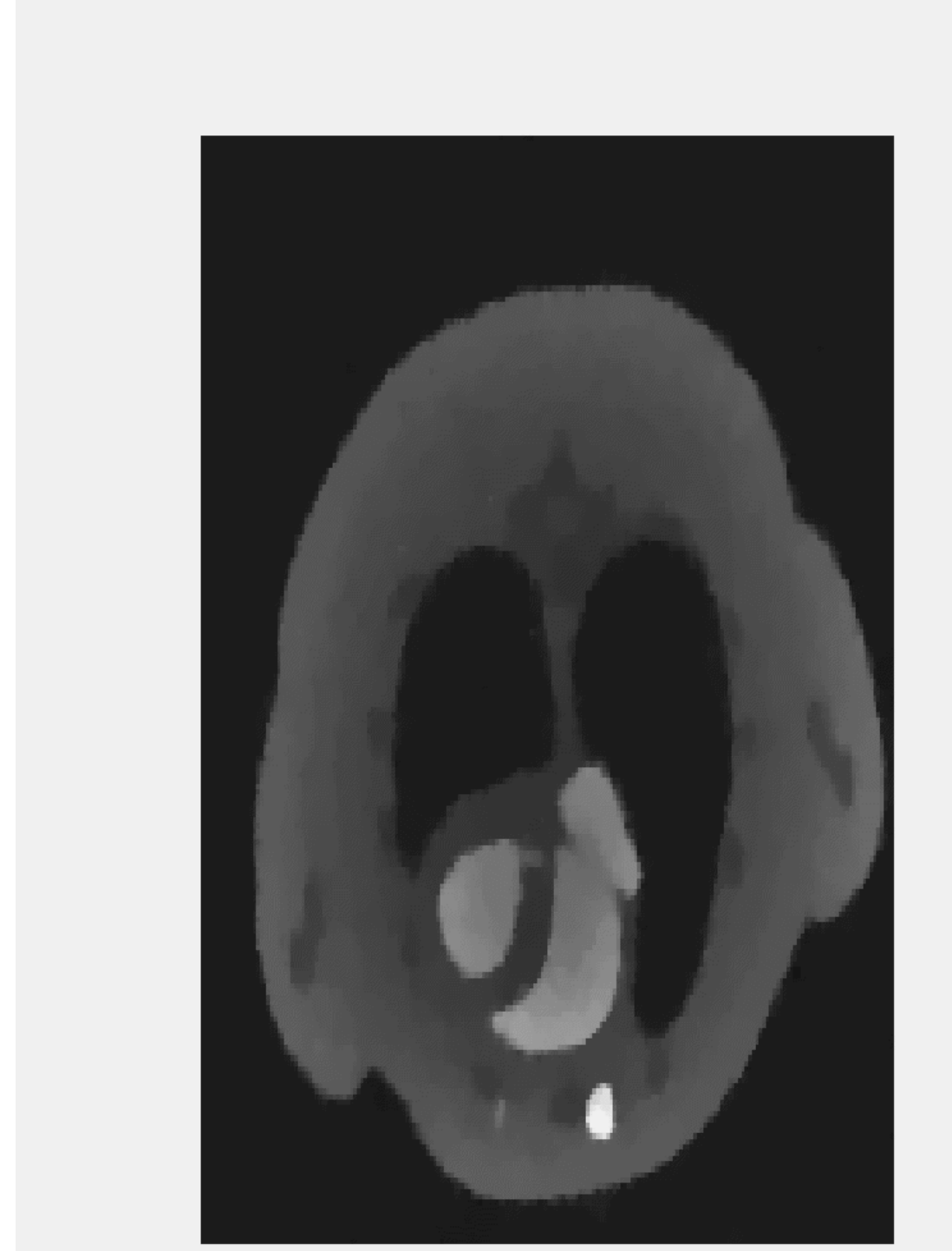}
    }
    
    \resizebox{.95\textwidth}{!}{\includegraphics[angle = 270, page = 2, trim={4.5cm 0 3cm 4cm },clip]{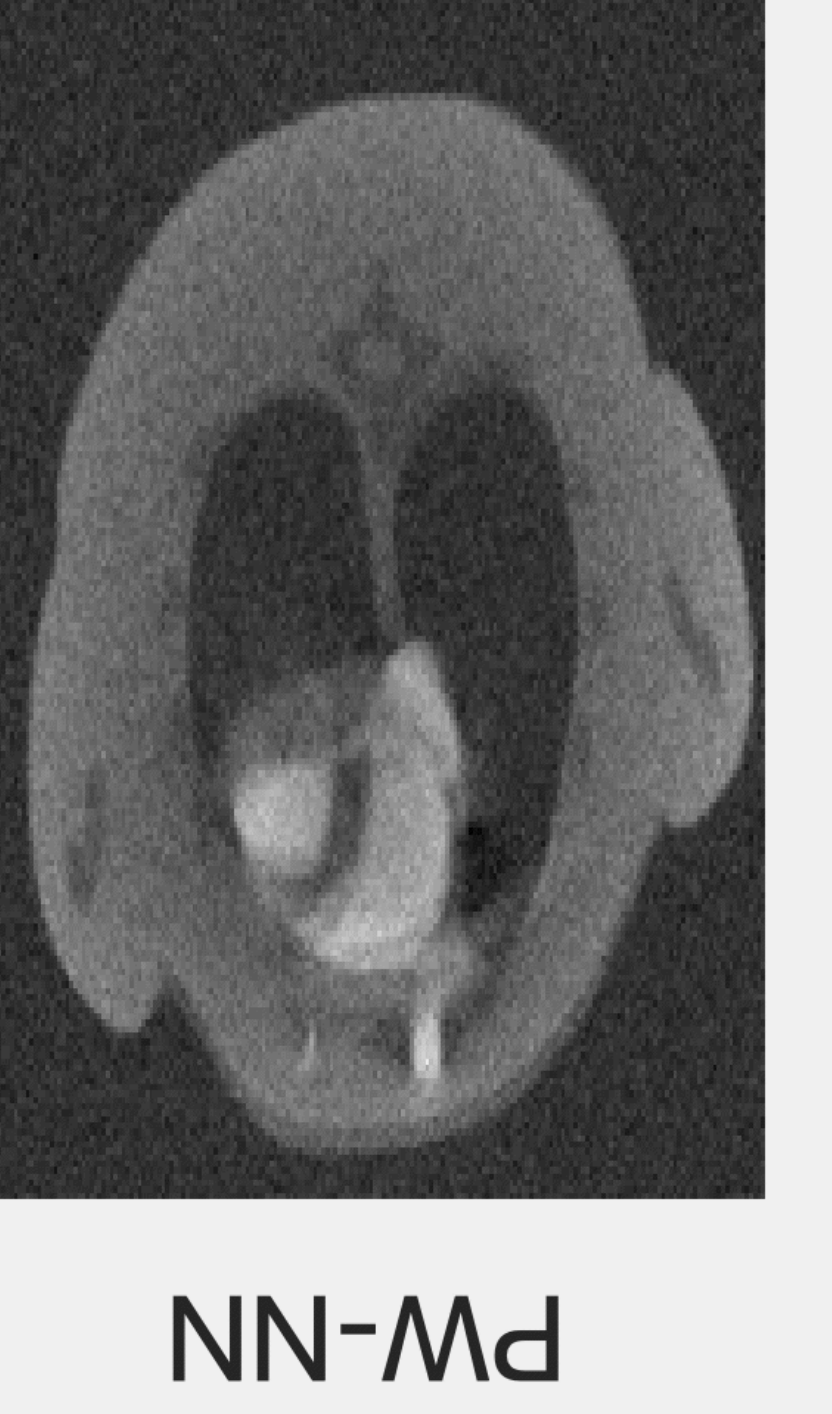}\includegraphics[angle = 270, page = 2, trim={4.5cm 3.5cm 3cm 4cm },clip]{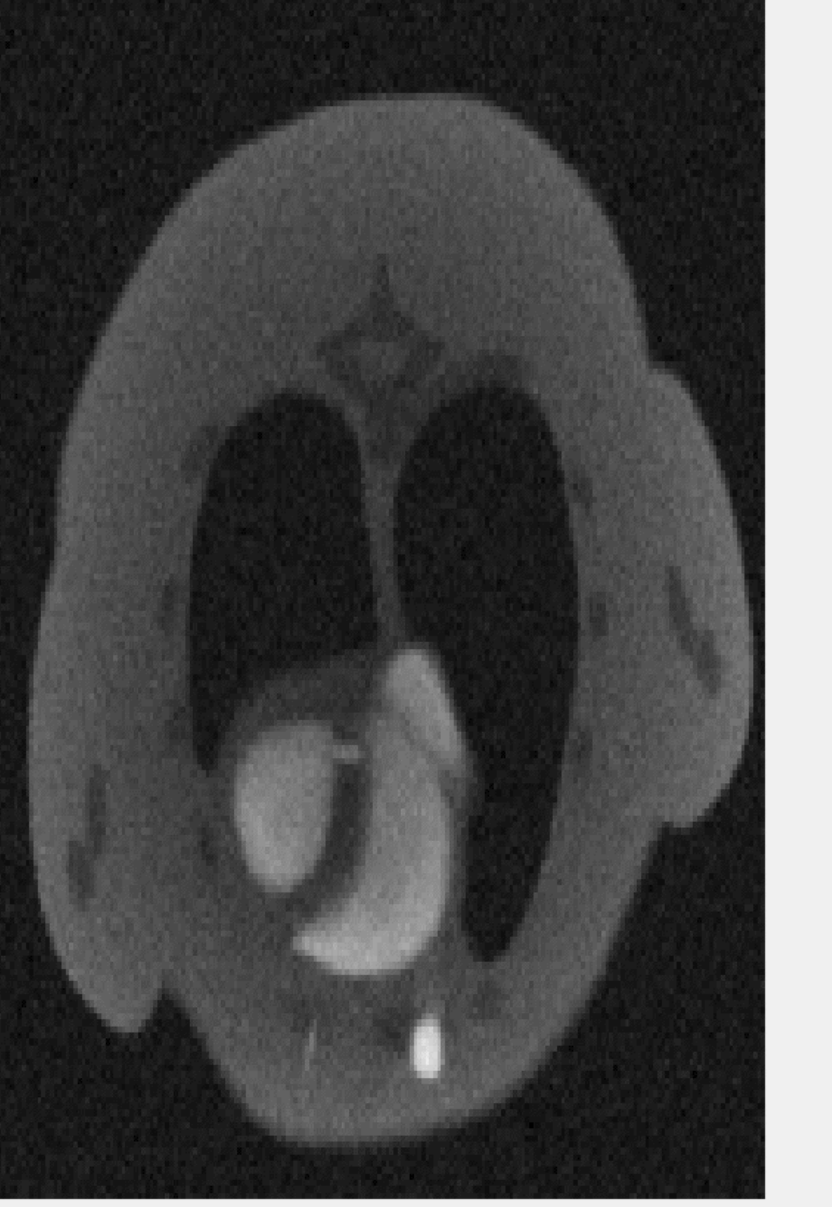}\includegraphics[angle = 270, page =2, trim={4.5cm 3.5cm 3cm 4cm },clip]{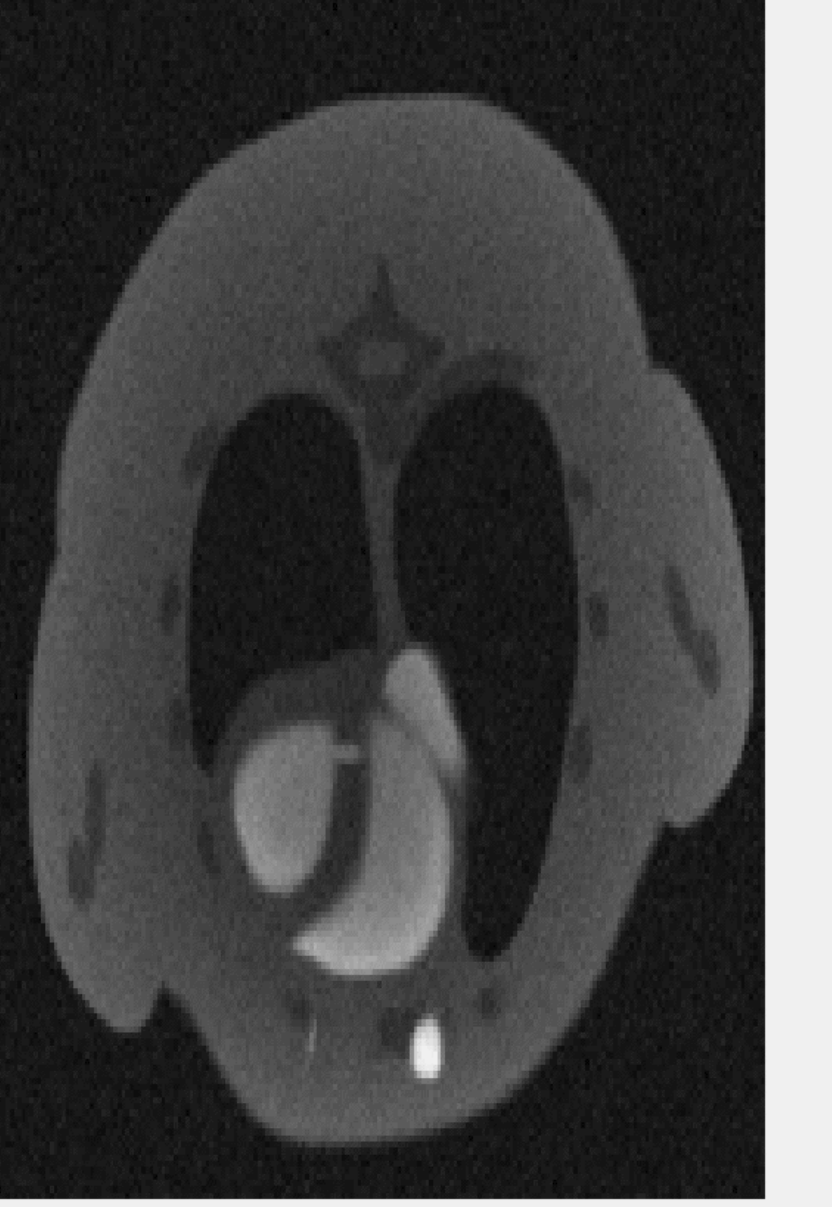} \hspace{4cm}  \includegraphics[angle = 270, page = 5, trim={4.5cm 3.5cm 3cm 4cm },clip]{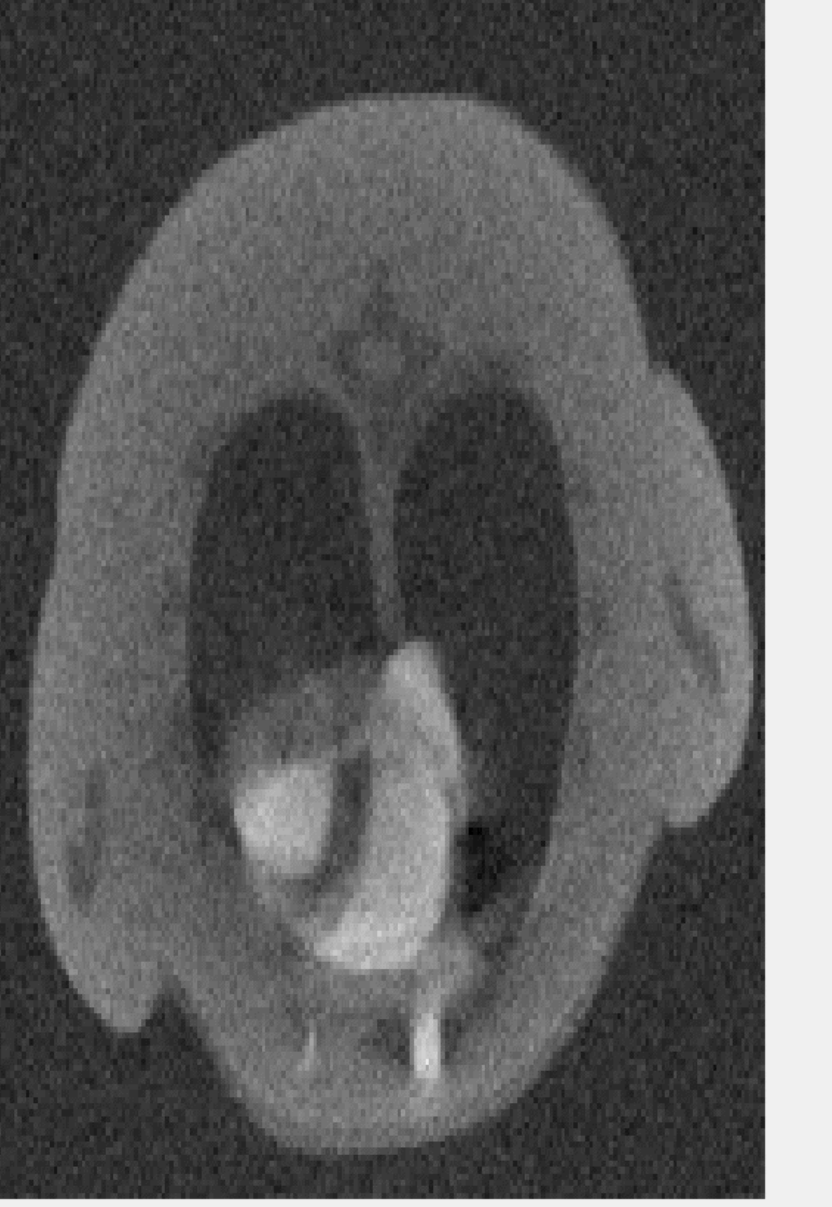}\includegraphics[angle = 270, page = 5, trim={4.5cm 3.5cm 3cm 4cm },clip]{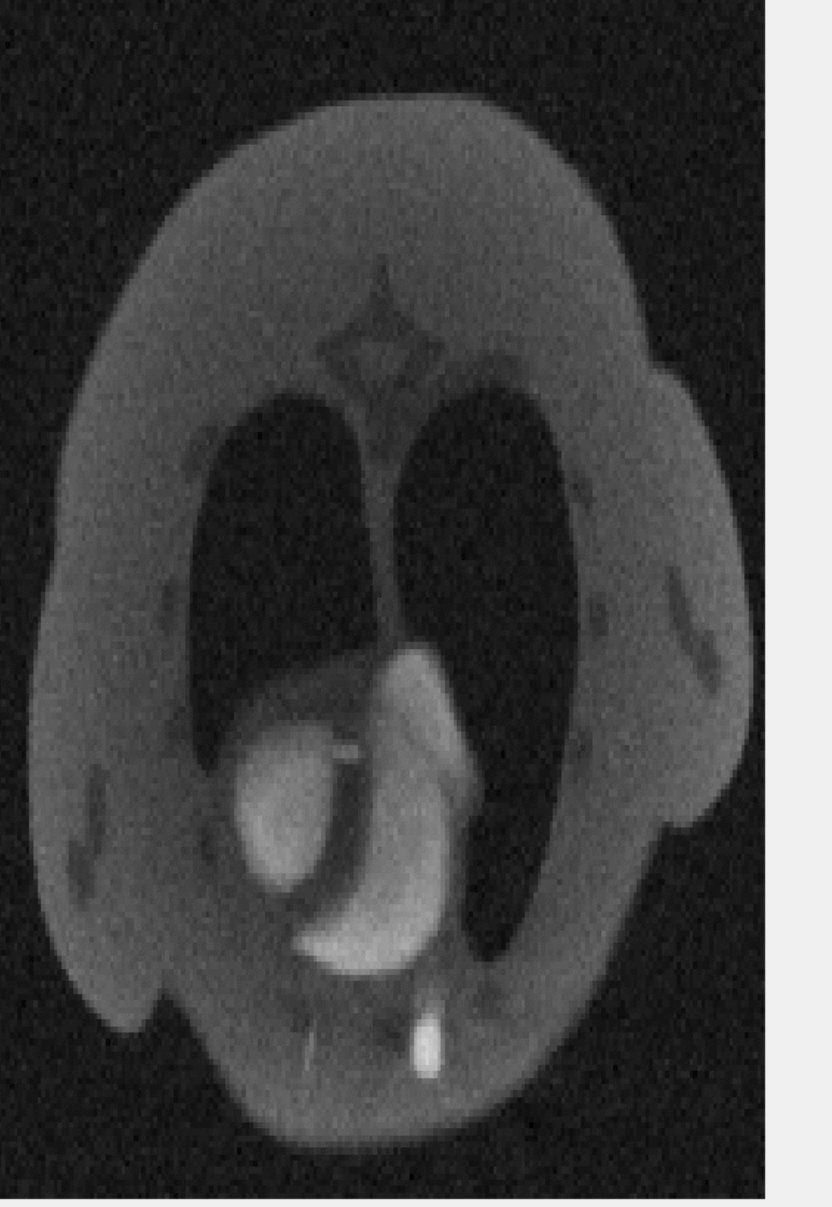}\includegraphics[angle = 270, page = 5, trim={4.5cm 3.5cm 3cm 0 },clip]{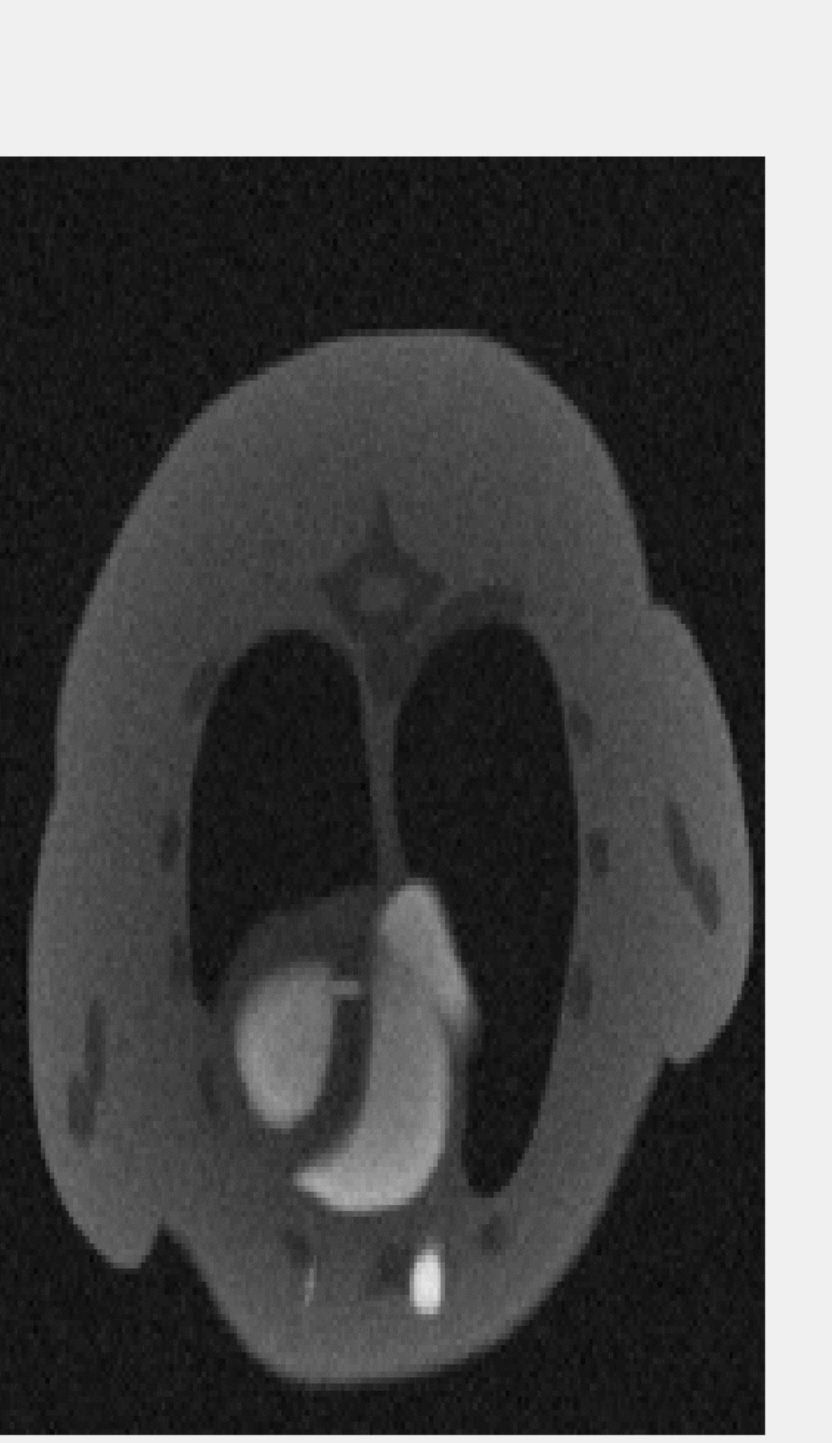}
    }
    
    \resizebox{.95\textwidth}{!}{\includegraphics[angle = 270, page = 2, trim={4.5cm 0 0 4cm },clip]{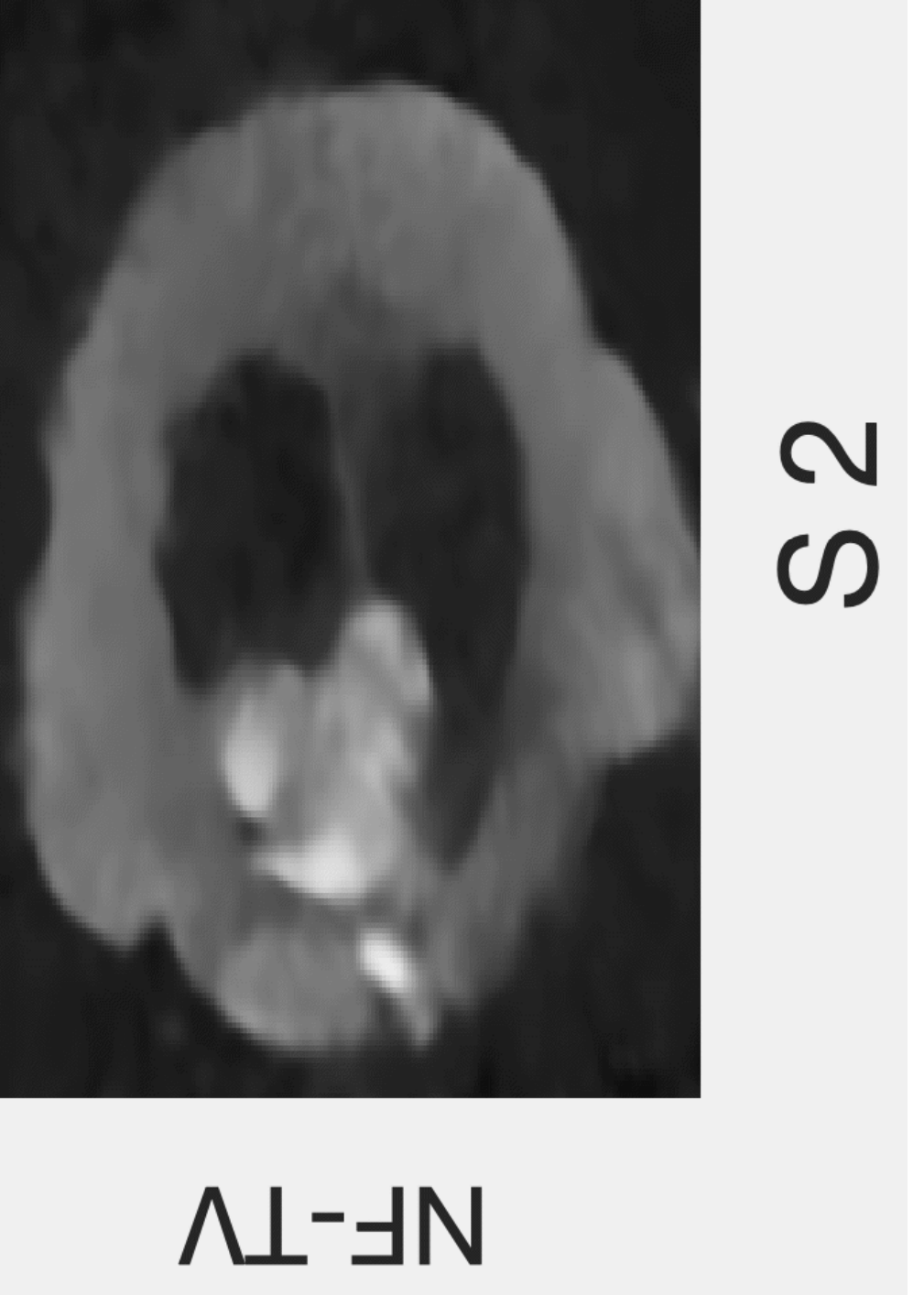}\includegraphics[angle = 270, page = 2, trim={4.5cm 3.5cm 0 4cm },clip]{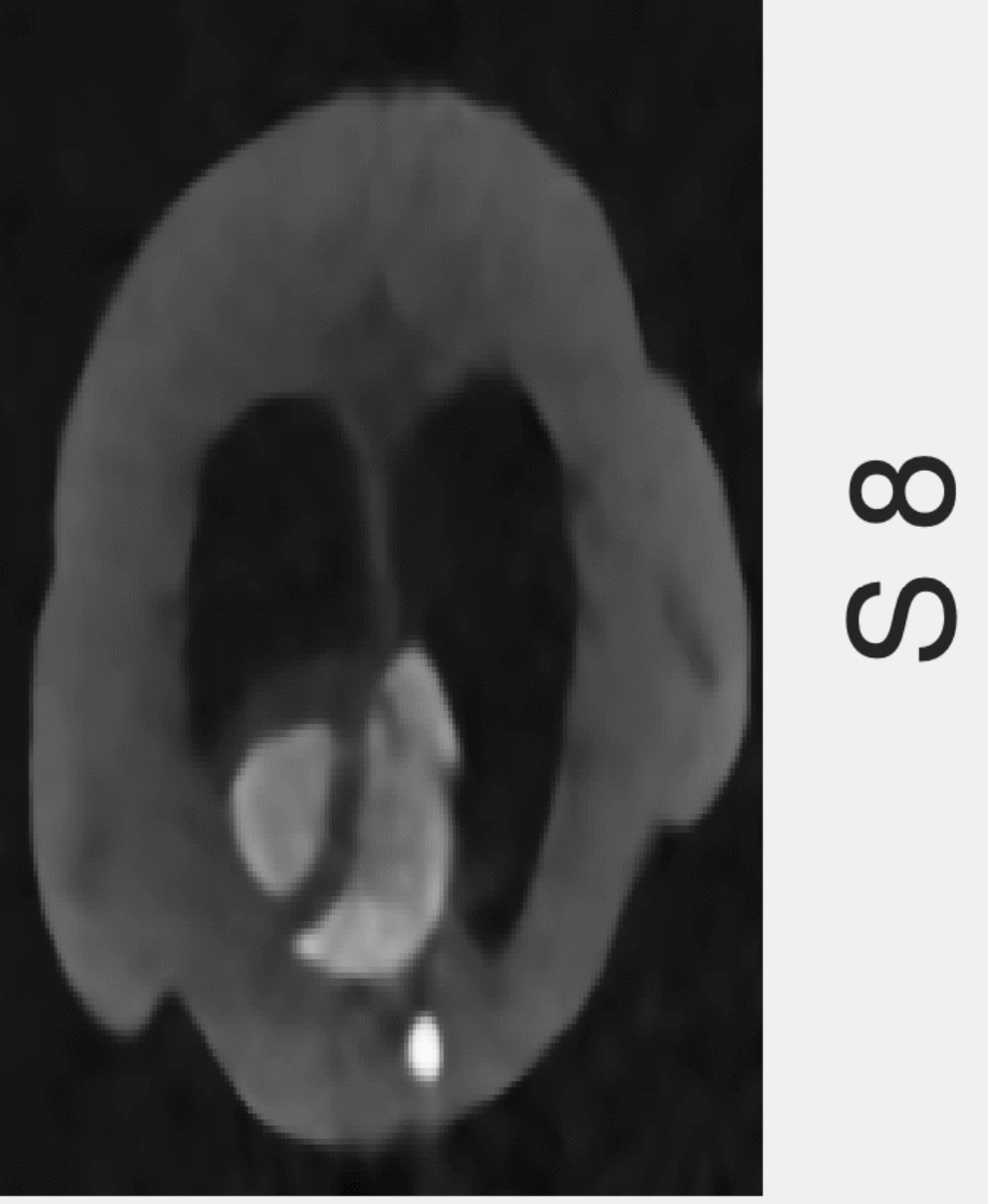}\includegraphics[angle = 270, page =2, trim={4.5cm 3.5cm 0 4cm },clip]{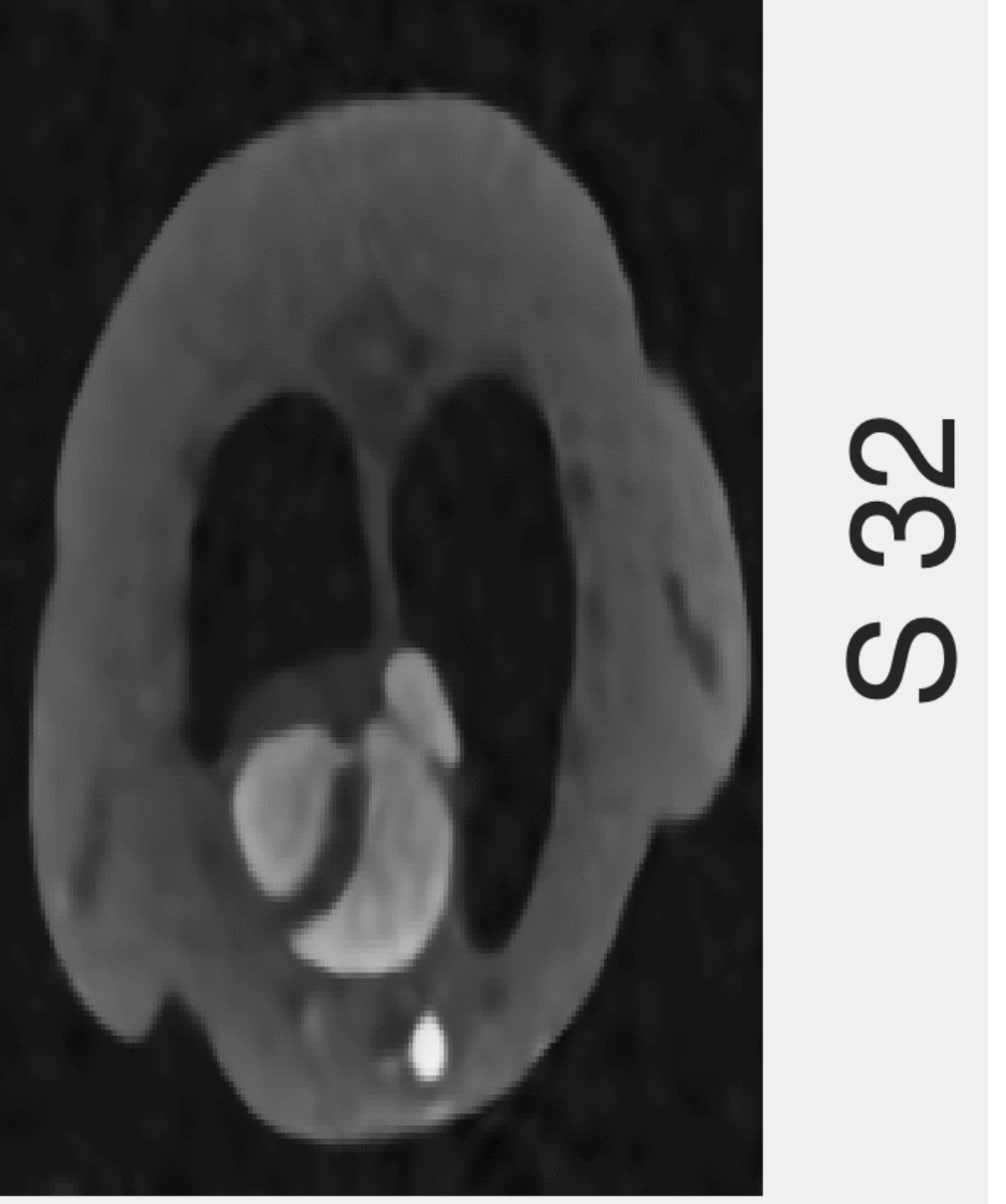}\hspace{4cm}  \includegraphics[angle = 270, page = 5, trim={4.5cm 3.5cm 0 4cm },clip]{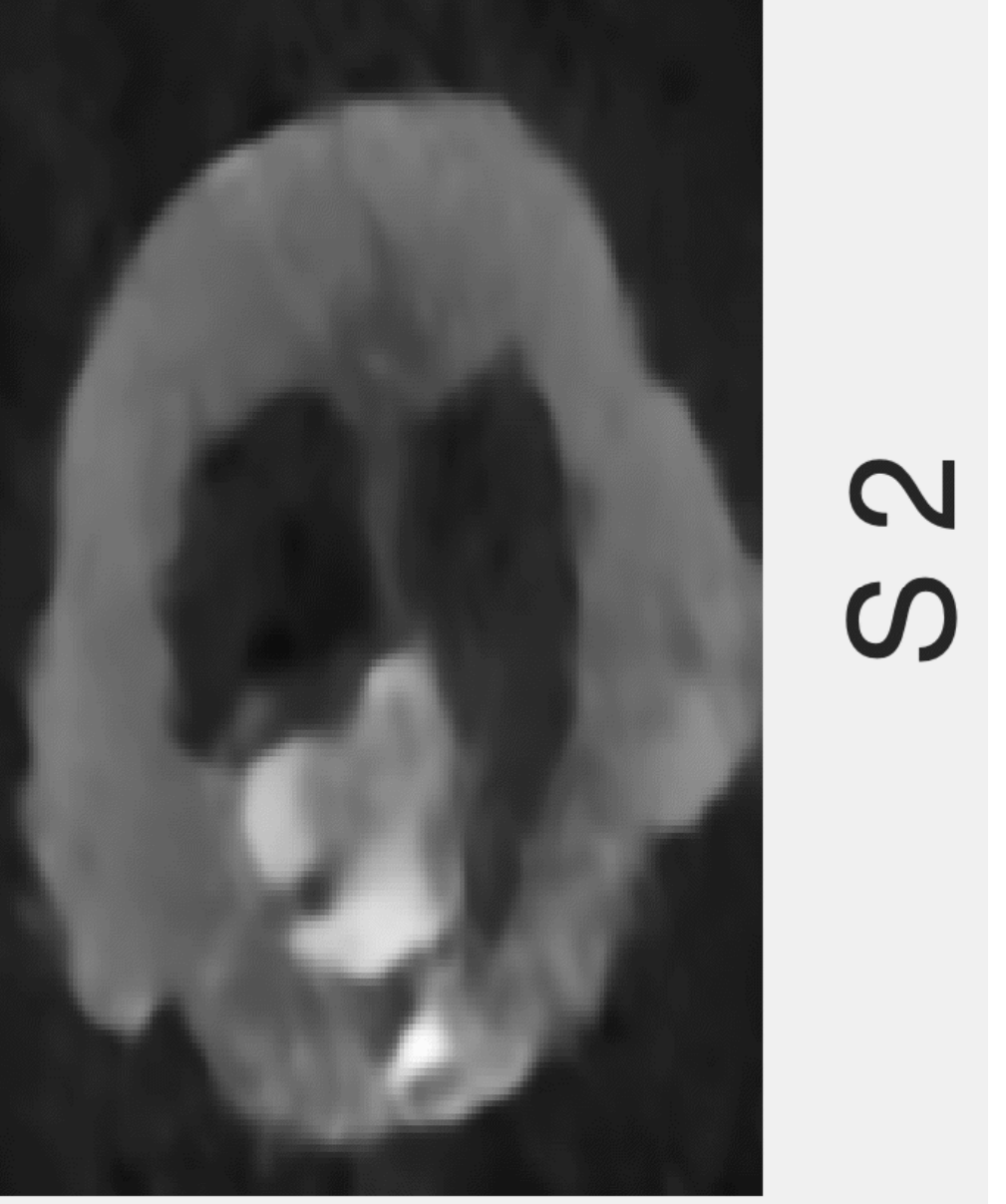}\includegraphics[angle = 270, page = 5, trim={4.5cm 3.5cm 0 4cm },clip]{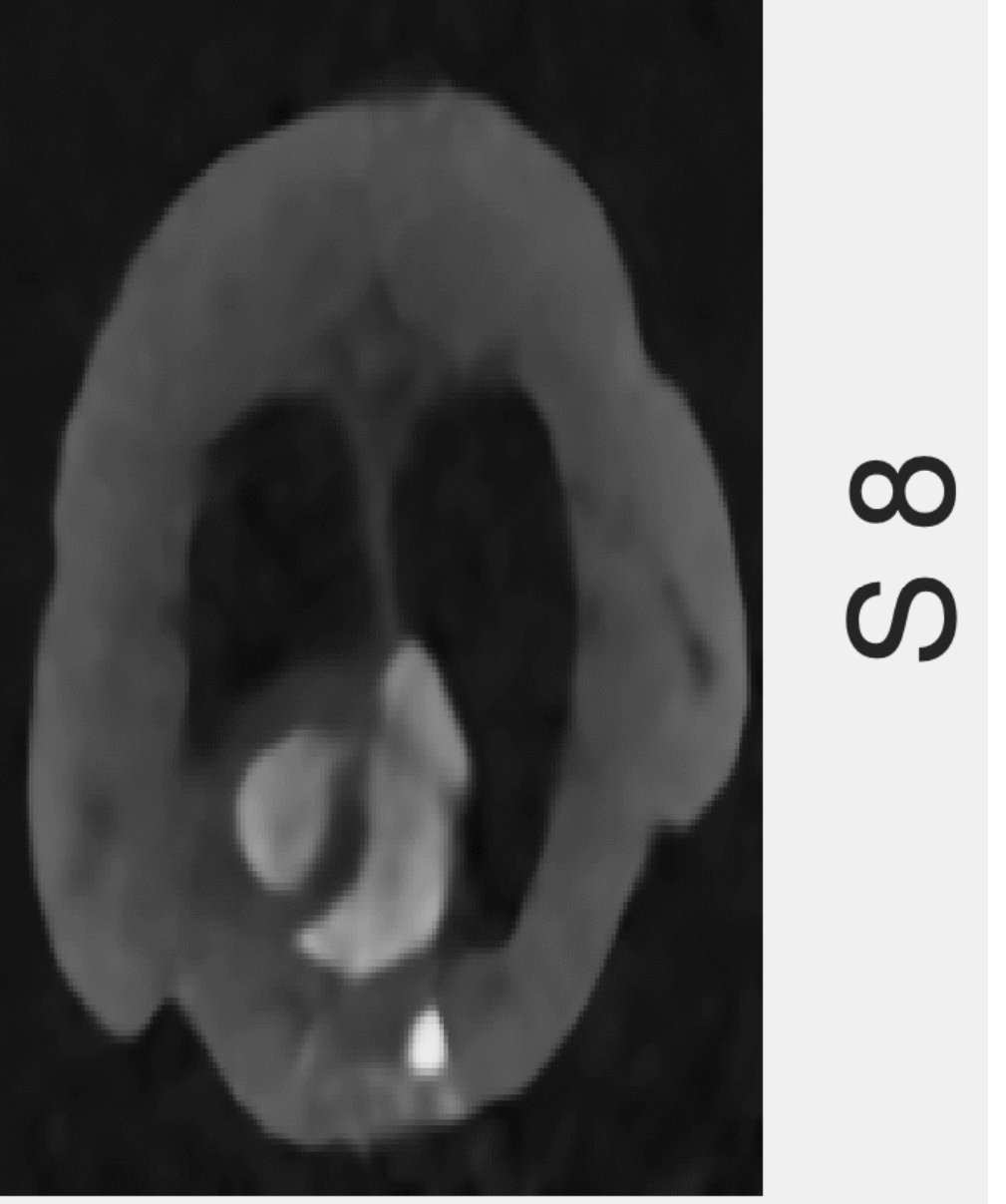}\includegraphics[angle = 270, page = 5, trim={4.5cm 3.5cm 0 0 },clip]{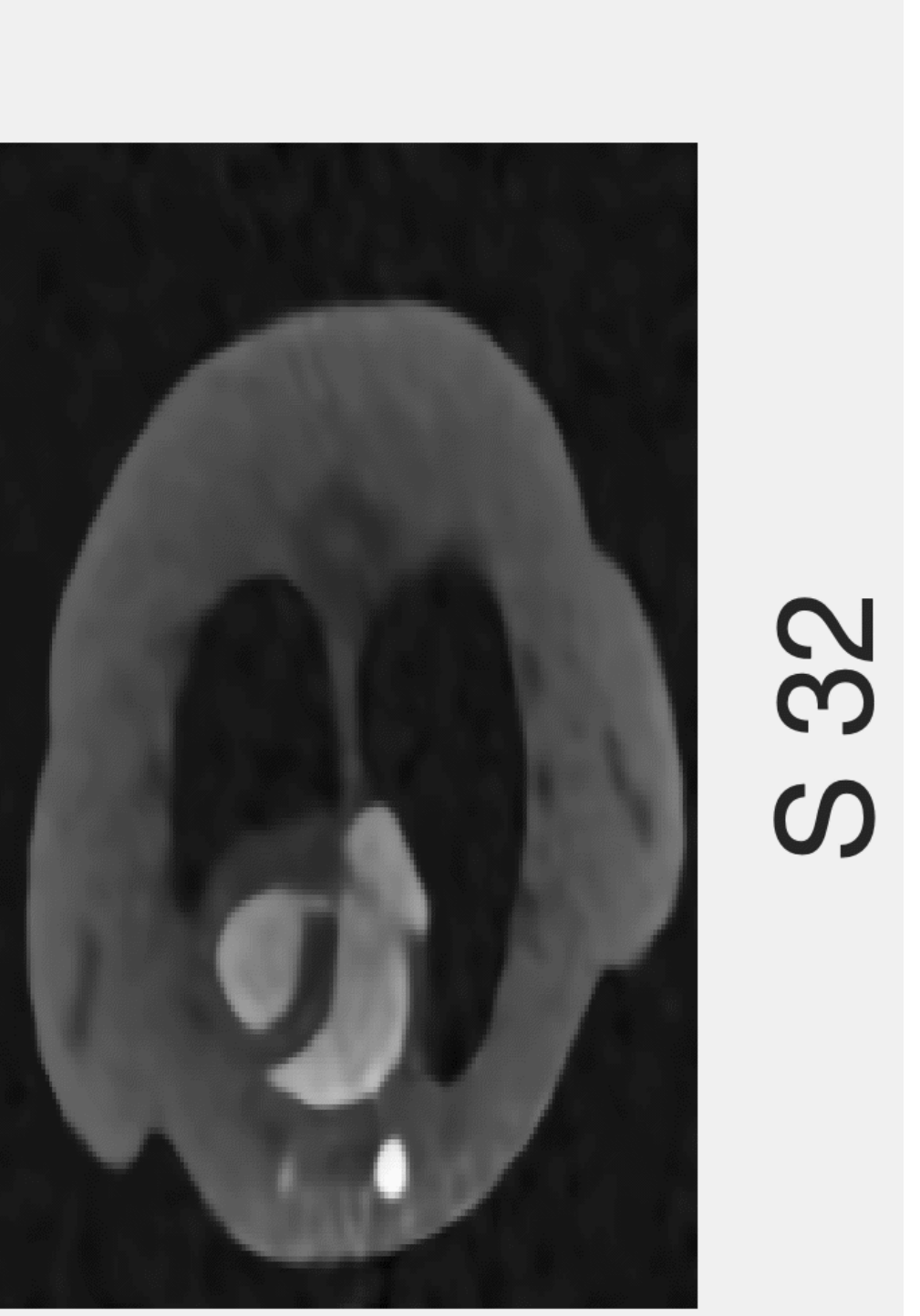}
    }

    \caption{Dynamic image reconstruction problem: Pixel-wise Total Variation (PW-TV), pixel-wise nuclear norm (PW-NN), and neural field Total Variation (NF-TV) reconstructions at $S=2$ views per frame, $S=8$ views per frame, and $S=32$ views per frame. NF-TV has robust performance with a varied number of views per frame, and a better reconstruction is only achieved with the PW-TV at a high number of views per frame. An animation of the dynamic reconstructions comparing the PW-TV, PW-NN, and NF-TV methods for different values of the number of views per frame $S$  is available in the supplemental multimedia materials (Video 4).}
    \label{fig:vpt_anims}
\end{figure*}

\begin{figure*}
    \centering
    \includegraphics[width = \textwidth]{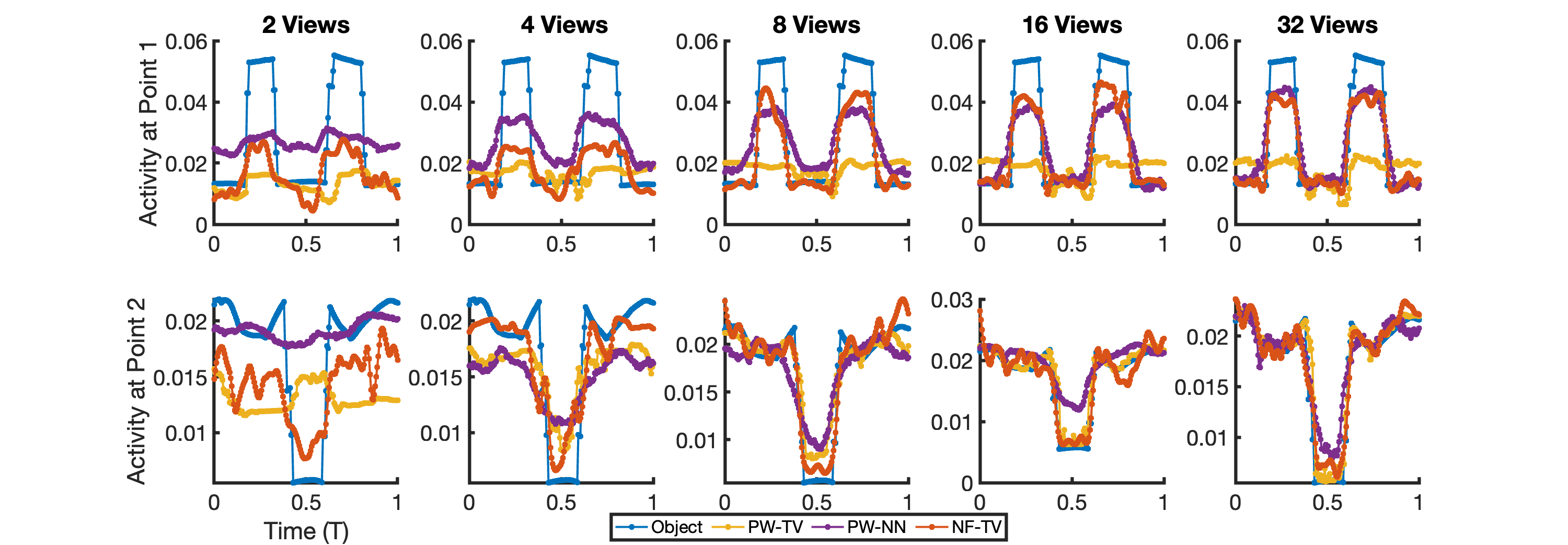}
    \caption{Dynamic image reconstruction problem: Time-activity curves for varying number of views per frame. The neural field captures the dynamic behavior of the object, while the pixel-wise approaches can only capture dynamic behavior with several views per time.} 
    \label{fig:vpt_activity}
\end{figure*}

The first experiment reconstructed the image for a CRT forward operator with a varying number of views per frame. The number of views per frame $S$ varied among 2, 4, 8, 16, and 32. The relative noise, defined as the ratio of the noise standard deviation and the infinity norm of the noiseless simulated data, was fixed to $0.025$ throughout the experiment. Estimates of the dynamic object reconstructed by the three methods using 2,8 and 32 views per frame are shown in Fig. \ref{fig:vpt_anims}. Figure \ref{fig:vpt_activity} displays the time-activity curves for the two points shown in Fig. \ref{fig:embedding_activity}.

Table \ref{tab:vpt_acc} displays the RRMSE, SSIM, optimal regularization values for each reconstruction, and the rank and number of parameters of the nuclear norm reconstruction. The NF-TV approach outperforms the classical PW-TV and PW-NN in terms of RRMSE and SSIM for all settings except the cases 16 and 32 views per frame ($S$ = 16 or 32), where PW-TV is slightly better. Based on RRMSE and SSIM, the neural field approach leads to better quality reconstructions for severely undersampled data ($S$ = 2, 4, or 8). Additionally, the neural field representation significantly reduced the number of stored parameters. For every reconstruction, the neural field required storing 86K parameters and the pixel-wise TV approach required 7.2M parameters. The nuclear norm required between 361K and 5.7M parameters depending on the number of views per frame.

\begin{table}
    \centering
    \caption{Accuracy metrics and optimal regularization parameters for different number of views per frame. The number of parameters for the PW-TV and NF-TV are 7.6M and 86K, respectively, independently of the number of views.}
    \begin{tabular}{c|ccccc}
    Views per Frame & 2 & 4& 8 & 16 & 32\\ 
    \hline
    RRMSE PW-TV & 0.5343 & 0.3202 & 0.2029 & \textbf{0.1545} & \textbf{0.1388} \\ 
    RRMSE PW-NN & \textbf{0.3158} & \textbf{0.2569} & 0.2184 & 0.1935 & 0.1815 \\
    RRMSE NF-TV &  0.3236 & 0.2574 & \textbf{0.1793} & 0.1758 & 0.1615 \\  
    \hline
    SSIM PW-TV & 0.2984 & 0.5628 & 0.7473 & \textbf{0.8712} & \textbf{0.8990} \\ 
    SSIM PW-NN & 0.2425 & 0.3153 & 0.3993 & 0.4720 & 0.4885 \\ 
    SSIM NF-TV & \textbf{0.5266}  & \textbf{0.6308}  & \textbf{0.8174}  &  0.8020  &  0.8020\\  
    \hline
    Optimal $\gamma$ PW-TV & 12.5 & 6.25 & 6.25 & 3.25 & 1.75 \\ 
    Optimal $\gamma$ PW-NN & 0.0500 & 0.0354 & 0.0250 & 0.0177 & 0.0088\\ 
    Optimal $\gamma$ NF-TV  & 2 & 1 & 0.5 & 0 & 0 \\
    \hline 
    PW-NN Rank $r$ & 9 & 12 & 13 & 15 & 142\\
    PW-NN parameters &  361K & 482K & 522K & 602K & 5.7M\\
\end{tabular}
    
    \label{tab:vpt_acc}
\end{table}

\subsubsection{Varying peak signal to noise ratio}

\begin{figure*}
    \centering
    \resizebox{.95\textwidth}{!}{\includegraphics[angle = 270, page = 2, trim={0 0 3cm 4cm },clip]{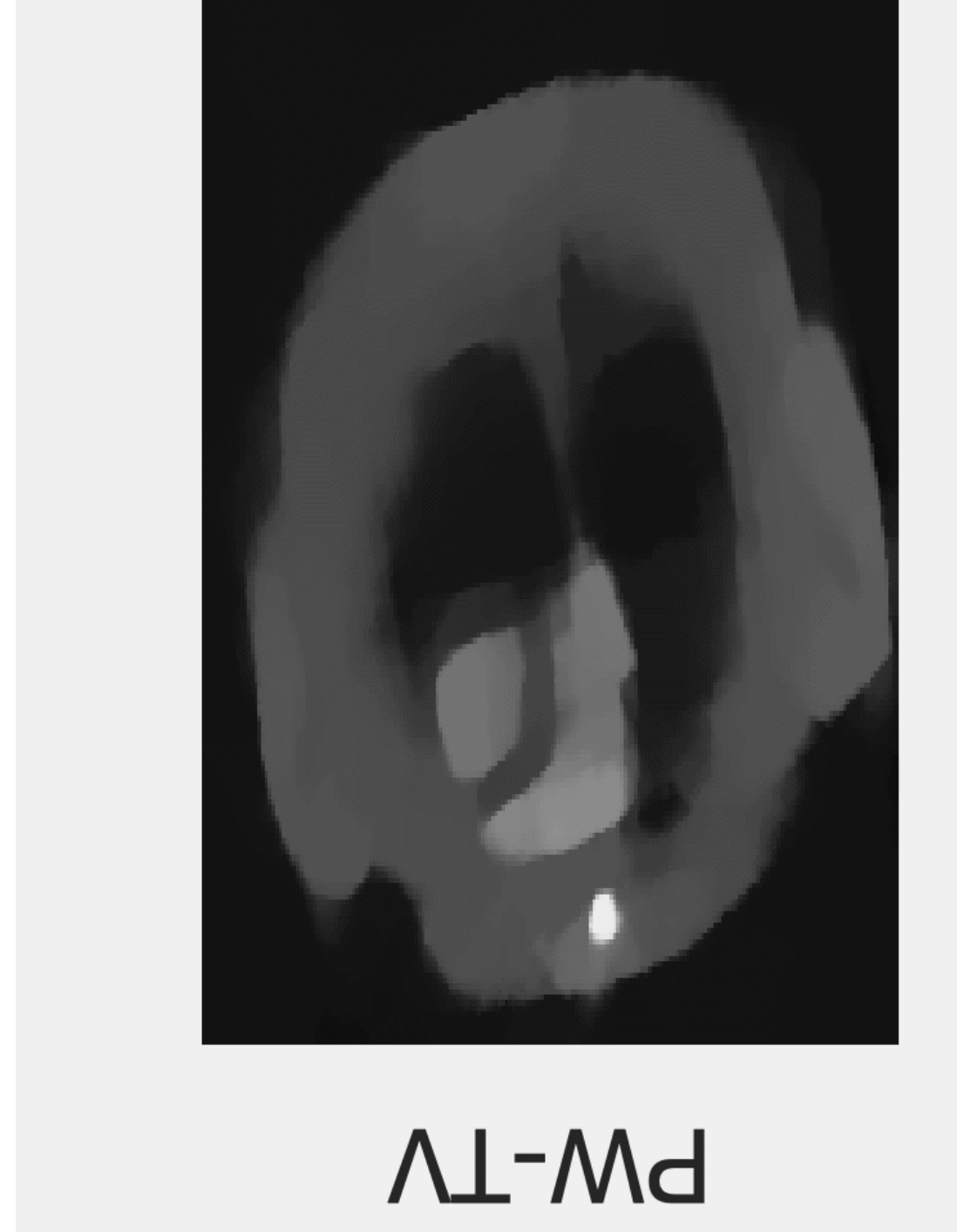}\includegraphics[angle = 270, page = 2, trim={0 3.5cm 3cm 4cm },clip]{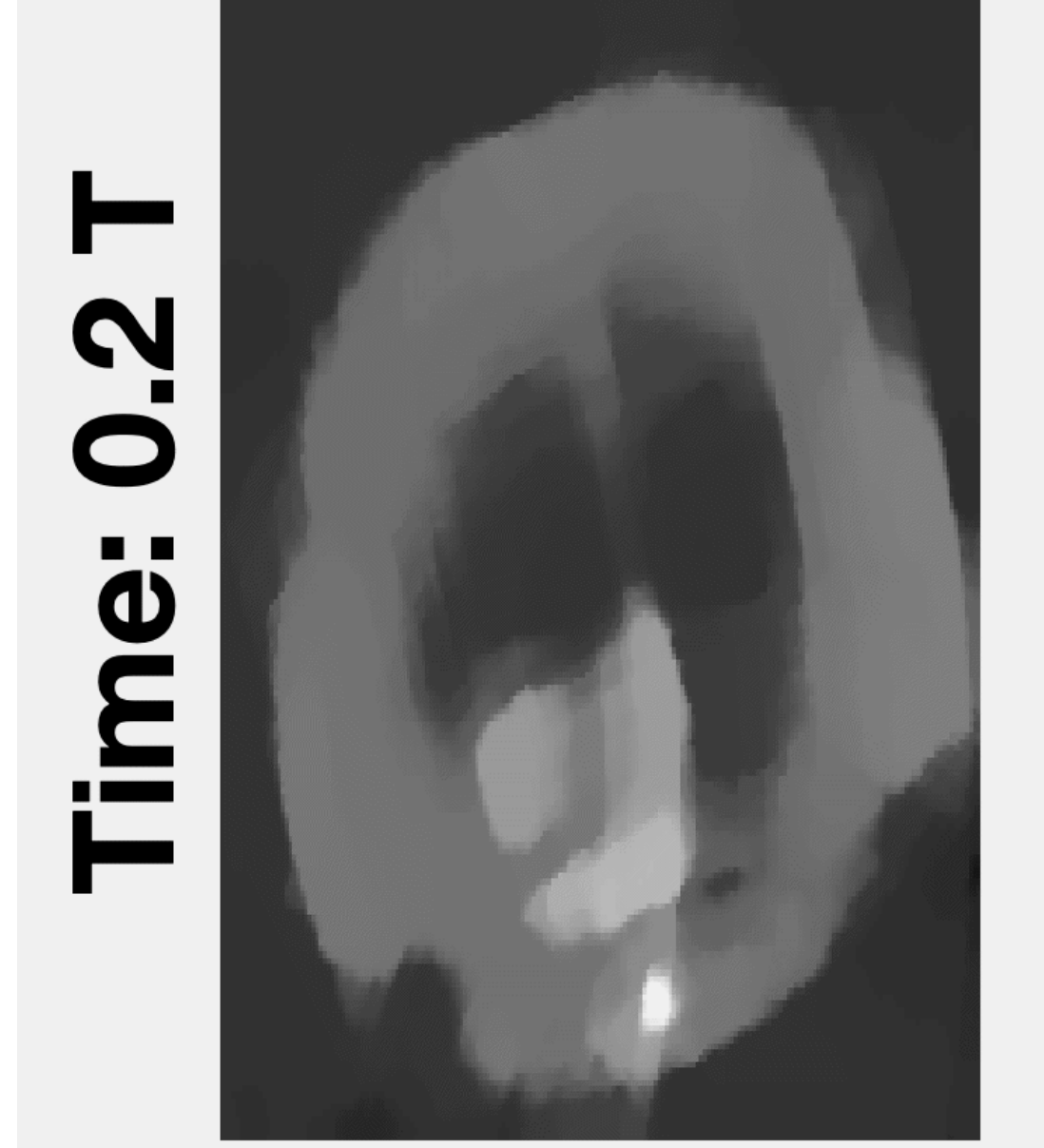}\includegraphics[angle = 270, page =2, trim={0 3.5cm 3cm 4cm },clip]{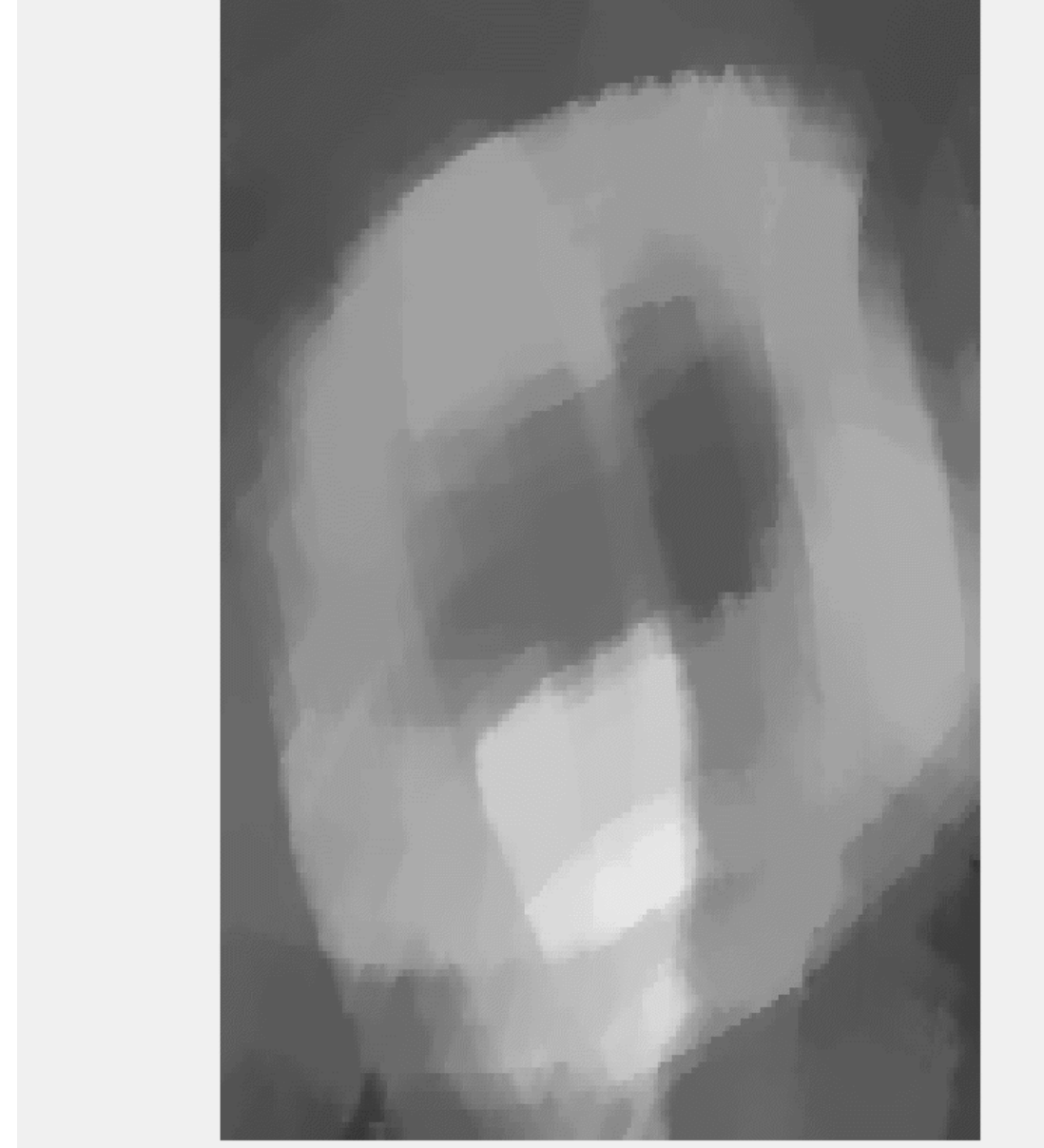}\hspace{4cm}  \includegraphics[angle = 270, page = 5, trim={0 3.5cm 3cm 4cm },clip]{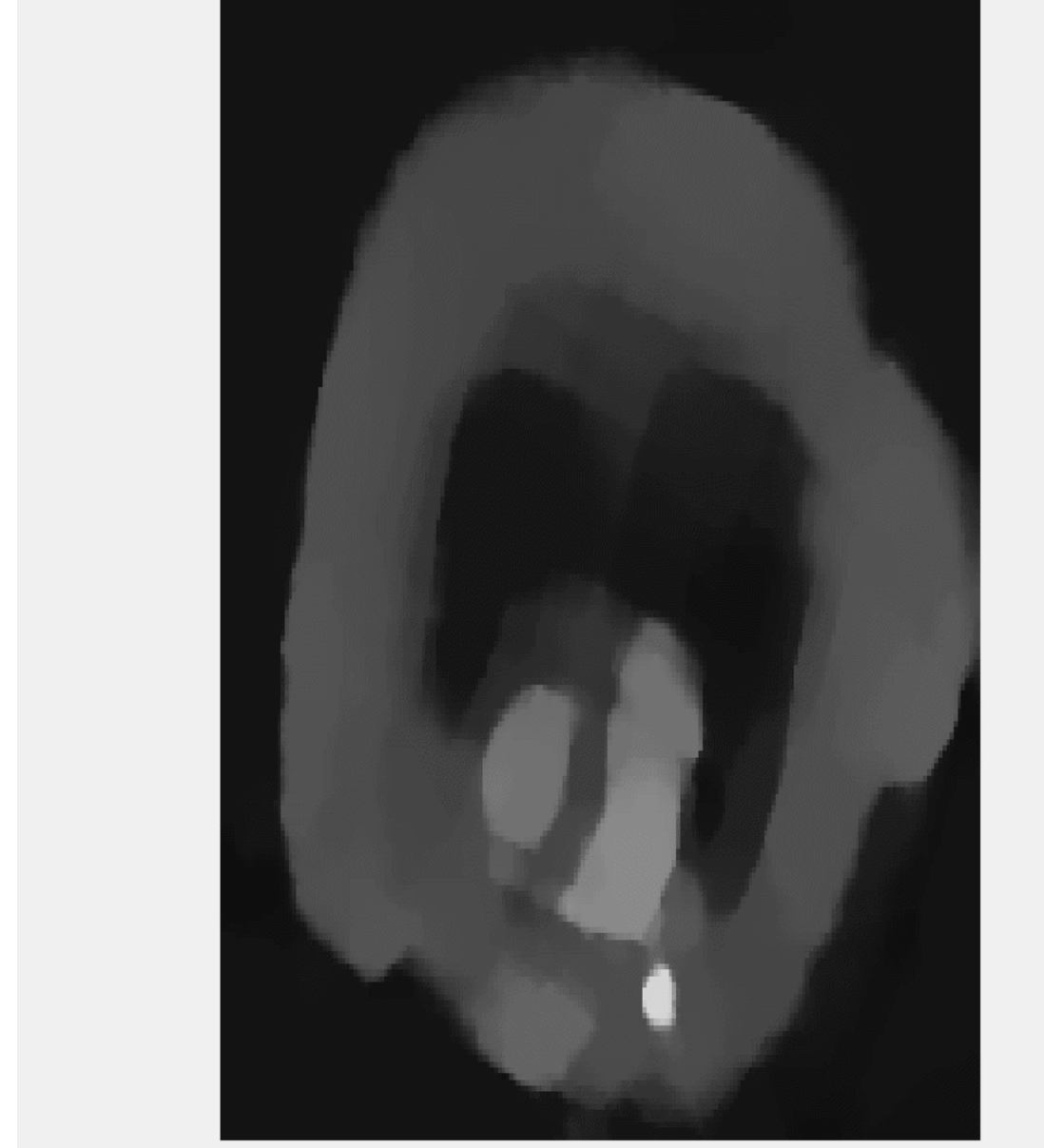}\includegraphics[angle = 270, page = 5, trim={0 3.5cm 3cm 4cm },clip]{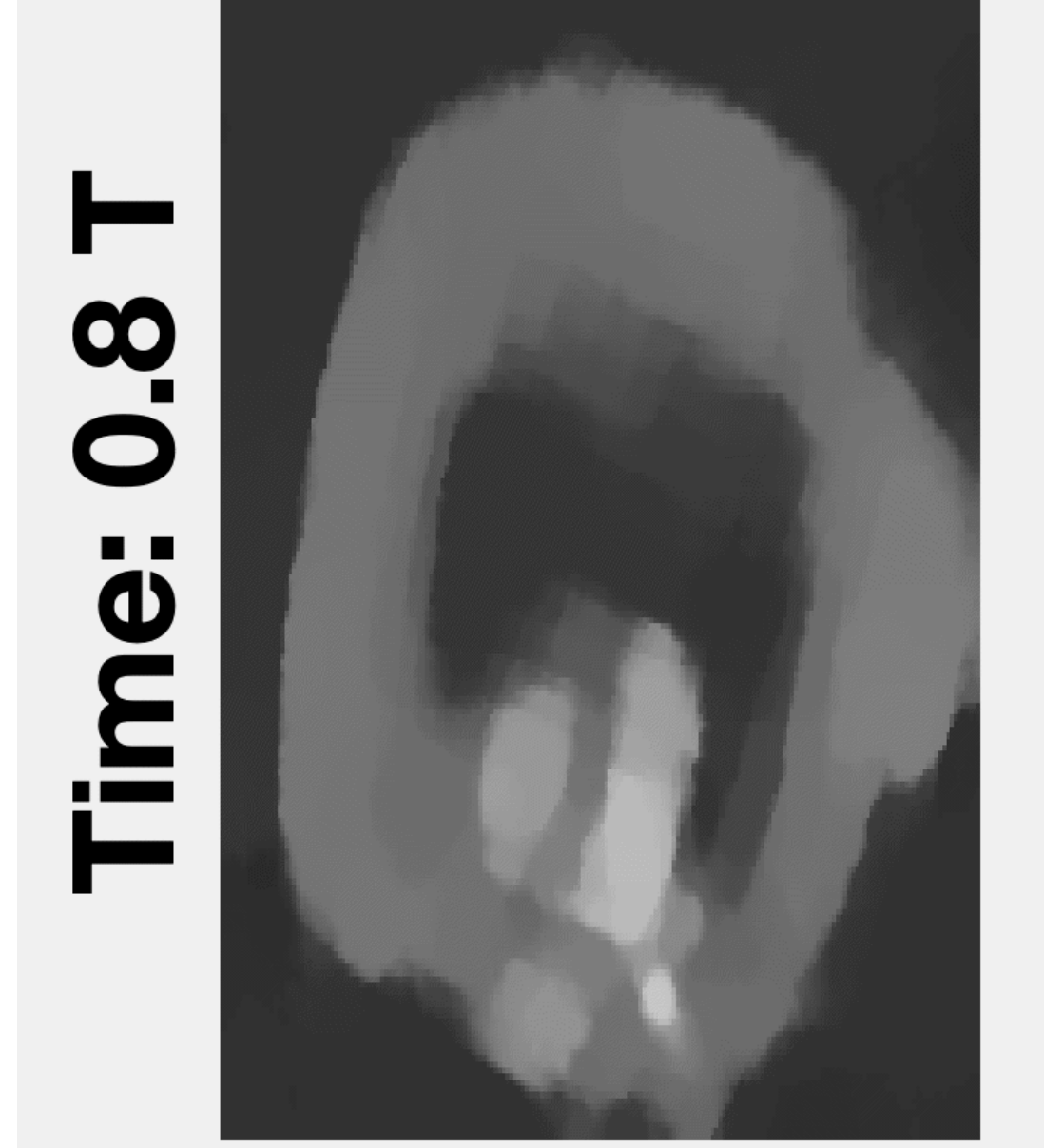}\includegraphics[angle = 270, page = 5, trim={0 3.5cm 3cm 0 },clip]{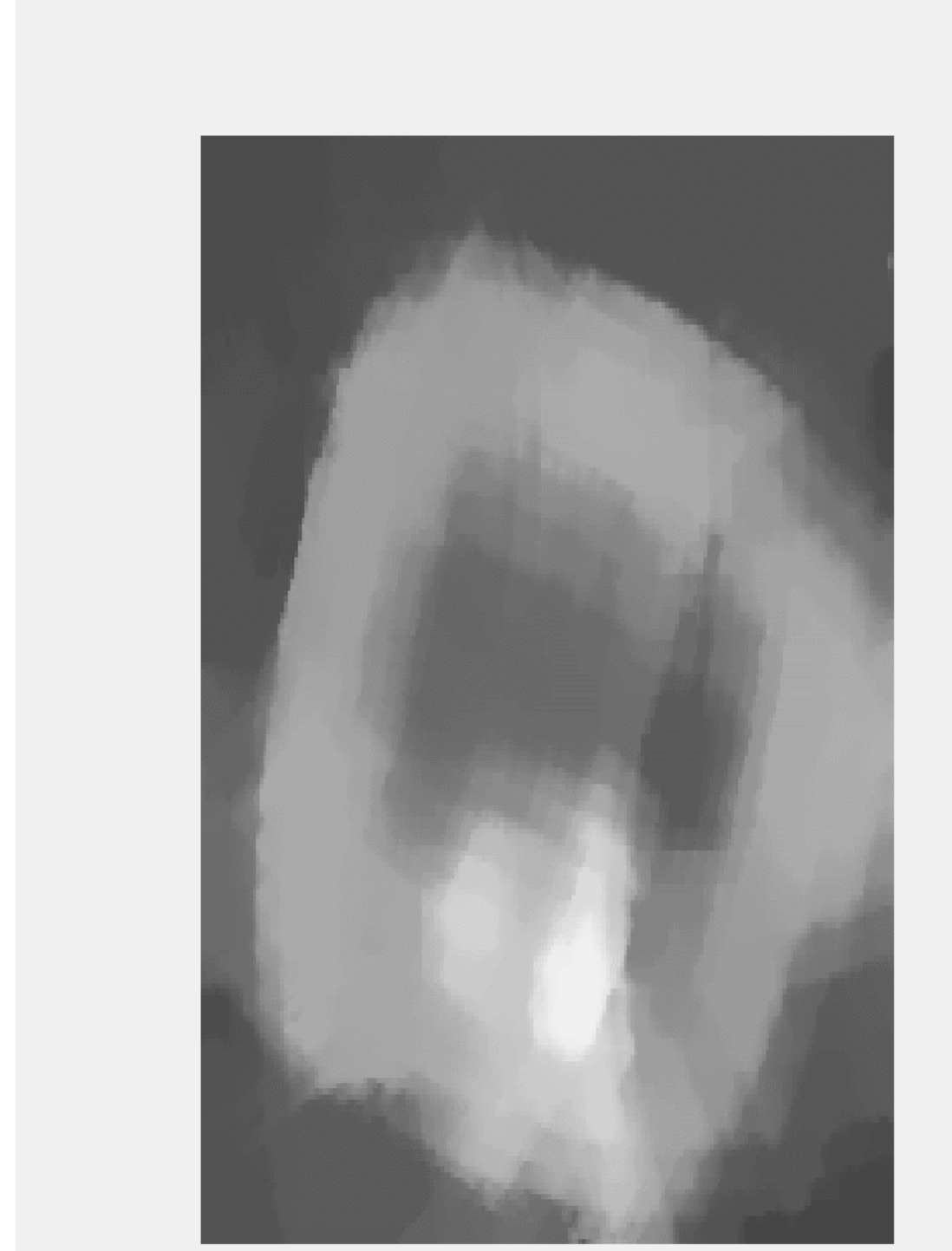}
    }
    
    \resizebox{.95\textwidth}{!}{\includegraphics[angle = 270, page = 2, trim={4.5cm 0 3cm 4cm },clip]{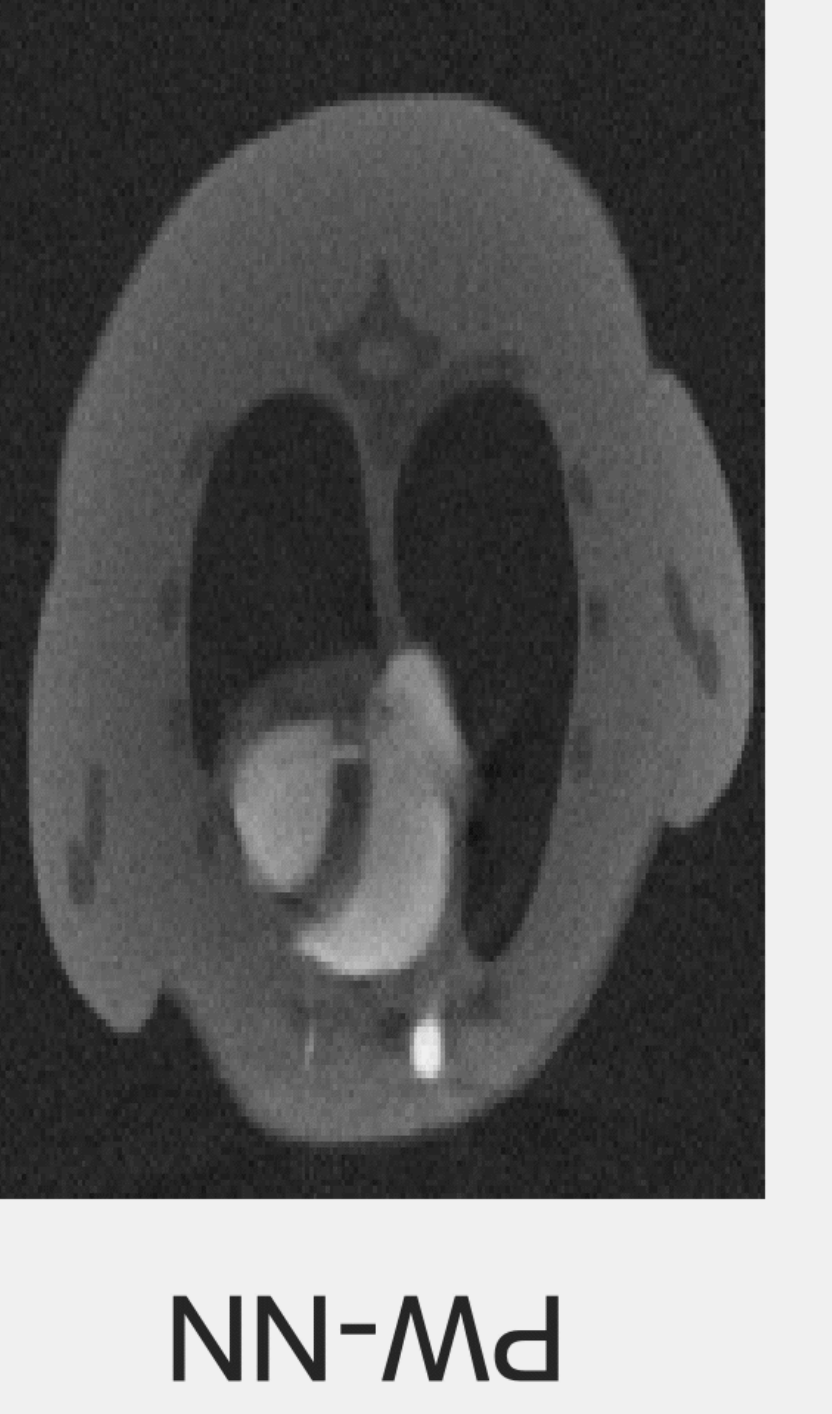}\includegraphics[angle = 270, page = 2, trim={4.5cm 3.5cm 3cm 4cm },clip]{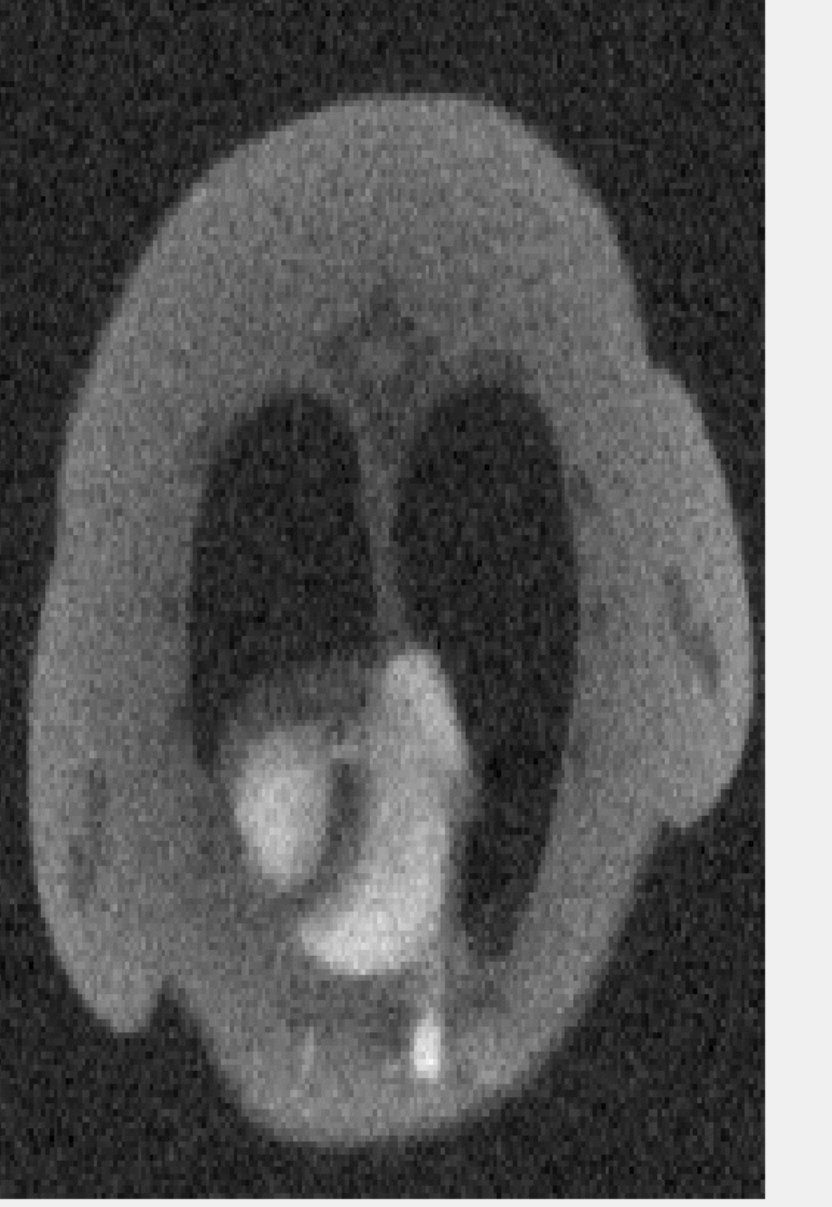}\includegraphics[angle = 270, page =2, trim={4.5cm 3.5cm 3cm 4cm },clip]{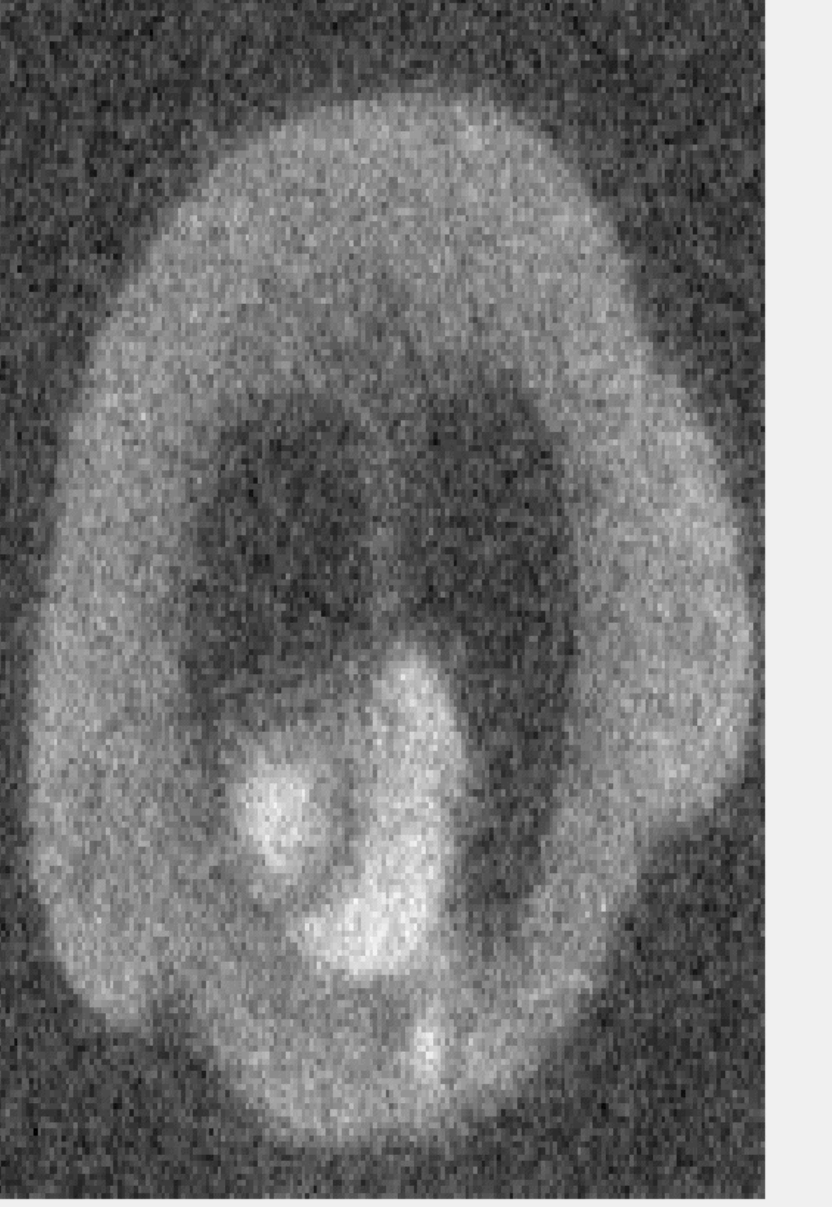}\hspace{4cm}  \includegraphics[angle = 270, page = 5, trim={4.5cm 3.5cm 3cm 4cm },clip]{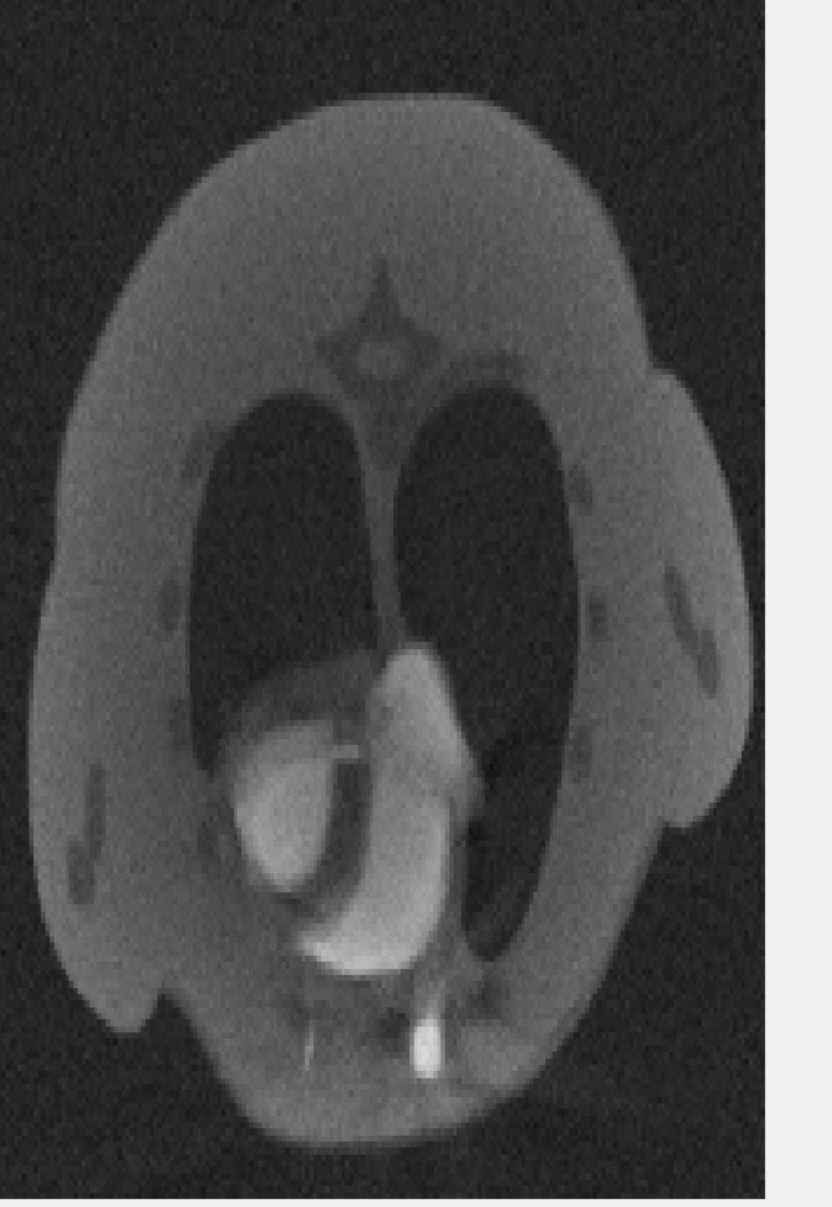}\includegraphics[angle = 270, page = 5, trim={4.5cm 3.5cm 3cm 4cm },clip]{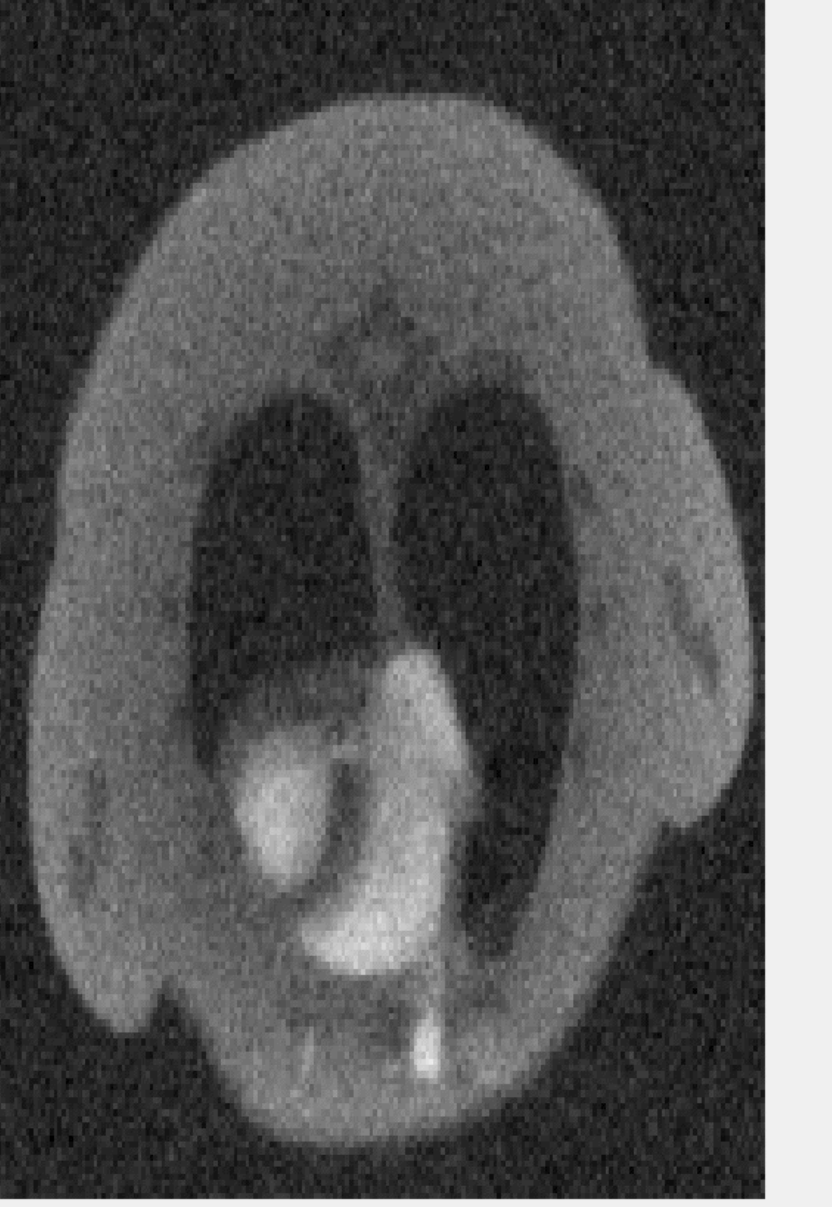}\includegraphics[angle = 270, page = 5, trim={4.5cm 3.5cm 3cm 0 },clip]{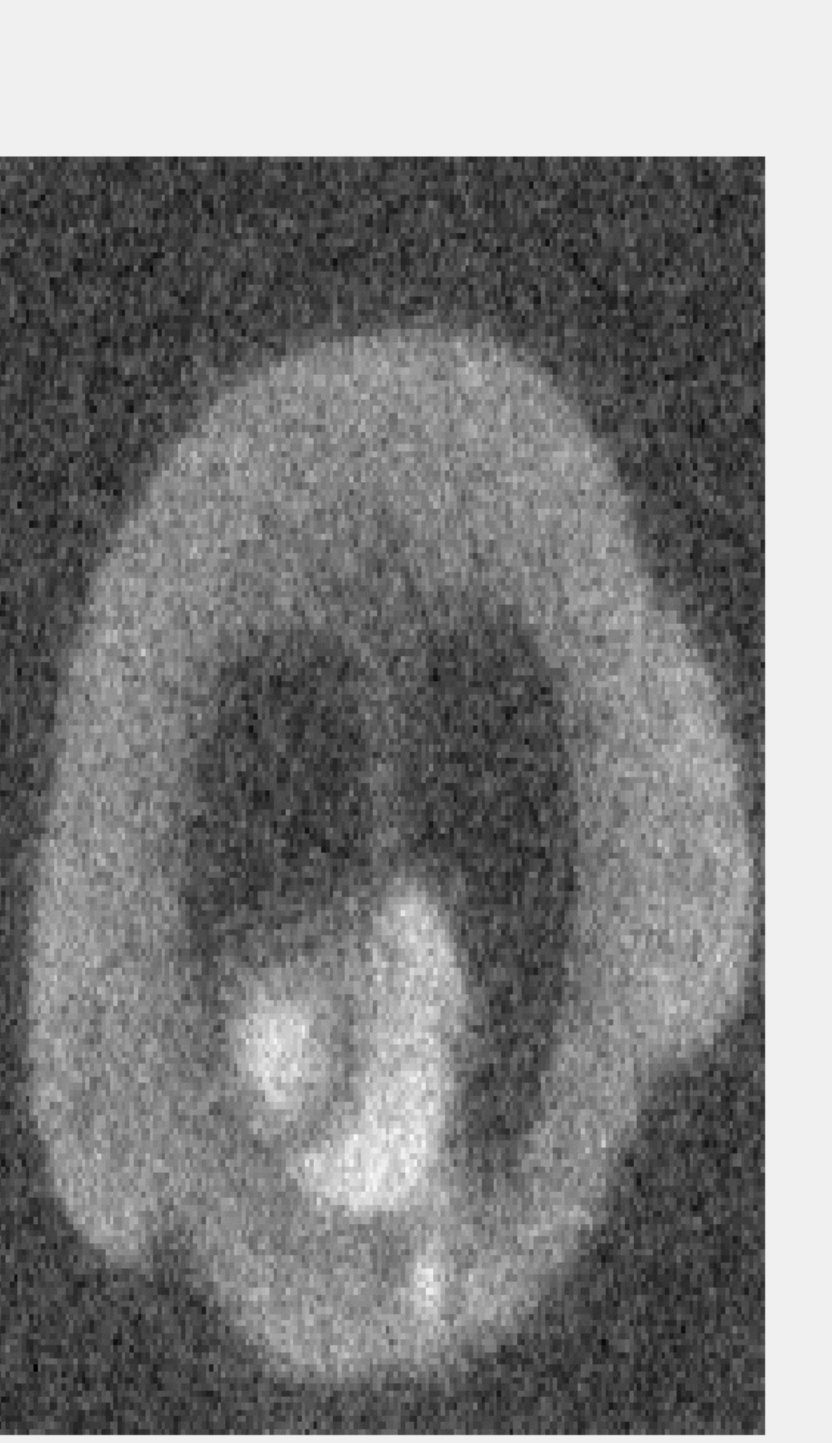}
    }
    
    \resizebox{.95\textwidth}{!}{\includegraphics[angle = 270, page = 2, trim={4.5cm 0 0 4cm },clip]{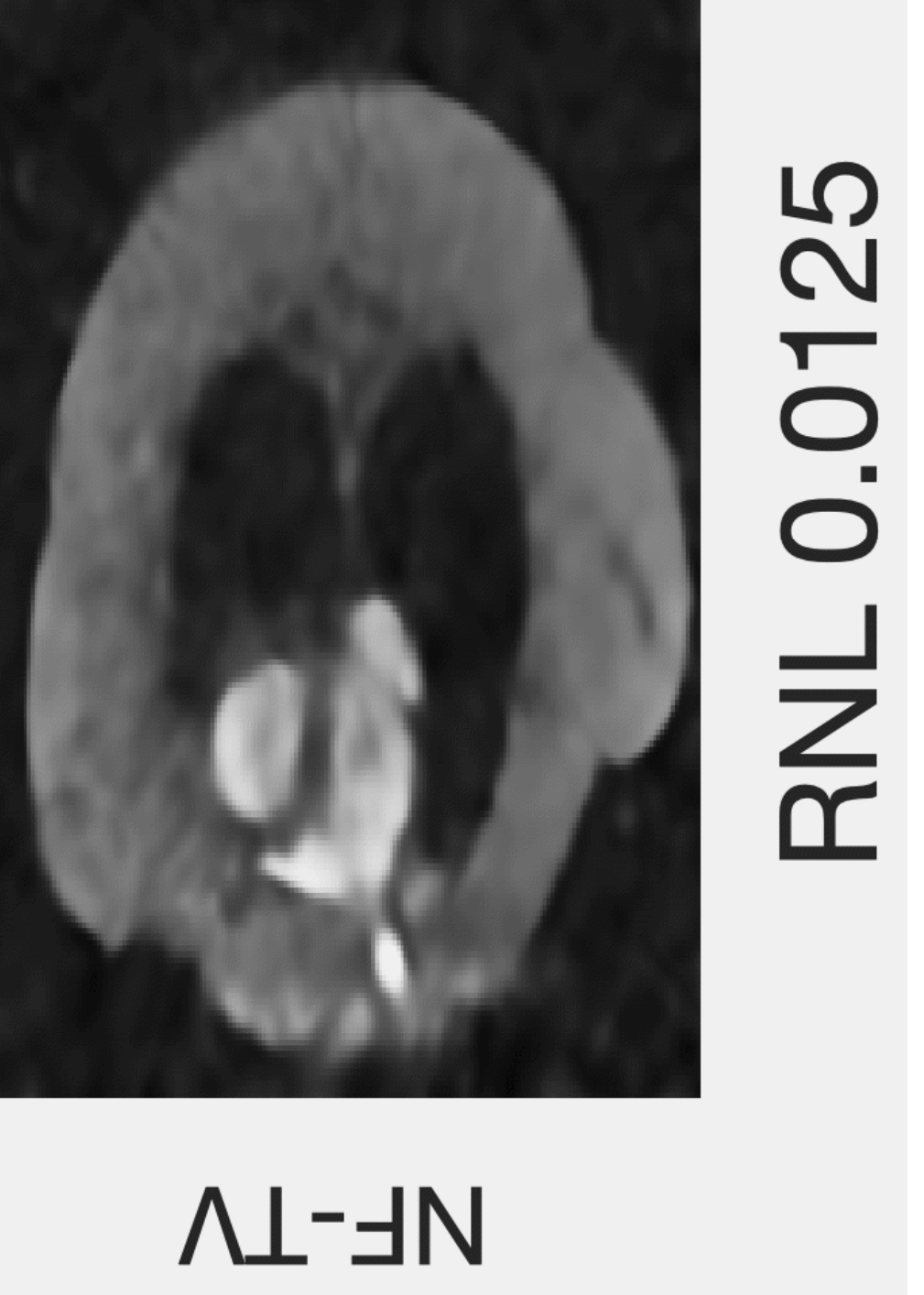}\includegraphics[angle = 270, page = 2, trim={4.5cm 3.5cm 0 4cm },clip]{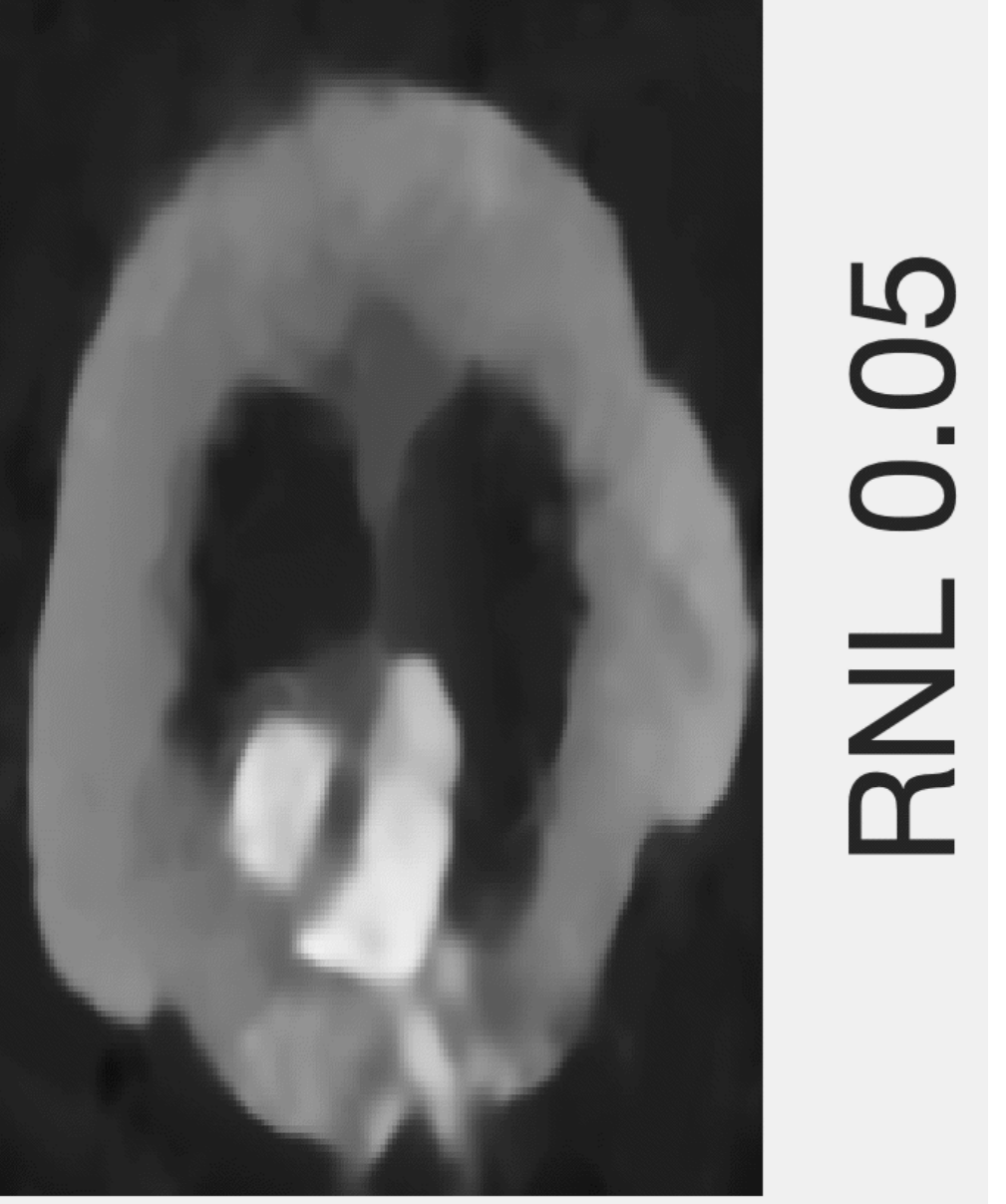}\includegraphics[angle = 270, page =2, trim={4.5cm 3.5cm 0 4cm },clip]{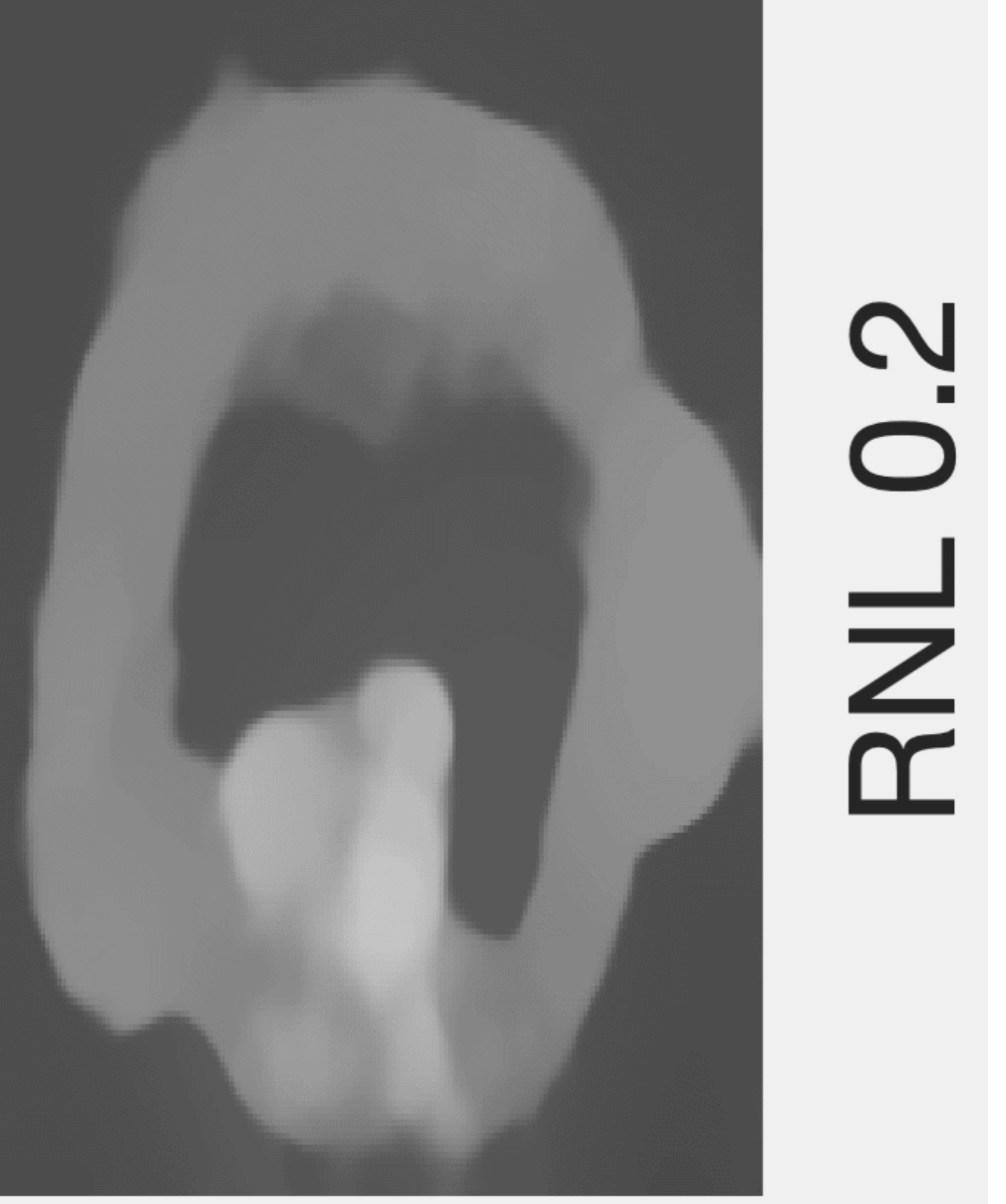} \hspace{4cm}  \includegraphics[angle = 270, page = 5, trim={4.5cm 3.5cm 0 4cm },clip]{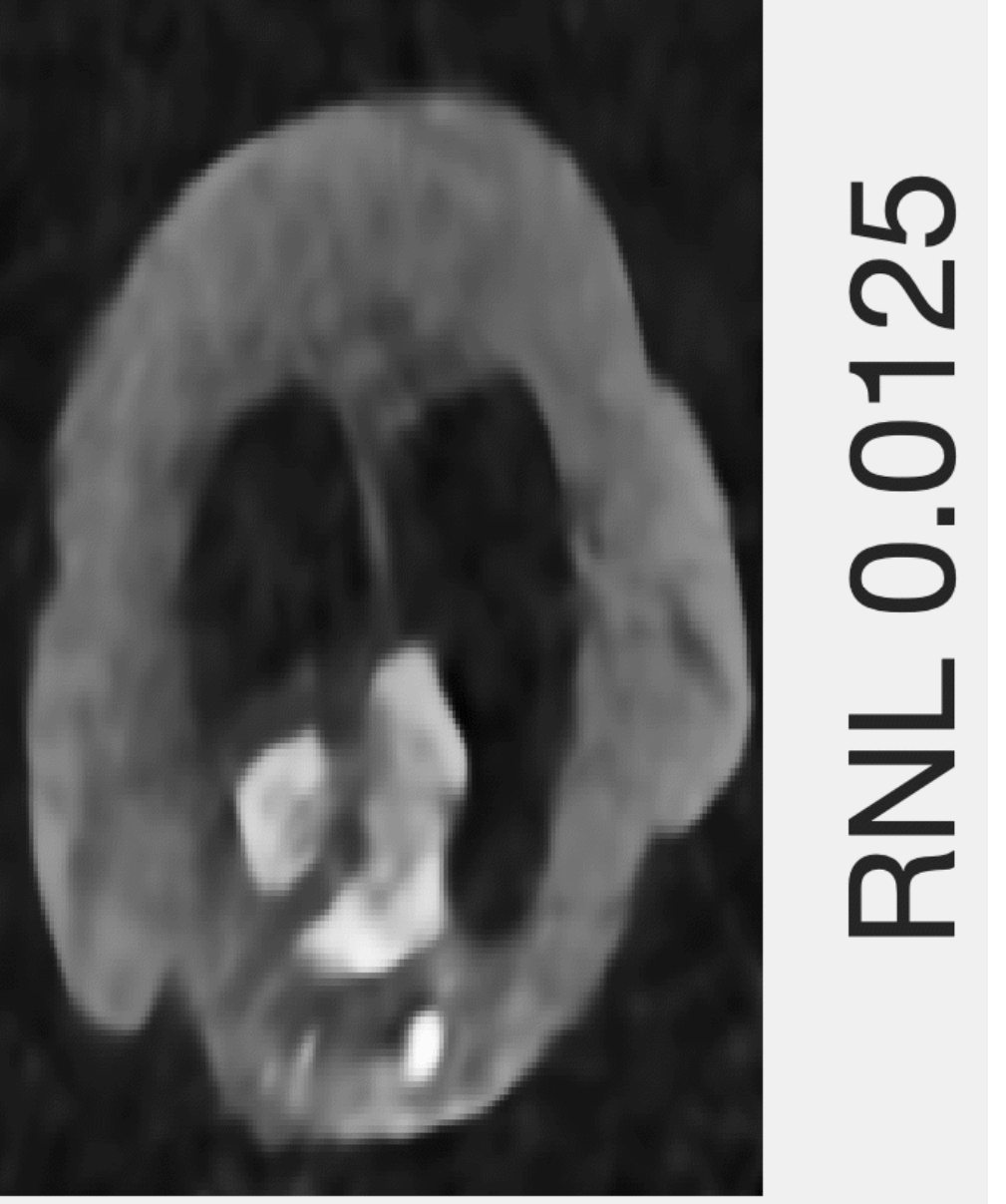}\includegraphics[angle = 270, page = 5, trim={4.5cm 3.5cm 0 4cm },clip]{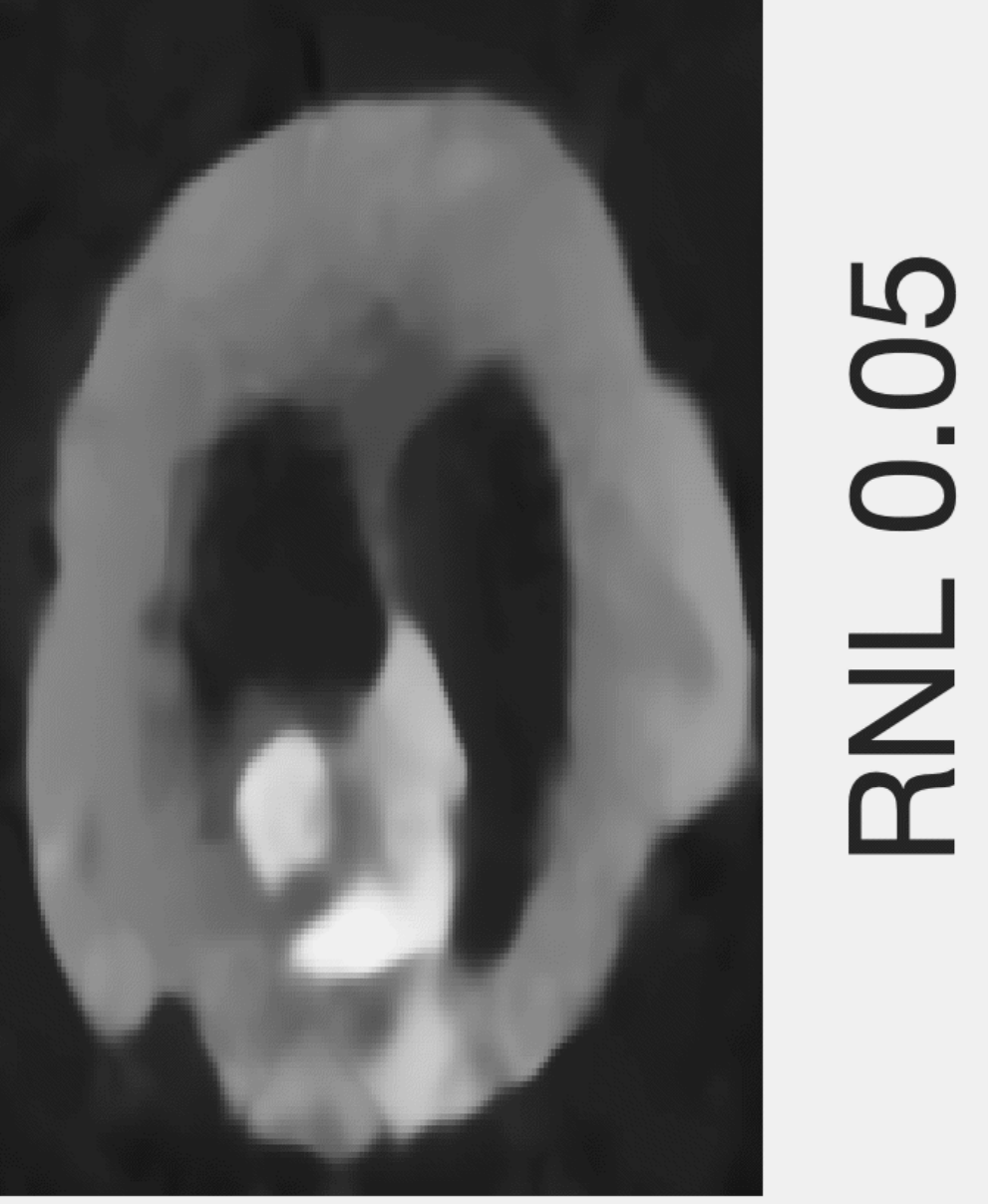}\includegraphics[angle = 270, page = 5, trim={4.5cm 3.5cm 0 0 },clip]{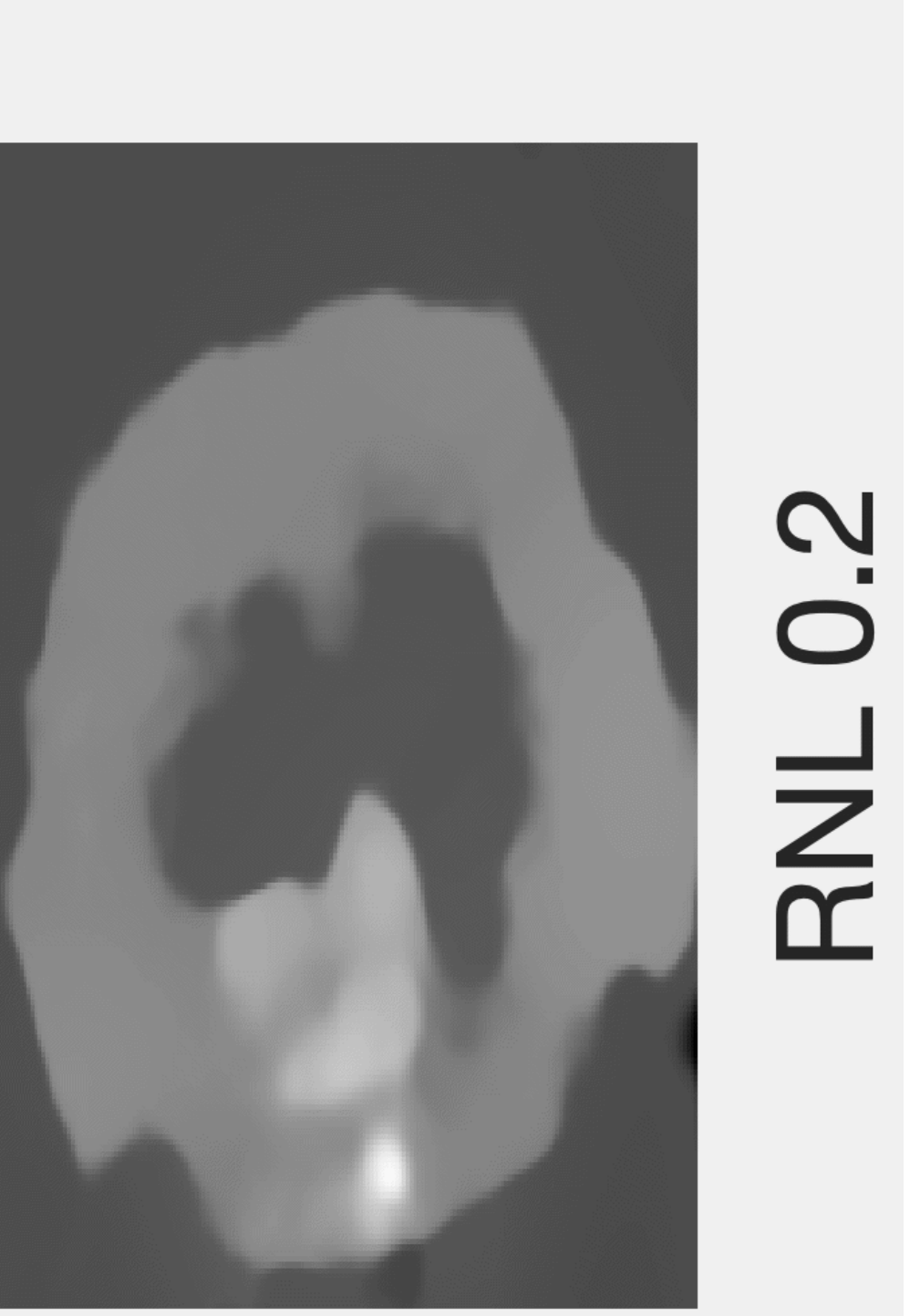}
    }

    \caption{Dynamic reconstruction problem: Pixel-wise TV, pixel-wise NN and neural field TV reconstructions with relative noise levels (RNL) $0.0125$, $0.05$, and $0.2$. The neural field displays robust performance in the presence of noise, while the performance of the classical methods degrades quickly with noise. An animation of the dynamic reconstructions comparing the PW-TV, PW-NN, and NF-TV methods for different noise levels  is available in the supplemental multimedia materials (Video 5).}
    \label{fig:psnr_anims}
\end{figure*}

\begin{figure*}
    \centering
    \includegraphics[width = \textwidth]{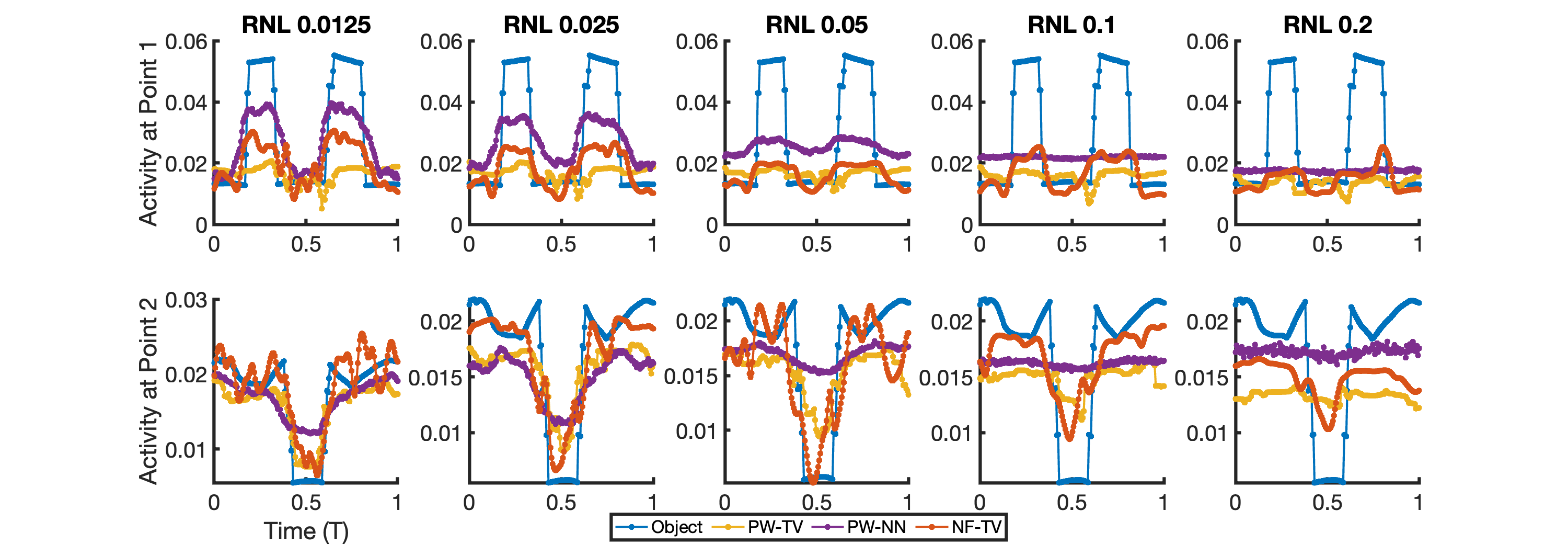}
    \caption{Dynamic reconstruction problem: Time-activity curves for varying relative noise levels.The neural field leads to dynamic smooth representation at higher noise levels while the classical methods are highly sensitive to noise and prioritize a static representation.}
    \label{fig:psnr_activity}
\end{figure*}

The second experiment reconstructed the image with the forward operator given by the CRT with four views per frame ($S=4$) and a varying noise level. The relative noise levels considered were 0.0125, 0.025, 0.05, 0.1, and 0.2. Figure \ref{fig:psnr_anims} displays the results of the reconstruction with relative noise levels $0.0125$, $0.05$, and $0.2$. Figure \ref{fig:psnr_activity} displays the activity curves for the two selected points. Table \ref{tab:psnr_acc} displays the RRMSE, SSIM, optimal regularization values for each reconstruction, and the rank of the nuclear norm reconstruction. 

The neural field approach led to reconstructions with low levels of variation in space and smoothly varying dynamic features. The classical reconstructions quickly degraded at higher noise levels, leading to images with high levels of variation in space and erratically changing dynamic features. Similar to the previous experiment, the neural field representation significantly reduced the number of stored parameters. The neural field representation required 86K parameters for each reconstruction and the pixel-wise TV approach required 7.2M parameters. The pixel-wise nuclear norm approach required between 40K (large noise) and 602K (small noise) parameters depending on the noise level.

\begin{table}
    \centering
    \caption{Accuracy metrics and optimal regularization parameters for different relative noise levels. The number of parameters for the PW-TV and NF-TV are set to 7.6M and 86K, respectively independently of the noise}
    \label{tab:psnr_acc}
    \begin{tabular}{c|ccccc}
    Relative noise level & 0.0125 & 0.025& 0.05 & 0.1 & 0.2\\ 
    \hline
    RRMSE PW-TV & 0.2937  &  0.3202  &  0.3441   & 0.3935  &  0.4851 \\ 
    RRMSE PW-NN & \textbf{0.2183}   & \textbf{0.2569}  &  0.3028  &  0.3767  &  0.4120 \\ 
    RRMSE NF-TV &  0.2463  &  0.2574  &  \textbf{0.2775} &   \textbf{0.3184}  &  \textbf{0.3803} \\  
    \hline
    SSIM PW-TV & \textbf{0.6195}  &  0.5628  &  0.4964  &  0.3513  &  0.1846 \\ 
    SSIM PW-NN & 0.4452  &  0.3153  &  0.2153  &  0.1429  &  0.0995 \\ 
    SSIM NF-TV & 0.5731  &  \textbf{0.6308}  &  \textbf{0.6088}  &  \textbf{0.5528}  &  \textbf{0.5550}\\  
    \hline
    Optimal $\gamma$ PW-TV & 12.5 &  6.25 & 3.25 & 1.75 & 0.875\\ 
    Optimal $\gamma$ PW-NN & 0.0707 & 0.0354 & 0.0250 & 0.0125 & 0.0088 \\ 
    Optimal $\gamma$ NF-TV  & 0.25 & 1 & 1 & 1 & 0.5 \\
    \hline 
    PW-NN Rank $r$ & 15 & 12 & 3 & 2 & 1\\
    PW-NN parameters &  602K & 482K & 120K & 80K & 40K\\
\end{tabular}
\end{table}

\section{Conclusion}\label{sec:conclusion}
This work presents a novel optimization-based framework for dynamic image reconstruction using neural fields. Neural fields are a special class of neural networks that provide accurate, differentiable, non-parametric representations of continuous objects, such as images, videos, and audio signals. The key advantages of the proposed approach are 1) relatively accurate reconstruction from severely undersampled data exploiting spatiotemporal redundancies in the object features, 2) ease of implementation by leveraging automatic differentiation, 3) computational efficiency and reduced memory requirements. The feasibility of the proposed approach was demonstrated with three virtual imaging studies of a cardiac-torso murine numerical phantom.

The first experiment displayed that a neural field can accurately represent a dynamic object with smaller memory requirements than semiseparable approximations. In fact, the neural field representation required 3 times fewer parameters than a semiseparable approximation with a comparable structural self-similarity index and 5 times fewer parameters than a semiseperable approximation with a comparable root mean square error. Furthermore, the neural field approach achieved a 3 fold reduction in root mean square error and a higher structural similarity than the semiseparable approximation with a comparable number of parameters.

The second experiment considered a dynamic image reconstruction problem from limited-views per frame circular Radon transform data with varying levels of data incompleteness. This experiment demonstrates that the proposed neural field image reconstruction strategy is well suited for image reconstruction with undersampled or incomplete data and outperforms classical methods in terms of structural self-similarity index and mean square error. Furthermore, it allows for a significant reduction (between 4X and 66X) in the memory requirements compared to a semiseparable representation of the reconstructed dynamic image.

The third experiment considered a dynamic image reconstruction problem from circular Radon transform data with limited views and varying noise levels. The neural field approach outperformed classical methods at all noise levels and was less sensitive to increasing levels of measurement noise. This stability with respect to noise may be due to the implicit regularization introduced by using a neural field; that is, the neural network is biased towards learning certain classes of functions.

In summary, this work established the feasibility of using neural field representation for dynamic image reconstruction problems and demonstrated computational advantages and image quality improvements with respect to classical dynamic image reconstruction algorithms for a stylized dynamic photoacoustic tomography imaging problem in two spatial dimensions.
Future work will explore further optimizations of the network architecture and reconstruction algorithms. In particular, the number of partitions and the layers in the POUnet was found with manual trial-and-error. In the future, there is a plan to use advanced neural architecture search (NAS) algorithms to optimize the network structure \cite{ZophLe16,ElskenMetzenHutter19} and apply deep image priors to further address data incompleteness \cite{ulyanov2018deep}. Ultimately, the goal is to extend the proposed method to 3D dynamic imaging of small-animal models using photoacoustic tomography.

\section*{Appendix A \\ Code and Data Availability}
The python code implementing the proposed dynamic image reconstruction method using neural field is available from \cite{Lozenski2022code} under GPLv3. The absorption coefficient and induced pressure maps of the dynamic object used in the numerical studies are available from \cite{Lozenski2022phantom} under CC-0 public dedication.

\section*{Appendix B \\  Supplementary Multimedia Material}
Supplementary multimedia materials include the following. Video 1 depicts the dynamic object (optical absorption and pressure distribution) shown in Fig. \ref{fig:moby}. Video 2 depicts the semiseparable and neural field representations of the object in the embedding numerical studies (c.f. Fig. \ref{fig:Representation_Test}). Video 3 represents the time evolution of the POUnet partition in the embedding numerical studies(c.f. Fig. \ref{fig:partitions}). Finally, Videos 4 and 5 depict the reconstructed objects in the two reconstruction studies (c.f. Fig.s , \ref{fig:vpt_anims}, and \ref{fig:psnr_anims}).

\bibliographystyle{IEEEtran}
\bibliography{local, references}

\end{document}